\documentclass[prr,aps,twocolumn,10pt,superscriptaddress,notitlepage,longbibliography]{revtex4-2}
\usepackage[margin=1.54cm]{geometry}
\usepackage[caption=false]{subfig}
\usepackage{graphicx}
\usepackage{amsmath,mathtools}
\usepackage{amssymb}
\usepackage{epstopdf}
\usepackage{siunitx}
\usepackage{color}
\usepackage[ngerman,english]{babel}
\usepackage{float}
\usepackage{bbold,bm}

\usepackage{hyperref}
 \hypersetup{
     colorlinks=true,
     linkcolor=blue,
     filecolor=blue,
     citecolor = magenta,      
     urlcolor=red,
     }

\usepackage{braket}
\definecolor{orange}{rgb}{1,0.5,0}

\usepackage{letltxmacro}
\LetLtxMacro{\ORIGselectlanguage}{\selectlanguage}
\makeatletter
\DeclareRobustCommand{\selectlanguage}[1]{%
  \@ifundefined{alias@\string#1}
    {\ORIGselectlanguage{#1}}
    {\begingroup\edef\x{\endgroup
       \noexpand\ORIGselectlanguage{\@nameuse{alias@#1}}}\x}%
}
\newcommand{\definelanguagealias}[2]{%
  \@namedef{alias@#1}{#2}%
}

\makeatother

\definelanguagealias{en}{english}
\definelanguagealias{EN}{english}
\definelanguagealias{eng}{english}
\definelanguagealias{de}{ngerman}

\usepackage{pythonhighlight}

\begin{document}
	
\title{Modelling quantum light-matter interactions
in waveguide-QED with retardation and a time-delayed feedback:  matrix product states versus
a space-discretized waveguide model}
\date{\today}
\author{Sofia Arranz Regidor}
\email{18sar4@queensu.ca}
\affiliation{Department of Physics, Engineering Physics and Astronomy,
Queen's University, Kingston, ON K7L 3N6, Canada}
\author{Gavin Crowder}
\affiliation{Department of Physics, Engineering Physics and Astronomy,
Queen's University, Kingston, ON K7L 3N6, Canada}
\author{Howard Carmichael}
\affiliation{The Dodd-Walls Centre for Photonic and Quantum Technologies, Department of Physics, University of Auckland, Private Bag 92019, Auckland, New Zealand}
\author{Stephen Hughes}
\affiliation{Department of Physics, Engineering Physics and Astronomy,
Queen's University, Kingston, ON K7L 3N6, Canada}

\begin{abstract}
We present two different methods for modelling
non-Markovian quantum light-matter interactions
in waveguide QED systems, using
matrix product states (MPSs) and a space-discretized waveguide (SDW) model.
After describing the general theory and implementation
of both approaches, we compare and contrast these methods directly
on three topical problems of interest in waveguide-QED,
including
(i)  a two-level system (TLS)
coupled to an infinite (one-dimensional) waveguide, (ii) 
a TLS coupled to a terminated waveguide with
a time-delayed coherent feedback,
and (iii) two spatially separated TLSs coupled within an infinite waveguide. 
Both approaches are shown to efficiently model multi-photon nonlinear dynamics  in highly non-Markovian regimes, and we  highlight the advantages and disadvantages of these methods for modelling waveguide QED interactions, including their implementation in {\sc Python}, computational run times, and ease of conceptual understanding.
We explore both vacuum dynamics as well as regimes of strong optical pumping, where a weak excitation approximation cannot be applied. The MPS approach 
scales better when modelling multi-photon dynamics and long delay times, and explicitly includes non-Markovian memory effects. In contrast, the  SDW model accounts for non-Markovian effects through space discretization, and solves Markovian equations of motion, yet  rigorously includes the effects of retardation. The SDW model, based on an extension of recent collisional pictures in quantum optics, is solved through quantum trajectory techniques, and can more easily add in additional dissipation processes, including off-chip decay and TLS pure dephasing. The impact of these processes is shown directly on feedback-induced population trapping and TLS entanglement between spatially separated TLSs.
\end{abstract}

\maketitle

\section{Introduction}
\label{sec:introduction}

Waveguide quantum electrodynamics (QED) deals with  quasi one-dimensional (1d) systems that couple
atoms and photons through waveguide geometries, where the atoms
and two-level systems (TLSs) are coupled to a continuum of  quantized field modes \cite{Hughes2004,PhysRevA.76.062709,PhysRevLett.98.153003,Zheng2010,Witthaut_2010,PhysRevA.83.063828,PhysRevLett.106.053601,PhysRevLett.113.263604,Calaj2016,cajitas}.
Such systems can also result in a significant enhancement of the spontaneous emission rate when coupled to slow light waveguide modes, with very little off-chip decay~\cite{PhysRevB.75.205437,PhysRevLett.101.113903,PhysRevX.2.011014}.
Many methods have been used to study light-matter interactions in quantum optical (e.g., Ref.~\cite{gardiner_zoller_2010}). However, only a few of them can be used  to model non-Markovian systems in the nonlinear regime, in which time delays
and retardation must be taken in account \cite{cajitas,Droenner2019,crowder_quantum_2020}. As the considered Hilbert space grows, the problems become very challenging from a computational perspective, leading to restricted analytic approaches \cite{Dorner2002,Tufarelli2013,PhysRevLett.110.013601,Nemet2019} or new 
approaches to model the complex system dynamics \cite{Grimsmo2015,Whalen_2017,PhysRevA.98.063832,crowder_quantum_2020}.
Many of the methods of choice can be overly complex, frequently lack an intuitive description for their implementation,
are too restrictive in what problems they can solve, or do not scale well numerically for a range of problems in waveguide QED.

Apart from fundamental interest in quantum optics
and light-matter interactions, 
the potential impact of exploiting waveguide QED and {\em coherent feedback} in quantum optics has  diverse applications, such as the possibility to more precisely control quantum optical (QO)  systems \cite{Kubanek2009,Gillett,Brandes,Balouchi2017,Calajo2019}, including improving the creation and control of the quantum entanglement \cite{Yao2009,Hein2016}. From a practical viewpoint, coherent feedback systems can now be 
realized in chip-based semiconductor systems with
semiconductor quantum dots (QDs) \cite{Buckley_2012,Matthiese,Heinze2015,Trschmann2019} and superconducting circuits  \cite{Gu2017,PhysRevLett.120.140404,Kannan2020}.

\begin{figure} [htb!]
\centering
\subfloat[Schematic of a single TLS embedded in an infinite waveguide. We assume the TLS couples asymmetrically to the waveguide, with rates $\gamma_L$ and $\gamma_R$.
]
{%
  \includegraphics[clip, width=0.9 \columnwidth,]{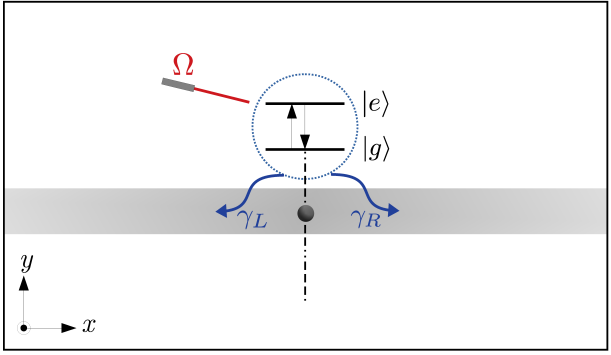}%
  }
\hspace{1pt}
\subfloat[Schematic of a single TLS embedded in a terminated waveguide,
with a time-delayed coherent feedback from a mirror.
The total length of one round trip is $L_0$, which causes a delay or memory time of $\tau$. We also consider a CW pump field with a Rabi frequency $\Omega$.
]{%
  \includegraphics[clip, width=0.9 \columnwidth,]{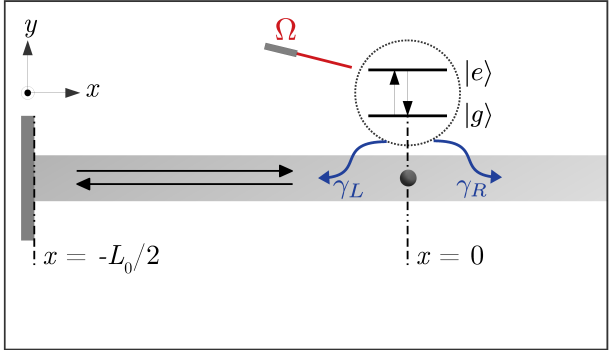}%
  }
\hspace{1pt}
\subfloat[Schematic of a two TLS embedded in an infinite waveguide
with a finite delay length/time between them. The total delay length is again $L_0$, causing a delay time  (between quantum emitters) of $\tau$. Both TLSs can be pumped separately. ]{%
  \includegraphics[clip, width=0.9 \columnwidth]{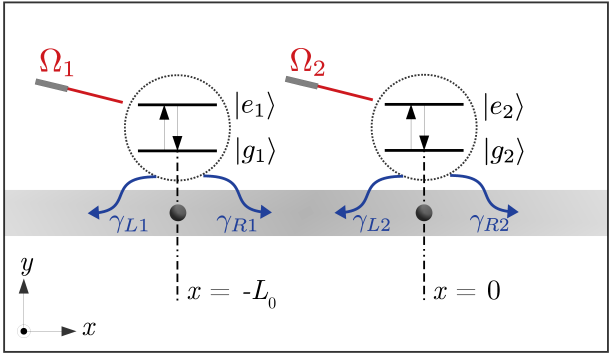}%
 }
 \captionsetup{justification=centering}
 \caption{Three systems of interest in waveguide QED, coupling one or two TLS with waveguides with and without a time-delayed coherent feedback and a possible CW pump field. 
 }
\label{schematics2}
\end{figure}

The aim of this paper is
to present and compare two 
powerful, but quite different, approaches for solving several classes of non-Markovian feedback and waveguide QED, which can be applied to 
study both vacuum dynamics and nonlinear (i.e., multi-photon)  excitation regimes. Specifically, these methods are based on: (a)  matrix product states (MPSs)~\cite{yang_matrix_2018,vanderstraeten_tensor_2017,Naumann2017,Droenner2019} and (b) a new space-discretized waveguide (SDW) model, using a collision approach for the 
waveguide environment~\cite{Ciccarello2017,PhysRevA.100.052113,GavinThesis,Cilluffo2020}.
The collision model is solved explicitly by allowing for 1 or 2 photons in the waveguide, which helps to show when several photons need to be included in the model, while the MPS model is not restricted in the number of photons. We present the theory of both of these approaches,
as well as their computational implementation, and subsequently investigate several important QED waveguide systems in which to compare them directly. 
With the SDW model, 
we also show the importance of including additional decoherence processes (such as pure dephasing), that are frequently difficult or impossible to account for in many of the current theoretical approaches
to waveguide QED, including MPSs.
While there are several  papers on MPSs for waveguide QED, they often lack the technical details for ease of implementation and understanding (or can be presented in an overly complex and intimidating manner), and the specific pros and cons are often not well documented
when compared directly to other alternative approaches.

For the main feedback systems of interest in our study, three systems are studied: (i) a two-level system (TLS) coupled to an infinite  1d waveguide, which can also be compared with the known analytical solution in the weak excitation approximation. In (ii), a TLS coupled to a truncated 1d waveguide with a time-delayed coherent feedback; this case introduces complex non-Markovian behavior as the time-delay from a distant mirror must include retardation. Finally, in (iii), two TLSs coupled within an infinite 1d waveguide are modelled, where a non-negligible time-delay is present between each TLS (see Fig.~\ref{schematics2}).
We also consider
optical pumping with a continuous wave (CW) field, though the extension to include pulsed
excitation in both models is straightforward. 
Some limiting cases for each method are also highlighted, e.g., MPSs have an advantage for modelling
long feedback delays and strong pump fields (multi-photons) and can be solved explicitly up to $N$ or $2N$ photons in the waveguide for the case with one or two TLSs respectively. 
While 
the SDW model is significantly easier to implement,
and can include additional important dissipation processes such as pure dephasing and off-chip 
spontaneous emission decay.
In both models, we show how to tackle the important non-Markovian problem
of modelling TLSs with a time-delayed feedback, including the role of multi photon scattering, which sets a limit on
the phase matching condition from the mirror for population trapping.

The rest of our paper is organized as follows:
In Sec.~\ref{sec:MPS}, we first present the MPS method, starting with a general introduction  for implementing the technique for waveguide QED systems. The MPSs form a practical application of tensor networks~\cite{orus_practical_2014} for studying 1D many-body quantum systems, in which the size of Hilbert space makes it difficult (or impossible) to solve  with other methods such as Quantum Monte Carlo. The essence
of the MPS method is exploited by limiting the entanglement between two parts of the entire system, which reduces considerably the Hilbert space considered and, thus, the computational cost. For our purpose,
we wish to solve the appropriate time-dependent
field operators and density matrix for the open waveguide system.
A diagrammatic representation for the MPS approach is presented in \ref{subsec:diag}, followed by  a description of how the MPS approach can be implemented
for various waveguide QED systems in \ref{subsec:hamil}. Specifically, in exploiting MPSs, a waveguide QED system can be considered as a many-body system in one dimension  \cite{cajitas,droenner_out--equilibrium_2019},
where 
the relevant waveguide modes are the ones close to the frequencies of interest. We will see that these continuous waveguide modes have to be discretized, and the basis will be transformed  to a time-discrete picture in order to solve the problem efficiently. 
Next, we introduce the Matrix Product Operators (MPOs) that will be used for solving our problem of interest (\ref{subsec:oper}). 
In \ref{subsec:comp}, MPS theory is applied to evolve our system using the time evolution operator for each time interval $\Delta t$, and example observables are computed. The advantage of this method lies in the fact that it only needs to be applied in a specific part of the MPS, which reduces the size of the computational space (Hilbert space) and optimizes the efficiency of the calculations~\cite{eduardo_one-dimensional_2017}.
In the last part of this section (\ref{subsec:impl}), we give a brief description of how the approach is implemented in 
{\sc Python}.

Next, in Sec.~\ref{sec:SDW}, we  present an alternative approach to MPSs, which we term the SDW model. This approach extends the recent approach introduced by Whalen~\cite{PhysRevA.100.052113}, and is substantially easier for a researcher to implement computationally than MPSs. In essence, the SDW model discretizes the waveguide field in the spatial domain over the waveguide length of interest and follows a ``collisional" model for the interaction with the quantum optic system of interest \cite{Brun2002,Kretschmer2016,Ciccarello2017,Cilluffo2020}. In contrast to MPS theory, the SDW model has a simple and intuitive implementation, 
without all the added complexities that come with MPSs and tensor networks in general.  We first show how to implement the SDW model for an open waveguide and describe the algorithm for evolving the waveguide in this regime. 
Then, in Sec.~\ref{sec:SDW_Schemes} we explain how to derive the interaction Hamiltonian between a general QO system and waveguide in the SDW picture as well as present the interaction Hamiltonians for our three schemes of interest. In Sec.~\ref{sec:Lindblad}, we extend the current SDW model to include additional Lindblad output channels following the formalism of quantum trajectory (QT) theory and, in Sec.~\ref{SDW_Python}, we discuss the computational implementation of the model.

In Sec.~\ref{sec:results}, results are shown and compared for both models, showing the advantages and disadvantages of each approach.
With our selected examples, we show how both models can accurately capture quantum light-matter interaction 
in waveguide QED, including the role of multi-photon
interactions with CW pumping. While the MPS approach does not make any approximation about maximum number of photons in the loop (or the waveguide),
the SDW model is explicitly solved to either a 
one-photon-in-the-loop or 
two-photons-in-the-loop approximation. This allows us to compare these methods directly to determine when one, two or even more photons need to be treated explicitly at the system level in quantum optics. Even under fairly extreme conditions, such as with very strong pumping fields and long delay lengths for coherent feedback, we find that both models agree extremely well under most situations. 
The SDW model
can also more easily add in additional and realistic dissipation processes, including 
off-chip decay and pure dephasing; although routinely neglected in most coherent feedback studies to date,  we show directly how such background decay processes influence well known feedback control phenomena, such as photon TLS population trapping and entanglement between two
spatially separated TLSs.
These additional processes are especially
important in modelling realistic quantum dots, where pure dephasing and electron-phonon scattering  are known to be key processes to 
understand~\cite{PhysRevB.98.045309,
IlesSmith2017,
Kuhlmann2013,
PhysRevLett.104.017402,
PhysRevLett.118.253602,
RevModPhys.87.347,
PhysRevB.66.165312,
PhysRevLett.91.127401,
PhysRevB.63.155307,Trschmann2019}.
We also show how the SDW waveguide model 
and QT theory leads to
delayed conditioning for single trajectories.
Finally, using MPSs, we investigate the entanglement entropy
for a two TLS system, and show the role of retardation (delay length).
Conclusions and closing discussions are presented  in Sec.~\ref{sec:conclusions}.

\section{System Hamiltonians}
\label{sec:Hamiltonian}

\subsection{Scheme (i): Single two level system in an infinite waveguide}
\label{subsec:scheme1}
First, we introduce the Hamiltonian modelling the interaction for one TLS coupled to an infinite waveguide (see Fig.~\ref{schematics2}(a)), in the rotating wave approximation:  
\begin{align}
    H& = H_{\rm TLS}+ H_{\rm pump} + H_{\rm W} +  H_{\rm I},
    \label{hamiltoniantotal}
\end{align}
where 
\begin{equation}
\label{HTLS}
    H_{\rm TLS}= \omega_0 \sigma^+\sigma^-,
\end{equation}
is the free term for the TLS,
with $\omega_0$ the resonance energy
and $\sigma^\pm$ the Pauli operator. The second term,
\begin{equation}
\label{pump}
    H_{\rm pump}=  \Omega_0(t)  (e^{-i\omega_{\rm L} t } \sigma^+ + e^{+i\omega_{\rm L} t}\sigma^- ),
\end{equation}
allows for a possible
pumping term for the TLS,
and $\omega_{\rm L}$
is the frequency of the laser drive;
below we will convert this to the standard form appropriate for a rotating wave approximation.
Natural units are used ($\hbar=1$) throughout our paper, and a
CW drive is considered, where $\Omega_0(t)=\Omega_0$; however, both techniques presented below can easily work with a time-dependent drive. 
The waveguide term,
\begin{equation}
    H_{\rm W} = \sum_{\alpha=L,R} \int_{-\infty}^{\infty} d\omega  \omega b_\alpha^\dagger (\omega)b_\alpha(\omega),
\end{equation}
is the free term for the waveguide modes (left and right propagating),
and
\begin{equation}
\label{HI}
    H_{\rm I}=  \int_{-\infty}^{\infty} d\omega \left[ \left(\kappa_L(\omega)\sigma^+ b_L(\omega) + \kappa_R(\omega)\sigma^+ b_R(\omega) \right) + \rm H.c \right],
\end{equation}
describes the TLW-waveguide interaction,
and H.c. is the Hermitian conjugate.
The field operators obey the usual commutation rules
for bosons, namely $[b_i(\omega),b_j^\dagger(\omega')]=\delta_{i,j}\delta(\omega-\omega')$.

Transforming to the interaction picture with respect to the TLS-
and waveguide-free Hamiltonians, and moving to a rotating frame at the frequency $\omega_L=\omega_0$, then 
 \begin{equation}
\begin{split}
     H&= \Omega_0 ( \sigma^+ + \sigma^- ) + 
     \int_{-\infty}^{\infty} d\omega \left[ \bigg( \kappa_L(\omega)\sigma^+ b_L(\omega) \right . \\  
     & \left . + \kappa_R(\omega)\sigma^+ b_R(\omega)  e^{-i(\omega - \omega_0)t} \bigg) + \rm H.c.\right].
\end{split}
\end{equation}

Next, we
define the time-dependent operators
\begin{equation}
    b_\alpha (t)=\frac{1}{\sqrt{2\pi}} \int_{-\infty}^{\infty} d\omega b_ \alpha(\omega) e^{-i(\omega - \omega_{0})t},
\end{equation}
and the TLS-waveguide 
decay rate,
\begin{equation}
\label{g}
\gamma_\alpha = {2\pi}\kappa_\alpha^2(\omega_0),
\end{equation}
where $\alpha=L,R$, and $\kappa_\alpha$ is assumed to be frequency independent as most of the coupling is close to TLS frequency; thus we have replaced
 $\kappa_\alpha(\omega)$ by 
 $\kappa_\alpha(\omega_0)=\sqrt{{\gamma_\alpha}/{2\pi}}$. The time-dependent field operators satisfy: $[b_i(t),b_j^\dagger(t')]=\delta_{i,j}\delta(t-t')$.
Subsequently, the interaction Hamiltonian can be written as 
\begin{align}
\label{int1}
     H_{\rm I}&=  \sqrt{\gamma_L} \left (\sigma^+ b_L(t)  +  \sigma^- b_L^\dagger(t)\right)
     \nonumber \\ &+
     \sqrt{\gamma_R} \left(\sigma^+ b_R(t)  +  \sigma^- b_R^\dagger(t)\right).
\end{align}

In the case where  there is equal coupling rates to both sides of the waveguide, $\gamma_L=\gamma_R=\gamma/2$, then one can introduce a single collective operator for the two waveguide modes, $b(\omega)$, so that
\begin{equation}
    b(t)=\frac{1}{\sqrt{2\pi}} \int_{-\infty}^{\infty} d\omega b(\omega) e^{-i(\omega - \omega_{0})t},
\end{equation}
and the total TLS-waveguide
decay rate
is 
$\gamma = {2\pi}\kappa^2(\omega_0)$.
Thus, with symmetric coupling, the Hamiltonian is now
 \begin{equation}
\begin{split}
     &H= \Omega_0 ( \sigma^+ + \sigma^- ) +\sqrt{\gamma} (\sigma^+ b(t)  +  \sigma^- b^\dagger(t)).
\end{split}
\end{equation}
Symmetry breaking, for example, can be achieved 
in photonic crystal waveguides, using
spin charged quantum dots coupled
to points of circular polarization~\cite{leFeber2015,PhysRevLett.115.153901,Sllner2015,Barik2018}, which can give rise to a number of interest effects in  chiral 
waveguide QED~\cite{Lodahl2017}.
Indeed, chiral field interactions can be found in many
photonic waveguide and resonator systems~\cite{PhysRevA.85.061801,Coles2016,MartinCano2019,PhysRevApplied.10.054069,Petersen2014}.

\subsection{Scheme (ii): Single two level system in a half open waveguide with a time-delayed coherent feedback}
\label{subsec:scheme2}

For our second waveguide system, we present the Hamiltonian for modelling the interaction of one TLS in a semi-infinite (half open) waveguide (see Fig.~\ref{schematics2}(b)), that is, when a mirror and coherent feedback is present. The Hamiltonian is again composed of a single  TLS and a pumping term as in Eq.~\eqref{hamiltoniantotal},  where $H_{\rm TLS}$ and $H_{\rm pump}$ follow Eq.~\eqref{HTLS} and Eq.~\eqref{pump}, respectively. 
The free term for the  waveguide modes  is now
\begin{equation}
\label{HW}
    H_{\rm W}=\int_{-\infty}^{\infty} d\omega \omega b^\dagger(\omega) b(\omega),
\end{equation}
which consists of a linear combination of modes that propagate to the left and to the right.
Note that $b(\omega)$ here 
 should not be confused with the
field operator we introduced in scheme (i) in the case of symmetric coupling; rather it is a generic field operator for the complete waveguide system.

Since the waveguide bath is modified by the feedback loop,  the interaction Hamiltonian takes the form:
\begin{equation}
     H_{\rm I} = \int d\omega \left[ G_{\rm fback}(\omega) b (\omega) \sigma^+ +\rm H.c. \right],
\label{feedbackhamil}
\end{equation}
which describes the interaction between the TLS and the mirror-modified reservoir, and $G_{\rm fback}(\omega)$ accounts for the boundary condition of the terminated side of the waveguide. The bath coupling term $G_{\rm fback}$ can in principle be solved formally
using scattering theory, e.g., using photon Green functions derived for a particular waveguide-cavity system~\cite{PhysRevB.80.165128}. Here we will adopt a simple model
to account for feedback from a perfect mirror (no losses):
\begin{align}
    G_{\rm fback}(\omega)&=
    \frac{1}{\sqrt{2\pi}} \left( 
    \sqrt{\gamma_L}e^{i(\omega \tau - \phi_{\rm M})/2} +
    \sqrt{\gamma_R}e^{-i(\omega \tau - \phi_{\rm M})/2}  \right),
\end{align}
where $\phi_{\rm M}$ is the mirror phase.
Note that a factor of 2 appears in the phase term as we define
$\tau$ as the total round-trip time from the TLS to the mirror and back. 
In the interaction picture,
\begin{equation}
    \begin{split}
     H_{\rm I} &= \int d\omega \left[ G_{\rm fback}(\omega) e^{-i(\omega - \omega_0)t} b(\omega) \sigma^+ +\rm H.c. \right] \\  
     &= \int d\omega \left[ 1/\sqrt{2\pi} \left( 
    \sqrt{\gamma_L}e^{i(\omega \tau - \phi_{\rm M})/2} +
    \sqrt{\gamma_R}e^{-i(\omega \tau - \phi_{\rm M})/2}  \right) \right . \\
    & \left .e^{-i(\omega - \omega_0)t}
    b(\omega) \sigma^+ +\rm H.c. \right].
    \end{split}
\end{equation}

Next, we can define
\begin{align}
    b(t-\tau/2)&=\frac{1}{\sqrt{2\pi}} \int d\omega b(\omega) e^{-i(\omega - \omega_0)(t-\tau /2)} \\ \nonumber
    &=\frac{1}{\sqrt{2\pi}} \int d\omega b(\omega) e^{-i(\omega t -\omega \tau /2 - \omega_0 t - \omega_0 \tau /2)},
\end{align}
and
\begin{align}
    b(t+\tau/2)&=\frac{1}{\sqrt{2\pi}} \int d\omega b(\omega) e^{-i(\omega - \omega_0)(t+\tau /2)} \nonumber \\
    &
    =\frac{1}{\sqrt{2\pi}} \int d\omega b(\omega) e^{-i(\omega t +\omega \tau /2 - \omega_0 t - \omega_0 \tau /2)},
\end{align}
to obtain
\begin{align}   
     &H_{\rm I} = \nonumber \\
     &\left( 
    \sqrt{\gamma_L}e^{-i\phi/2} b(t-\tau /2) 
    +\sqrt{\gamma_R}e^{i\phi/2} b(t+\tau /2)  \right) \sigma^+  +\rm H.c.,
\end{align}
where $\phi= \phi_{\rm M} -\omega_0 \tau$.
Finally, redefining
\begin{align}
    b(t+\tau /2)e^{i\phi/2} \rightarrow b(t), \\
    b(t-\tau /2)e^{i\phi/2} \rightarrow b(t-\tau),
\end{align}
we obtain
\begin{equation}
    \begin{split}
     H_{\rm I} &=  \left( 
    \sqrt{\gamma_L}e^{-i\phi} b(t-\tau) +
    \sqrt{\gamma_R} b(t)  \right) \sigma^+  +\rm H.c.,
    \end{split}
\end{equation}
 consistent with the form of Pichler and Zoller~\cite{Pichler11362}. 
It is clear that $\tau$ dependence of the
$b(t-\tau)$ 
includes the effects of retardation (memory), and 
this term makes the problem  
non-Markovian, in contrast to the case without feedback.
This is precisely why we must treat
 the circulating waveguide photons at the system level,
 since the usual Markov approximations would fail.

Commonly, symmetrical coupling rates are assumed for this feedback setup with $\gamma_L=\gamma_R=\gamma/2$. This changes the form of $G_{\rm fback}$ in the interaction picture to
\begin{equation}
    G_{\rm fback}(\omega) = \sqrt{2} \sqrt{\frac{\gamma}{2\pi}} \sin \left( \frac{\omega \tau + \phi'}{2} \right),
\end{equation}
with $\phi' = \omega_0 \tau + \pi - \phi_{\rm M}$. Note that this differs slightly from other common bath functions in this form \cite{Droenner2019,crowder_quantum_2020} because we have defined $\gamma$ as the full decay rate (i.e., a population decay rate) to the waveguide rather than the half rate and here $\phi'$ explicitly contains the phase introduced from the mirror.

\subsection{Scheme (iii): Two coupled two-levels systems separated with some finite distance and time delay}
\label{subsec:scheme3}

Lastly, two spatially-separated TLSs in an infinite waveguide are considered (see Fig.~\ref{schematics2}(c)). The Hamiltonian is 
\begin{equation}
    H = \sum_{n=1,2} (H_{\rm TLS}^{(n)} + H_{\rm pump}^{(n)}) + H_{\rm W} + H_{\rm I},
    \label{hamil2tls}
\end{equation}
where the TLS free terms are now
\begin{equation}
    H_{\rm TLS}^{(n=1,2)}= \omega_n \sigma_n^+ \sigma_n^-
    \label{TwoTLS_H},
\end{equation}
and the pump terms are
\begin{equation}
    H_{\rm pump}^{(n=1,2)}= 
    \frac{1}{2}\left[\Omega_n \sigma_n^- e^{i\omega_{\rm L} t} + \rm H.c. \right],
    \label{TwoTLS_P}
\end{equation}
where $\Omega_n$ is the Rabi frequency of a driving field (at TLS $n$),
$\omega_n$  is the transition frequency of each TLS, and $\sigma_n^+$, $\sigma_n^-$ are the Pauli operators for each TLS. 

Since the two TLS case results in symmetry breaking (namely, with the finite retardation phase effects), the waveguide Hamiltonian must now be separated into left and right going channels,
\begin{equation}
    H_{\rm W} = \sum_{\alpha=L,R} \int_{-\infty}^{\infty} d\omega  \omega b_\alpha^\dagger (\omega)b_\alpha(\omega),
\end{equation}
and  the interaction Hamiltonian is
\begin{equation}
\begin{split}
    H_{\rm I} &=  \frac{1}{\sqrt{2\pi}}\int_{-\infty}^{\infty} d\omega \Bigg \{ 
    \left( \sqrt{\gamma_{L1}} e^{i\omega x_1/c}\, b_L(\omega)\sigma^+_1 \right .\\
    & \left . +  \sqrt{\gamma_{R1}} e^{-i\omega x_1/c} b_R(\omega)\sigma^+_1 
     \right) + \rm H.c. \\
    &+ \left(  \sqrt{\gamma_{L2}} e^{i\omega x_2/c} b_L(\omega)\sigma^+_2 \right . \\
    & \left . +  \sqrt{\gamma_{R2}} e^{-i\omega x_2/c} b_R(\omega)\sigma^+_2 
     \right) + \rm H.c. \Bigg \}.  ,
\end{split}
\end{equation} 
where $x_n$ with $n=1,2$ is the position of each TLS.

In the interaction picture, again at the frequency of $\omega_{\rm L}=\omega_0$, we have
\begin{equation}
\label{twoTLS_interaction}
\begin{split}
    H_{\rm I} &=  \frac{1}{\sqrt{2\pi}} \int_{-\infty}^{\infty} d\omega \Bigg \{ 
    \left(  \sqrt{\gamma_{L1} } e^{i\omega x_1/c}\, b_L(\omega)\sigma^+_1 \right .\\
    & \left . +  \sqrt{\gamma_{R1} } e^{-i\omega x_1/c}\, b_R(\omega)\sigma^+_1 
     \right) e^{-i(\omega-\omega_0)t} + \rm H.c. \\
    &+ \left(  \sqrt{\gamma_{L2} } e^{i\omega x_2/c}\, b_L(\omega)\sigma^+_2 \right . \\ 
    & \left . +  \sqrt{\gamma_{R2}} e^{-i\omega x_2/c}\, b_R(\omega)\sigma^+_2 
     \right)e^{-i(\omega-\omega_0)t} + \rm H.c. \Bigg \}.    
\end{split}
\end{equation} 
Defining the operators,
\begin{equation}
\begin{split}
    &b_L(t-x_n/c)=\frac{1}{\sqrt{2\pi}} \int d\omega\, b_L(\omega) e^{-i(\omega - \omega_0)(t-x_n /c)} \\
    &=\frac{1}{\sqrt{2\pi}} \int d\omega\, b_L(\omega) e^{-i(\omega t -\omega x_n/c - \omega_0 t + \omega_0 x_n/c)},
    \end{split}
\end{equation}
\begin{equation}
\begin{split}
    &b_R(t+x_n/c)=\frac{1}{\sqrt{2\pi}} \int d\omega b_R(\omega) e^{-i(\omega - \omega_0)(t+x_n /c)} \\
    &=\frac{1}{\sqrt{2\pi}} \int d\omega b_R(\omega) e^{-i(\omega t +\omega x_n/c - \omega_0 t - \omega_0 x_n/c)},
    \end{split}
\end{equation}
then we obtain
\begin{equation}
\begin{split}
    H_{\rm I} &=  
    \left(  \sqrt{\gamma_{L1} } e^{-i\omega_0 x_1/c}\, b_L(t-x_1/c)\sigma^+_1 \right .\\ 
    & \left . +  \sqrt{\gamma_{R1} } e^{i\omega_0 x_1/c}\, b_R(t+x_1/c)\sigma^+_1 
     \right)  + \rm H.c. \\
    &+ \left( \sqrt{\gamma_{L2} } e^{-i\omega_0 x_2/c}\, b_L(t-x_2/c)\sigma^+_2 \right .\\
    & \left . +  \sqrt{\gamma_{R2} } e^{i\omega_0 x_2/c}\, b_R(t+x_2/c)\sigma^+_2 
     \right) + \rm H.c..    
\end{split}
\end{equation} 

Next,  we redefine the following terms:
\begin{equation}
\begin{split}
    &b_R(t+x_2/c)e^{i\omega_0 x_2/c} \rightarrow b_R(t), \\
    &b_R(t+x_1/c)e^{i\omega_0 x_2/c} \rightarrow b_R(t+x_1/c-x_2/c) = b_R(t-\tau), \\
    &b_L(t-x_1/c)e^{-i\omega_0 x_1/c} \rightarrow b_L(t), \\
    &b_L(t-x_2/c)e^{-i\omega_0 x_1/c} \rightarrow b_L(t+x_1/c-x_2/c) = b_L(t-\tau), \\
\end{split}
\end{equation}
where $\tau = (x_2-x_1)/c$. The interaction Hamiltonian,
in the time domain, is now 
\begin{equation}
\begin{split}
    H_{\rm I} & = \left(  \sqrt{\gamma_{L1} }\,  b_L(t) 
    +  \sqrt{\gamma_{R1}} e^{-i\omega_0 \tau} b_R(t-\tau) 
     \right)\sigma^+_1  + \rm H.c. \\
    &+ \left(  \sqrt{\gamma_{L2} } e^{-i\omega_0 \tau} b_L(t-\tau) 
    +  \sqrt{\gamma_{R2} } \, b_R(t) 
     \right)\sigma^+_2 + \rm H.c.
\end{split}
\end{equation} 

Finally, by defining the phase $\phi=-\omega_0 \tau$, we obtain the desired interaction Hamiltonian:
%
\begin{equation}
\begin{split}
    H_{\rm I} &= \left(  \sqrt{\gamma_{L1} }   b_L(t)
    +  \sqrt{\gamma_{R1} }  e^{i\phi} b_R(t-\tau)
     \right)\sigma^+_1  + \rm H.c. \\
    &+ \left(  \sqrt{\gamma_{L2} } e^{i\phi} b_L(t-\tau)
    +  \sqrt{\gamma_{R2}}\, b_R(t) 
     \right)\sigma^+_2 + \rm H.c.
\end{split}
\end{equation} 
Note that in the limit of only one TLS,
we recover the $H_{\rm I}$ result of Eq.~\eqref{int1}.


\section{Matrix Product States}
\label{sec:MPS}


\subsection{Quantum states, diagrammatic representation and canonical form of 
matrix product states}
\label{subsec:diag}
A quantum state for  many-body problems can quickly  have an
impractically large Hilbert space, e.g., for a system of $N$ spins with spin 1/2, the dimension of the Hilbert space is 2$^N$. The MPS method takes advantage of the significance of some quantum states compared to others; this can be shown in the entanglement between the states composing the system \cite{droenner_out--equilibrium_2019}. As the Hamiltonian evolves the system in time, these states will become entangled. Furthermore, by choosing the {\it significant entangled states} appropriately, the total Hilbert space will be restricted to a smaller and more efficient subspace \cite{vanderstraeten_tensor_2017,mcculloch_density-matrix_2007}. 

The quantum state for a 1D spin-chain, with $N$ spins, is given by \cite{woolfe_matrix_2015},
\begin{equation}
    \ket{\psi}= \sum_{i_1,...,i_N}^{d} c_{i_1,...,i_N} \ket{i_1,...,i_N},
    \label{eq:psi}
\end{equation}
where $i_k$ (with $k \ \in \ \{1,...N\}$) represents each state with a dimension of $d$, and $c_i$ are the coefficients of the corresponding state. 

\begin{figure}[H]
    \centering
    \includegraphics[width=0.5 \columnwidth]{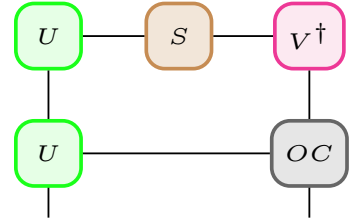}
    \caption{Diagrammatic representation of a right contraction after a SVD of a matrix; $U$ is a left normalized matrix, $S$ is the diagonal with the Schmidt coefficients, $V^\dagger$ is a right normalized matrix and $OC$ is the orthogonality center.}
    \label{occontract}
\end{figure}

The MPS algorithm relies on the Schmidt decomposition of a quantum system, which considers  
the bipartition state of the system 
as 
 a tensor product \cite{woolfe_matrix_2015}. 
 In practice, 
the state can be transformed using the singular-value decomposition (SVD) or Schmidt decomposition. The SVD theorem states that any matrix can be factorized, decomposing it into 3 new matrices  \cite{woolfe_matrix_2015,orus_practical_2014}. 
The SVD decomposition of a matrix $M$ of dimension $N_A \times N_B$ is \cite{schollwock_density-matrix_2011},
\begin{equation}
\label{usv}
    M = U S V^\dagger,
\end{equation}
where $S$ is a diagonal matrix containing the Schmidt coefficients in descendent order (i.e., largest to smallest), $U$ is {\em left-normalized} and $V$ is {\em right-normalized}~\cite{schollwock_density-matrix_2011}. Then, one of the side matrices can be multiplied by the one containing the Schmidt coefficients, which receives the name of ``Orthogonality Center'' (OC) \cite{droenner_out--equilibrium_2019}, and we end up with 2 new matrices (see Figure~\ref{occontract}).

Assuming a system can be divided into two subsystems $A$ and $B$, then 
\begin{equation}
\label{svd}
     \ket{\psi}= \sum_{ij} C_{i,j} \ket{i}_A \ket{j}_B,
\end{equation}
where $\ket{i}_A$ and $\ket{j}_B$ form the new orthogonal basis, $C_{i,j}$ describes the matrix containing the $c_i$ defined in Eq.~\eqref{eq:psi}, and $i$ and $j$ contain several indices. In the case of a TLS in a waveguide, the first basis would correspond to the TLS with just one state, and the second one would correspond the rest of the states for the waveguide describing the number of photons. Furthermore, for two TLSs we will have one basis where both TLSs are included and a second one including the waveguide. In general, a system can be divided in different subsystems. If some parts of the entire system can be written in terms of the same basis, they can belong to the same subsystem (e.g. the two TLSs); if not, they will form a different subsystem (e.g., the waveguide). 

\begin{figure}[H]
    \centering
    \includegraphics[width=0.7 \columnwidth]{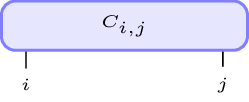}
    \caption{Diagrammatic representation of a matrix ($C_{i,j}$) with physical indices (physical dimensions of system) $i$ and $j$.}
    \label{diag1}
\end{figure}

\begin{figure}[H]
    \centering
    \includegraphics[width=0.5 \columnwidth]{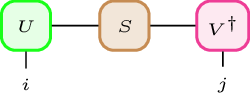}
    \caption{Diagrammatic representation of the SVD of a matrix, where $U$ is a left normalized matrix, $S$ is the Schmidt coefficients and $V^\dagger$ is a right normalized matrix (see Eq.~\eqref{usv}).}
    \label{diag2}
\end{figure}

The diagrammatic representation for MPSs is usually used to better
visualize a simpler representation of the operations performed on a state \cite{droenner_out--equilibrium_2019,eduardo_one-dimensional_2017,iblisdir_matrix_2007}. For example, an arbitrary matrix can be represented as shown in Fig.~\ref{diag1}. 
Another simple example is to represent Eq.~\eqref{usv} in its diagrammatic form, where we  obtain the scheme shown in Fig.~\ref{diag2}.
The indices are divided in ``physical indices'', which correspond to the physical dimensions of our system and are represented as open vertical links, and the bond or virtual indices related to the decomposition of the MPS, are represented as horizontal links between tensors and will store the entanglement information. The number of indices of a tensor defines its rank, for example, a vector will have one index and a matrix will have two.  
Decomposing $C_{ij}$ through a SVD, and taking into account that $U$ and $V^\dagger$ are orthonormal, the total state can be written as
\begin{equation}
    \ket{\psi}=\sum_{\alpha=1}^{{\rm min}(N_A,N_B)} s_{\alpha} \ket{\alpha}_A \ket{\alpha}_B,
\end{equation}
where we have introduced a new basis and $s_\alpha$ are the elements of the diagonal matrix, and the sub-states are
\begin{align}
    &\ket{\alpha}_A = \sum_i U_{i\alpha} \ket{i}_A;
    &\ket{\alpha}_B = \sum_j V^\dagger_{j\alpha} \ket{j}_B.
\end{align}
The Schmidt rank, which we label with $r$, is defined as the number of non negligible values for the Schmidt coefficients, hence, $r \leq {\rm min}(N_A,N_B)$. Performing 
a SVD, we can truncate the matrix for values greater than the rank and thus reduce the number of columns of $U$ and the rows of $V^\dagger$. Now, when the entanglement between the states is small, many values of $s_\alpha$ tend to zero and this approximation is excellent and considerably reduces the dimensions of the system.

The entanglement can be quantified by the von Neumann entropy  \cite{pavarini_emergent_2013},
\begin{equation}
    \label{vonneumann}
    S_{A|B}=-\sum_{\alpha = 1}^r s_\alpha^2 \ln \ s_\alpha^2. 
\end{equation}
However, in practice, one can simply
limit the number of Schmidt coefficients considered 
and check that the number is sufficient for numerical convergence.

After several SVDs, the general expression for a MPS follows \cite{mcculloch_density-matrix_2007},
\begin{equation}
\label{general}
    \ket{\psi}=\sum_{i_1...i_N} A_{a_1}^{i_1}A_{a_1,a_2}^{i_2} \hdots A_{a_{N-2},a_{N-1}}^{i_{N-1}}A_{a_{N-1}}^{i_{N}}\ket{i_1,\hdots,i_N},
\end{equation}
where each term is a tensor, and $a_1,\hdots,a_{N-1}$ are the ``bond'' lengths or auxiliary dimensions of each element and $i_1,\hdots,i_N$ represent the physical dimensions of the system. Here, in the bonds dimensions, is where we approximate our system. By limiting this value, we limit the number of values of the Schmidt coefficient considered. This will make the method more efficient, while keeping a high precision on the results.  

As an example, in the case of a TLS in a waveguide, after a first SVD,  Eq.~\eqref{svd} will follow, where on one side we have the TLS and on the other the entire waveguide. Hence, for getting every site separated, we have to continue applying the SVD to the waveguide part until we decompose it in $N$ sites, obtaining the following form:
\begin{equation}
\label{waveguidecase}
    \ket{\psi}=\sum_{i_s i_1...i_N} A_{a_1}^{i_s}A_{a_1,a_2}^{i_1} \hdots A_{a_{N-1},a_{N}}^{i_{N-1}}A_{a_{N}}^{i_{N}}\ket{i_s, i_1,\hdots,i_N},
\end{equation}
where the first term represents the TLS, and the remaining $N$ terms represent the discretized waveguide, representing the possibility of at least $N$ photons in the waveguide. In addition, this $N$ can become arbitrarily large as the waveguide can be divided in as many time bins as needed.

\begin{figure}[H]
    \centering
    \includegraphics[width=0.8 \columnwidth]{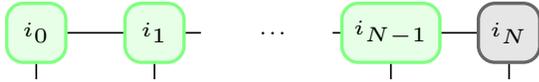}
    \caption{Diagrammatic representation of a left-canonical MPS, where the OC is situated at the right of the system (black) and the rest of the bins are left normalized (green).}
    \label{left}
\end{figure}

\begin{figure}[H]
    \centering
    \includegraphics[width=0.8 \columnwidth]{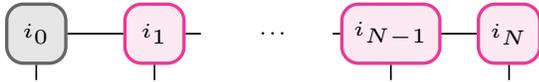}
    \caption{Diagrammatic representation of a right-canonical MPS, where the OC is situated at the left of the system (black) and the rest of the bins are right normalized (magenta).}
    \label{right}
\end{figure}

\begin{figure}[H]
    \centering
    \includegraphics[width=0.8 \columnwidth]{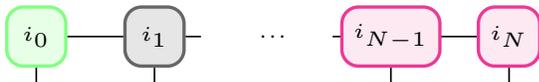}
    \caption{Diagrammatic representation of a mixed-canonical MPS, where the OC is situated at an arbitrary position in the system (black), the bins on its right are right normalized (magenta) and the ones on its left are left normalized (green).}
    \label{mixed}
\end{figure}

There is no single (unique) way of performing the number of SVDs required, as the system is divided in two subsystems each time a SVD is done \cite{orus_practical_2014}. One way is to start from the left and take $i_0$ as the first subsystem and ${i_1,...,i_N}$ as the second one. This process is repeated from the left to the right until the orthonormal matrix is on right side (OC). This method  is called a left-canonical MPS (Fig.~\ref{left}) \cite{vanderstraeten_tangent-space_2019}. On the other hand, a right-canonical MPS will be the one made in the opposite direction, with the OC at the left (Fig.~\ref{right}). Finally, in a mixed-canonical MPS, the OC is situated in an arbitrary position (see Fig.~\ref{mixed}). The mixed-canonical case will be the one used in the systems studied below, as the OC will be moving in the system to keep track of the observables. This will be explained in more detail in Section~\ref{subsec:comp}. 


\subsection{Hamiltonian of the systems in the picture of matrix product states}
\label{subsec:hamil}

We consider a waveguide that is coupled to the TLSs (one or two), and here we will treat these systems as a many-body system which will be solved using the MPS formalism. 

To write the Hamiltonians in terms of the MPS formalism, the frequency-dependent creation and annihilation operators for the waveguide can be transformed to the time domain as follows,
\begin{equation}
    b_\alpha(t)=\frac{1}{\sqrt{2\pi}} \int d\omega b_\alpha(\omega) e^{-i(\omega - \omega_{0})t},
    \label{continuous-time-ops}
\end{equation}
and are defined in terms of the time-bin noise operators,
\begin{align}
    \label{noise1}
    &\Delta B_\alpha(t_k) = \int_{t_k}^{t_{k+1}} dt' b_\alpha(t'), \\
    \label{noise2}
    &\Delta B_\alpha^\dagger (t_k) = \int_{t_k}^{t_{k+1}} dt' b_\alpha^\dagger(t').
\end{align}

 These time-bin noise operators form a time-discrete and orthogonal basis which is 
 normalized with the commutator proportional to $\Delta t$,
\begin{equation}
    \left[ \Delta B_\alpha(t_k), \Delta B_{\alpha'}^\dagger(t_{k'}) \right] = \Delta t \delta_{k,k'} \delta_{\alpha,\alpha'}.
\end{equation}
Consequently, a time-discrete number basis is created as follows:
\begin{equation}
\label{timebasis}
    \ket{i^\alpha_k} = \frac{(\Delta B_\alpha^\dagger (t_k))^{i^\alpha_k}}{\sqrt{i^\alpha_k ! (\Delta t)^{i^\alpha_k}}} \ket{\rm vac},
\end{equation}
where $\sqrt{ (\Delta t)^{i_k^\alpha}}$ appears in the denominator for normalization. The state $\ket{i_k^\alpha}$ is referred to as the `time-bin' and represents the number of photons created in the waveguide at time interval $\Delta t$. Subsequently, we can write $\ket{\psi}$ in the time-discrete basis and operate on it with the time-evolution operator.

Note in the above treatment, the spatial dependency of the photons is absorbed. Hence, it is hidden in the model and there is no explicit information about the position of each photon in the waveguide.

\subsubsection{Scheme (i): Single two level system in an infinite waveguide}

The Hamiltonian modelling this system is described in \ref{subsec:scheme1}. The expression of the time evolution operator, for a time step in terms of the noise operators, is
\begin{equation}
    U(t_{k+1},t_k) =  \exp{ \bigg( -i \int_{t_k}^{t_{k+1}} dt' H(t')\bigg)}. 
\end{equation}
For the general case when the TLS decay rates are different, and considering the right and left moving photons separately, then we have
\begin{equation}
    \begin{split}
        U(t_{k+1},t_k)&= \exp \bigg[ -i\Omega_0 \Delta t \left( \sigma^+ +  \sigma^-\right) \\
        &-i \sqrt{\gamma_L}\left( \sigma^+ \Delta B_L(t_k) + \sigma^- \Delta B_L^{(\dagger)}(t_k) \right)\\
        &-i  \sqrt{\gamma_R}\left( \sigma^+ \Delta B_R(t_k) + \sigma^- \Delta B_R^{(\dagger)}(t_k) \right) \bigg].     
    \end{split}
\end{equation}

If we consider equal (symmetric) coupling and one waveguide mode, we can write 
this as 
\begin{equation}
\label{mpsU1}
    \begin{split}
        U(t_{k+1},t_k)&= \exp \bigg[ -i\Omega_0 \Delta t \left( \sigma^+ +  \sigma^-\right) \\
        &-i \sqrt{\gamma} \left( \sigma^+ \Delta B(t_k) + \sigma^- \Delta B^{(\dagger)}(t_k) \right) \bigg],     
    \end{split}
\end{equation}
and the noise operators represent the creation or annihilation of a photon in a time interval $\Delta t$ (time-bin).

\subsubsection{Scheme (ii): Single two level system in a half open waveguide with a time-delayed coherent feedback}

Starting from the equations shown in \ref{subsec:scheme2}, the complete Hamiltonian is transformed into a rotating frame with respect to the free evolution of the system and the waveguide reservoir. Following the same procedure as above,  the time-evolution operator, written in terms of the noise operators, is 
\begin{equation}
\begin{split}
    &U(t_{k+1},t_k)=\exp\Bigg \{ \bigg[ -i\Delta t\Omega_0 (\sigma^+ + \sigma^-) \\
    & - i \left( \sqrt{\gamma_L} \Delta B(t_{k-l})e^{-i\phi} + \sqrt{\gamma_R} \Delta B(t_{k}) \right) \sigma^+ + \rm H.c.\bigg]\Bigg \},    
\end{split}
\end{equation}
where $t_k=k\Delta t$ and $t_{k-l} = t_k - \tau$. 

Considering symmetric coupling, $\gamma_L=\gamma_R=\gamma/2$, then
\begin{equation}
\begin{split}
    &U(t_{k+1},t_k)=\exp\Bigg \{ \bigg[ -i\Delta t\Omega_0 (\sigma^+ + \sigma^-) \\
    & - i\left( \sqrt{\frac{\gamma}{2}} \Delta B(t_{k-l})e^{-i\phi} + \sqrt{\frac{\gamma}{2}} \Delta B(t_{k}) \right) \sigma^+ + \rm H.c.\bigg]\Bigg \}.    
\end{split}
\end{equation}
As expected, we see that this system is intrinsically
non-Markovian through the time delay $\tau$,
which has a memory of the past quantum dynamics that are introduced through feedback.

\subsubsection{Scheme (iii): Two two levels
separation with some finite time delay}

In the third waveguide QED system of interest (see \ref{subsec:scheme3}), the time evolution operator is also obtained from the discretization of the time bins;   using the quantum noise operators already defined in 
Eqs.~\eqref{noise1} and \eqref{noise2}, we obtain
\begin{equation}
    \begin{split}
    &U(t_{k+1},t_k) =  \exp \Bigg \{ -i H_{\rm TLS}^{(n=1,2)}(t_k) \Delta t \\
    &-i \left[ \left( \sqrt{\gamma_{L1}} \Delta B _L (t_k) + \sqrt{\gamma_{R1}} \Delta B_R (t_{k-l}) e^{i\phi}\right)\sigma^+_1 +\rm H.c. \right] \\
    &-i \left[ \left( \sqrt{\gamma_{L2}} \Delta B_L(t_{k-l})e^{i\phi} + \sqrt{\gamma_{R2}} \Delta B_R(t_k) \right)\sigma^+_2 + \rm H.c. \right] \Bigg \} ,        
    \end{split}
    \label{u2tls}
\end{equation}
where $l=\tau/\Delta t$, which now represents the number of sites (time bins) between the two TLSs.

\subsection{Matrix Product Operators}
\label{subsec:oper}

\begin{figure}[H]
    \centering
    \includegraphics[width=0.4 \columnwidth]{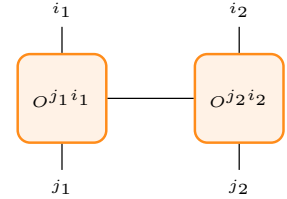}
    \caption{Diagrammatic representation of $\hat{O}$ as a MPO acting on two sites  with physical indices $i_1, \ j_1 $ and $i_2, \ j_2$.}
    \label{diag3}
\end{figure}

The time evolution is computed through the time operator ${U}$. This operator can be seen as a projector which projects one physical index $i$ to another $j$ with some coefficients $U^{ji}$. For example, a MPO operating on two sites, $1$ and $2$, can be written as follows (see Fig.~\ref{diag3}),
\begin{equation}
    O=\sum_{\boldsymbol{j,i}} O^{j_1,i_1} O^{j_2,i_2}\ket{\boldsymbol{j}}\bra{\boldsymbol{i}}  , 
\end{equation}
where $\boldsymbol{j}=j_1,j_2$ and $\boldsymbol{i}=i_1,i_2$ are the labels for the physical indices of the corresponding bra and ket. 
The MPOs have two physical indices per site. The main advantage is that the whole state does not need to be computed when an operator is applied, since it will only affect the corresponding sites~\cite{pavarini_emergent_2013, genericmpo}.

We can construct the time evolution operator as a local operator operating on two sites in the no feedback case (the TLS bin and the corresponding time bin), and on three sites when the feedback is included (the TLS bin, time bin and feedback bin) or when the system is made of two TLS (see \ref{subsec:hamil} for more details). 

The quantum noise operators involved in the time evolution are represented as one site operators, defined through 
\begin{align}
    \Delta B_\alpha =  
    \begin{bmatrix}
            0 \ \sqrt{\Delta t} \\
            0 \ 0 \\
    \end{bmatrix}&, \ \
    \Delta B^\dagger_\alpha = 
    \begin{bmatrix}
            0 \ 0 \\
            \sqrt{\Delta t} \ 0 \\
    \end{bmatrix},
\end{align}
with $\alpha=L,R$. Here we truncate the Hilbert space for each time bin to one excited photon state; this of course can be generalized, e.g., if we allowed up to two photon states per time bin, then we would have a 3 by 3 matrix representation for the quantum noise operators.

On the other hand, the expectation of the TLS atom population operator will be a single site MPO ($n_{a}$), operating on the TLS bin: 
\begin{equation}
\label{population}
    n_{a,n}=\bra{\psi} \sigma^+_n \sigma^-_n \ket{\psi},
\end{equation}
where $\sigma^+_n$ $\sigma^-_n$ are defined as,
\begin{align}
    \sigma^+_n =
    \begin{bmatrix}
            0 \ 0 \\
            1 \ 0 \\
    \end{bmatrix}, \  
    \sigma^-_n =
    \begin{bmatrix}
            0 \ 1 \\
            0 \ 0 \\
    \end{bmatrix}. 
\end{align}

Finally, a swap operator will be applied to switch the position of two bins in the chain. Therefore, it will be a 2-site MPO operating on the sites involved. This MPO will depend on the dimensions of the sites to be swapped. In order to have the same dimensions in every bin and be able to use this operator, we limit the number of photons per time bin to one. This approximation is very accurate as the time steps considered are small.

The swap operator can be defined as \cite{Suba2019},
\begin{equation}
    V_{\rm swap} = \sum_{j,k} \ket{jk} \bra{kj},
\end{equation}
where $j$ and $k$ can be made of one or more subsystems. For example, in the case of swapping the TLS bin and one time bin for one TLS we can have,
\begin{equation}
    \label{swap1}
    V_{\rm swap}^{i_t,i_{s}} = \sum_{i_t,i_{s}} \ket{i_t i_{s}} \bra{i_{s} i_t},
\end{equation}
with $i_s$ corresponding to the TLS bin and $i_t$ a time bin. In this case both bins have dimension of 2, $i_t \otimes i_{s}$ will have dimension of 4 and the operator then will be a matrix 4x4. This can be also applied when swapping a time bin with the feedback bin,
\begin{equation}
    V_{\rm swap}^{i_t,i_{\tau}} = \sum_{i_t,i_{\tau}} \ket{i_t i_{\tau}} \bra{i_{\tau} i_t},
\end{equation}
having the same dimensions in this case.

In both cases the following matrix representation will follow,
\begin{equation}
    V_{\rm swap} =
    \begin{bmatrix}
            1 \ 0 \ 0 \ 0 \\
            0 \ 0 \ 1 \ 0 \\
            0 \ 1 \ 0 \ 0 \\
            0 \ 0 \ 0 \ 1 \\
    \end{bmatrix}.
\end{equation}
This turns out to be the same as in the case of swapping two spins with the usual 
spin-1/2 Pauli operators \cite{nielsen_quantum_2003,band_quantum_2013},
\begin{equation}
    V_{\rm swap} = 1/2 \left( I \otimes I + \sigma_x \otimes \sigma_x + \sigma_y \otimes \sigma_y + \sigma_z \otimes \sigma_z \right),
\end{equation}
where $I$ is the identity matrix, and $\sigma_x$, $\sigma_y$ and $\sigma_z$ are the Pauli gates. 

For two TLSs in the waveguide, we can write the swap operator in a similar form as Eq.~\eqref{swap1}, but now the TLSs bin includes both TLSs, 
\begin{equation}
\label{2TLSsystemEQ}
    i_s=i_{TLS1} \otimes i_{TLS2},
\end{equation}
giving a dimension equal to 4, and the time bin includes the photons moving to the right and left in each time step,
\begin{equation}
\label{2TLStimeEQ}
    i_t=i_{tL} \otimes i_{tR},
\end{equation}
hence it also has $d=4$. Now we can have up to 2 photons per time bin, and then the limit in the waveguide is $2N$. The tensor product of both systems has a length of 16 and, consequently, the swap operator in this case will be a much larger matrix of $16\times 16$. As before, it can be also applied between time bins, and it will keep the same dimensions:
\begin{equation}
\begin{split}
    &V_{\rm swap}^{i_t,i_{s}} = \sum_{i_t,i_{s}} \ket{i_t i_{s}} \bra{i_{s} i_t}= \\
    &\sum_{i_{tL},i_{tR},i_{TLS1},i_{TLS2}} \ket{i_{tL} i_{tR} i_{TLS1} i_{TLS2}} \bra{ i_{TLS2} i_{TLS1} i_{tR} i_{tL}},
\end{split}
\end{equation}
\begin{equation}
\begin{split}
    &V_{\rm swap}^{i_t,i_\tau} = \sum_{i_t,i_{\tau}} \ket{i_t i_{\tau}} \bra{i_{\tau} i_t}= \\
    &\sum_{i_{tL},i_{tR},i_{\tau L},i_{\tau R}} \ket{i_{tL} i_{tR} i_{\tau L} i_{\tau R}} \bra{i_{\tau R} i_{\tau L} i_{tR} i_{tL} }.
\end{split}
\end{equation}


\subsection{Time Evolution and Observables}
\label{subsec:comp}

\begin{figure} [htb!]
\centering
\subfloat[MPS schematic of a single TLS embedded in an infinite waveguide.]{%
  \includegraphics[clip, width=0.9 \columnwidth,]{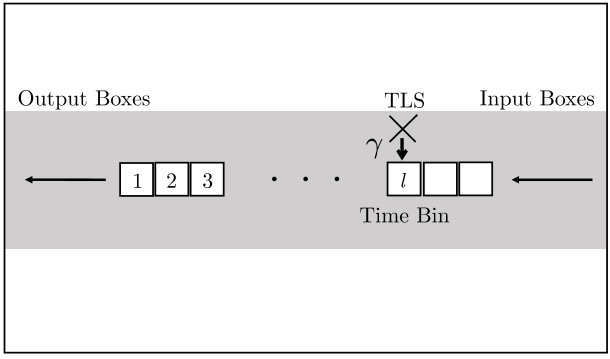}%
  }
\hspace{1pt}
\subfloat[MPS schematic of a single TLS embedded in a terminated waveguide,
with a time delayed feedback ($\tau=L_0/c$).]{%
  \includegraphics[clip, width=0.9 \columnwidth,]{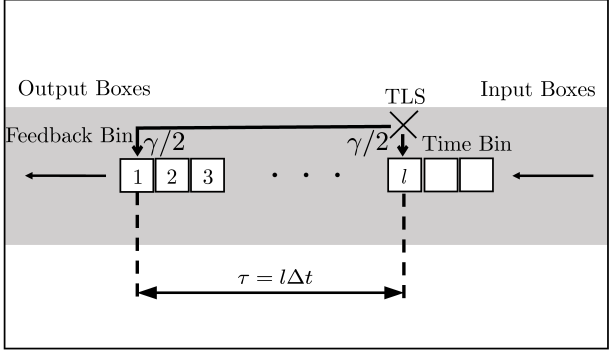}%
  }
\hspace{1pt}
\subfloat[MPS schematic of  two TLSs embedded in an infinite waveguide
with a delay length/time between them.]{%
  \includegraphics[clip, width=0.9 \columnwidth]{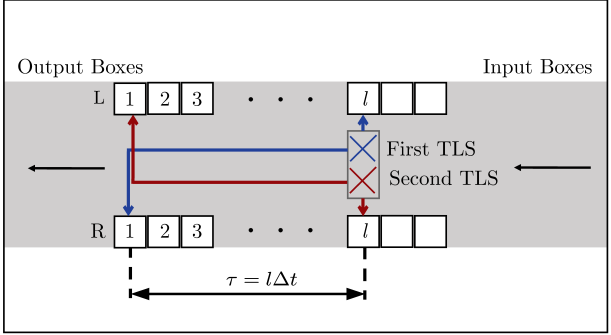}%
 }
 \captionsetup{justification=centering}
 \caption{Three systems of interest in waveguide QED, coupling one or two TLS to waveguides with and without a time-delayed coherent feedback. Each case is represented in the time frame, where the movement of one bin represents one time step. The TLSs are interacting with their corresponding time bin as time evolves. In the cases where a feedback is considered, the TLS (or TLSs) will interact with two different time bins at the same time step, as indicated with the arrows between them. 
 }
\label{schematicsMPS}
\end{figure}

The initial state for one TLS can be represented in the time-bin basis as a product state of the system (with basis ($\ket{g},\ket{e}$)) and the discretized time (with a subspace ($\ket{0},\ket{1}$),  if considering a maximum of one photon per time interval). As the initial state $\ket{\psi(0)}$ is a product state, there is no initial entanglement and there are no virtual links at the beginning.

\begin{figure}[htbp]
    \centering
    \includegraphics[width=0.45\columnwidth]{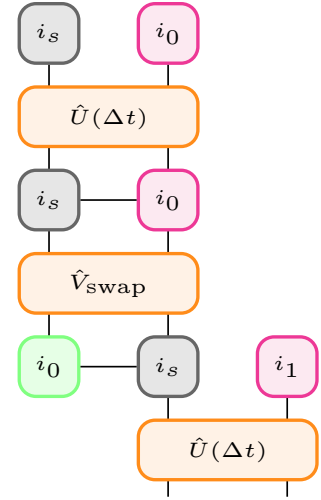}
    \caption{Diagrammatic representation of the first time step in which the time evolution operator is applied. After that, the swap operator is applied to bring the TLS bin to the right, leaving it ready for the next time step as can be seen in the last line. Green boxes represent left-normalize bins, magenta boxes represent right-normalized boxes, and the grey one is the OC. The operators are represented in orange and with the addition of a hat.}
    \label{timestepnofeedback}
\end{figure}

In the case with no  feedback  (see Fig.~\ref{schematicsMPS}(a)), we can apply directly the time evolution operator $\hat{U}$ written as a MPO on the system and the first time bin as in Fig.~\ref{timestepnofeedback}. After that, a SVD has to be done and both sites might get entangled. We also need to switch their positions in order to have the system on the left, ready for repeating the same procedure with the second time bin. This will be done with a swap MPO, ${V}_{\rm swap}$, applied on both sites, together with another SVD \cite{band_quantum_2013,nielsen_quantum_2003}. It is important to point out that the OC is kept in the system bin, as it must be in one of the bins involved in each operation. Furthermore, it 
is also necessary for computing the TLS population. Iterating this process, we can see the evolution on the TLS in the waveguide, computing the TLS population for each time step (time bin).

\begin{figure}[htbp]
    \centering
    \includegraphics[width=1 \columnwidth]{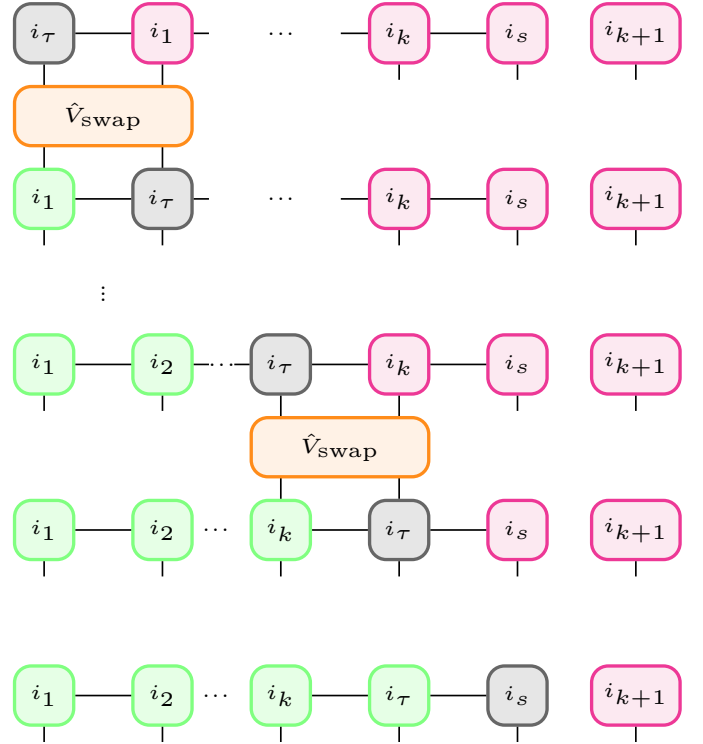}
    \caption{Diagrammatic representation of the first part of a single time step, in which the swap operator is applied to bring the feedback bin next to the system bin, and the OC is brought to the system bin. Green boxes represent left-normalize bins, magenta boxes represent right-normalized boxes, and the grey one is the OC.}
    \label{swapping}
\end{figure}

\begin{figure}[htbp]
    \centering
    \includegraphics[width=1 \columnwidth]{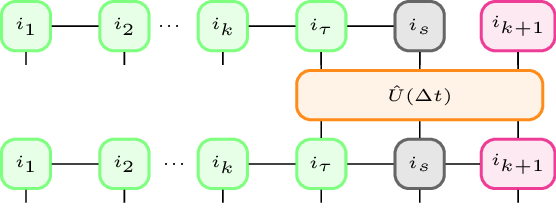}
    \caption{Diagrammatic representation of the application of the local operator U on the feedback, system and time bins. The OC is left in the system bin.}
    \label{applyingU}
\end{figure}

\begin{figure}[htbp]
    \centering
    \includegraphics[width=1 \columnwidth]{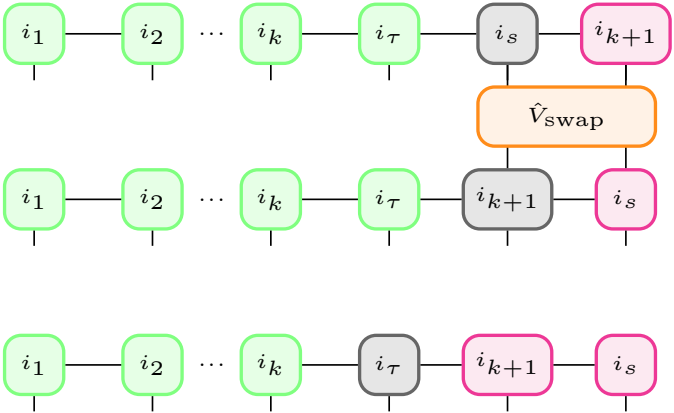}
    \caption{Diagrammatic representation of the swap operation for leaving the system bin on the right, ready for the next time step. The OC is moved back to the feedback bin.}
    \label{swapsystem}
\end{figure}

The evolution becomes more complicated once a time-delayed feedback is introduced (see Fig.~\ref{schematicsMPS}(b)). 
Now, there are three bins involved in the evolution: the system bin, the current time bin and the time bin involving the feedback, called the feedback bin. This last bin is not situated next to the other two, which means that now our Hamiltonian has a long range interaction. In order to avoid contracting all the bins between the feedback bin and the system bin for operating the time evolution operator, the feedback bin is brought next to the system bin using ${V}_{\rm swap}$ (Fig.~\ref{swapping}). After each swap, a SVD must be done in order to keep the canonical form. In addition, the OC must be kept in the feedback bin to apply the swap operator. 
Once the feedback bin is next to the system one, the three bins are contracted and ${U}$ is applied on them. After that, two SVD are performed to recover the three different bins and the OC is brought to the system bin in order to compute the TLS population (Fig.~\ref{applyingU}). 
Then ${V}_{\rm swap}$ is applied to the system and time bin to leave the system ready for the next time step and the OC is changed to the time bin (Fig.~\ref{swapsystem}). 
On the other hand, a series of ${V}_{\rm swap}$ are also applied to bring back the feedback bin to its corresponding position. 
Each operation is followed by a SVD and after the first operation the OC is kept another time in the feedback bin. This procedure is then repeated for each time step.

\begin{figure}[htbp]
    \centering
    \includegraphics[width=1 \columnwidth]{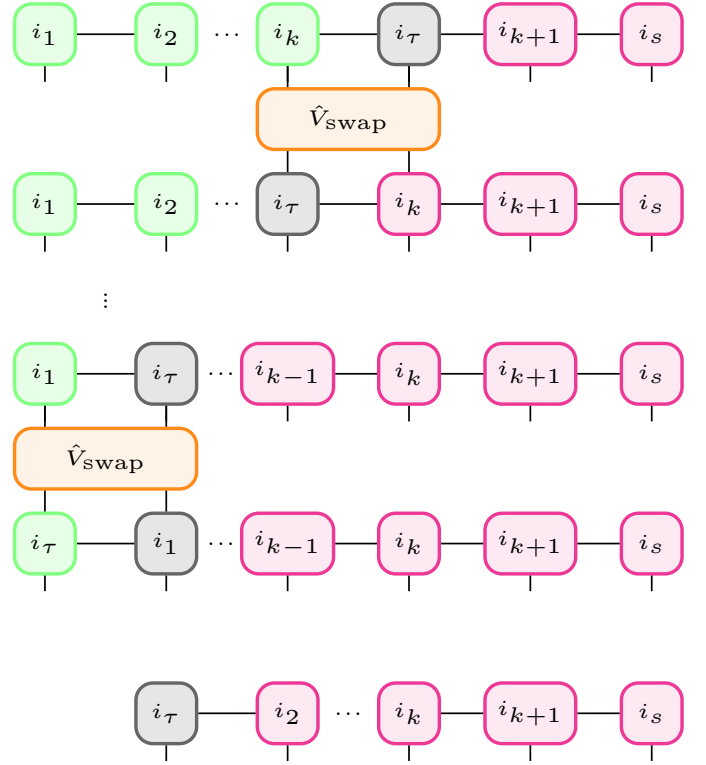}
    \caption{Diagrammatic representation of the swap back steps using the swap operator ${V}_{\rm swap}$ for leaving the feedback bin in its original position, that is, on the left. After finishing these operations the OC is left in the first time bin which is renamed as $i_\tau$ becoming the new feedback bin for the next time step.}
    \label{swapback}
\end{figure}

Finally, when working with two TLSs (see Fig.~\ref{schematicsMPS}(c)), the procedure is similar to the one described for the feedback case, following also the steps shown in Figs.~\ref{swapping}-\ref{swapback}, but now with the time evolution operator shown in Eq.~\eqref{u2tls}. In this case, the two time bins involved correspond to the two TLSs. The main difference is that a new basis is introduced as the system bin now includes the two TLSs; hence, it will have a dimension $d=4$ (Eq.~\eqref{2TLSsystemEQ}), and each time bin will also have a dimension $d=4$ (Eq.~\eqref{2TLStimeEQ}), as it counts the left and right moving photons. This can be seen in Fig.~\ref{schematicsMPS}(c), where the TLSs are represented together. In addition, each time bin contains both boxes (from $R$ and $L$) labelled with the same number.   

\subsection{Implementing the MPS algorithm  in {\sc Python}}
\label{subsec:impl}

The systems described are implemented in {\sc Python} (specifically 3.7) to obtain their time evolution and compare later with the SDW model results (described below). The general descriptions below can of course also be adapted to other programming environments such as {\sc C},  {\sc C++}, and {\sc Matlab}. 

Firstly, the one site operators are written as matrices. This includes the noise operators of the time bins and the creation and annihilation operators for the TLS.

Once these basic operators are defined, the time evolution, swap operator and the TLS population are defined, following the equations shown in \ref{subsec:oper} and \ref{subsec:hamil}. As described before, these operators are represented as tensors.   
The time evolution operator is easily defined in {\sc Python} using the exponential function included in the \pyth{scipy linalg} package. This function is very efficient, allowing us to avoid any approximation or expansions
that are frequently done for approximating the exponential of an operator~\cite{droenner_out--equilibrium_2019}. The terms in the exponential can be created by making use of the basic operators already defined which will be multiplied via the Kronecker product of arrays included in the \pyth{numpy} package.

For the initialization at time zero, the initial state is defined as follows,
\begin{equation}
    \psi(0)=A^{i_0} \otimes A^{i_s},
\end{equation}
or in index notation,
\begin{equation}
    \psi(0)= \sum{a,b,c,d} A^{i_0}_{a,b} A^{i_s}_{c,d},
\end{equation}
where $a,b,c,d=1,bond$. Here, $a,b$ and $c,d$ correspond to the virtual links of the time bin and the TLS, respectively, and $i_0$ and $i_s$ to the physical links. The dimension of the physical indices is $d=2$ in the case of one TLS and $d=4$ in the case of two TLSs, and the initial bond dimension is 1 as there is no initial entanglement. The bond dimension will increase as the entanglement appears, and it will be limited to a determinate dimension which will vary depending on the precision of each case. A chosen value will be used, and increased manually if necessary.

For example, for one TLS initialized in the ground state and the waveguide in vacuum, we have
\begin{align}
    &A^{i_s}=\begin{bmatrix}
            1 \\
            0 \\
    \end{bmatrix}, \
    A^{i_0}=\begin{bmatrix}
            1 \\
            0 \\
    \end{bmatrix},
\end{align}
where $A^{i_s}$ represents the TLS bin, and $A^{i_0}$ represents the first time bin.
Alternatively, if the TLS is initially in the excited state, then 
\begin{equation}
\label{excitedTLS}
    A^{i_s}=\begin{bmatrix}
            0 \\
            1 \\
    \end{bmatrix}.
\end{equation}
In the case of two TLSs, the initial state will be the outer product of each TLS. If both start in the ground state, then
\begin{equation}
    A^{i_s}=\begin{bmatrix}
            1 \\
            0 \\
    \end{bmatrix}
    \otimes
    \begin{bmatrix}
            1 \\
            0 \\
    \end{bmatrix}=
    \begin{bmatrix}
            1 \\
            0 \\
            0 \\
            0 \\
    \end{bmatrix}.
\end{equation}

Note that when we have one TLS in the MPS scheme, we shall refer to the TLS system bin, and for two TLSs, we shall refer to the TLSs bin, in the sense that we consider a common bin labelling with both TLSs. Thus, later, when we explore entanglement
between the TLSs bin and the waveguide, we are treating the two TLSs as a common system (which of course, also become entangled in their reduced Hilbert space). 

If any of the TLSs starts in an excited state, then the excited TLS will be written as in Eq.~\eqref{excitedTLS}, e.g.,
\begin{equation}
    A^{i_s}=\begin{bmatrix}
            1 \\
            0 \\
    \end{bmatrix}
    \otimes
    \begin{bmatrix}
            0 \\
            1 \\
    \end{bmatrix}=
    \begin{bmatrix}
            0 \\
            1 \\
            0 \\
            0 \\
    \end{bmatrix}.
\end{equation}

Similarly, the waveguide is in vacuum which means there will not be moving photons in either direction of the waveguide, giving the following initial state,
\begin{equation}
    A^{i_0}=\begin{bmatrix}
            1 \\
            0 \\
    \end{bmatrix}
    \otimes
    \begin{bmatrix}
            1 \\
            0 \\
    \end{bmatrix}=
    \begin{bmatrix}
            1 \\
            0 \\
            0 \\
            0 \\
    \end{bmatrix}.
\end{equation}

Although it will not be studied in this paper, having a initial state of waveguide photons is also possible,
e.g., see Refs.~\cite{SnchezBurillo2015,Xu2018}. 

In order to operate the MPO on the MPS, we make use of the function
\pyth{ncon}~\cite{pfeifer_ncon:_2015}, which is
a tensor network contractor that contracts the common indices in each case, and summed indices are reduced to a single tensor or a number by evaluating the index sums. 
After each contraction a SVD must be performed. This is done with the SVD function predefined in the 
\pyth{scipy linalg} package for {\sc Python}. It will give us the left normalized matrix, the right normalized one and the Schmidt coefficients. These last ones will be contracted with any of them (depending the case) to recover the OC after each operation.

With those functions and following the steps shown in Section~\ref{subsec:comp}, the evolution of each system is computed, with various initial conditions
and pumping strengths.

\section{Space-Discretized Waveguide Model}
\label{sec:SDW}

An alternative, but less well developed,  approach to modelling waveguide QED is to use a ``collision model" where the system repeatedly interacts with discrete slices (or bins) of the environment \cite{Brun2002,Kretschmer2016,Ciccarello2017,Cilluffo2020}. This model can be integrated with the physical insight of QT theory \cite{Dalibard92,Carmichael92,Dum92a} and an intuitive strategy for modelling non-Markovian dynamics emerges. Namely, by expanding the non-Markovian system to include the waveguide it is interacting with (represented by say small spatial boxes), the whole system dynamics become Markovian again, similar to how one solves Maxwell's equation on a finite-size space grid. Thus, we can model the waveguide by slicing it into discrete time/space bins and simulate the full system dynamics using QT theory. This general approach was recently used by Whalen \cite{PhysRevA.100.052113} to model the dynamics of a single TLS with time-delayed coherent feedback. We further explain how to implement this model for two TLSs in an open waveguide as well as expand the model to include Lindblad output channels required for realistic simulation of relevant experimental setups such as with semiconductor QED systems and circuit-QED systems.
Also, since the SDW model uses QT theory as its backbone, it can give insight into individual realizations of the system stochastic dynamics on top of the ensemble average \cite{Dalibard92,Carmichael92,Dum92a}. Indeed,
as we will show below, the SDW model leads to a picture
of delayed conditioning for simulating the emission of a photon. This is an effect that is not captured in standard QT theory.
In addition, from a computational perspective, the independent nature of the individual trajectories allows for the SDW approach to be completely parallelizable and make use of modern computational infrastructures such as single machine multi-threading or computational clusters

To model an open waveguide over some space interval $[-L_0,0]$, we would typically describe it using the annihilation operators for the discrete frequency domain modes of the waveguide propagating to the left ($L$) or right ($R$), $b_{k,\alpha}$ with $\alpha \in \{ L,R \}$. Instead, the collision model transforms these operators into their time domain representations, $B_{n,\alpha}$, using a discrete Fourier transform. This gives the explicit relationship
\begin{align}
\label{discrete-Ops}
    B_{n,\alpha} ={} & \frac{1}{\sqrt{N}} \sum_{k=0}^{N-1} b_{k,\alpha} e^{(-1)^{m_{\alpha}} i \omega_k n \Delta t}, \\
    b_{k,\alpha} ={} & \frac{1}{\sqrt{N}} \sum_{n=0}^{N-1} B_{n,\alpha} e^{(-1)^{m_{\alpha} +1} i \omega_k n \Delta t}, \nonumber
\end{align}
where $\Delta t = L_0/N$ is the time domain sampling (i.e., the corresponding time bin step for each spatial slice), $\omega_k = 2 \pi k/L_0$, which assumes linear dispersion in the waveguide, and $m_{L,R} = 1,2$, which ensures the correct direction of propagation for the boxes in Eq.~\eqref{SDWPropagation}. With this representation, $B_{n,\alpha}$ can be thought of as representing all frequency modes of the field across the spatial length $-n \Delta t$ to $- (n+1) \Delta t$, with the commutator
\begin{equation}
    [B_{n,\alpha} , B^{\dagger}_{n',\alpha}] = \delta_{n,n'}.
\end{equation}

\begin{figure}[tbh]
    \includegraphics[width=0.48\textwidth]{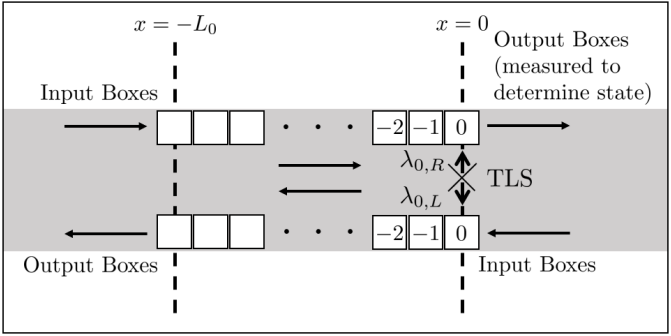}
    \caption{Schematic of a single TLS embedded in an open waveguide in the SDW model. The boxes are labelled by the negative of their index to emphasize that the spatial location of box $n$ is $x_n = -n \Delta t$, i.e. as the box number increases the corresponding location moves in the negative $x$ direction.}
    \label{SDW_Schematic_OpenWaveguide}
\end{figure}

It is important to note that these $B_{n,\alpha}$ operators are distinct from the $\Delta B_{\alpha} (t_k)$ operators introduced in the MPS description of the waveguide system. In the MPS formalism, the complete waveguide is described by the continuous time domain operators (Eq.~\eqref{continuous-time-ops}) and $\Delta B_{\alpha} (t_k)$ is a discrete time step of the interaction between the system and waveguide. Notably, the number of $\Delta B_{\alpha} (t_k)$ operators can grow as large as needed to reach the desired end time for the model. The growing Hilbert space is then truncated through the SVDs to keep this exact approach to the waveguide numerically tractable. On the other hand, in the SDW model, there are a finite number of $B_{n,\alpha}$ operators set at the beginning of the simulation by choice of $\Delta t$. The waveguide is then exactly described over the length of interest covered by these spatial operators,  and the evolution out of this length (and into the rest of the open waveguide) is dealt with through stochastic measurements of the outgoing boxes as described later. This trades out the exact approach of MPS with SVDs for a stochastic approach where individual realizations of the system must be averaged over. An advantage of this approach is that the ket state includes the waveguide section being modelled and contains the full entanglement between the waveguide and interacting QO system (e.g., a TLS). This is schematically shown in Fig.~\ref{SDW_Schematic_OpenWaveguide}, where a TLS is coupled to an open waveguide in the SDW model.

Before including a waveguide QO system, it is important to understand how the waveguide evolves in the SDW model. As in the MPS description, the waveguide undergoes free evolution of the frequency modes under the discrete form of the free Hamiltonian in Eq.~\eqref{HW}, which is now written as
\begin{equation}
    H_{\rm W}(\omega) = \sum_{\alpha \in \{ L,R \}} \sum_{k=0}^{N-1} \omega_k b_{k,\alpha}^{\dagger} b_{k,\alpha}.
\end{equation}
Then the evolution of the waveguide over a single time step is given by the operator $U_{\rm W} (\Delta t) = e^{-i H_{\rm W} \Delta t}$. This gives the evolution of the spatial operators for the right propagating modes to be
\begin{align}
\label{SDWPropagation}
    U_{\rm W}^{\dagger}(\Delta t) B_{n,R} U_{\rm W} (\Delta t) & {}= \frac{1}{\sqrt{N}} \sum_{k=0}^{N-1} b_{k,R} e^{i \omega_k (n-1) \Delta t} \\
    & {}= B_{n-1,R}, \nonumber
\end{align}
while for the left propagating modes:
\begin{align}
\label{SDWPropagationb}
   U_{\rm W}^{\dagger}(\Delta t) B_{n,L} U_{\rm W} (\Delta t) = B_{n+1,L}.
\end{align}
Therefore, over each time step, the waveguide evolves by passing along one ``box" of the waveguide to the next. The boxes moving to the right flow from the $N-1$'th box to the $0$'th box, and the boxes moving to the left flow from the $0$'th box to the $N-1$'th box, as shown in Fig.~\ref{SDW_Schematic_OpenWaveguide}.

Explicitly, the ket vector for the waveguide is now
\begin{equation}
    \ket{\psi_{\rm W}} = \prod_{\alpha \in \{ L,R \}} \ket{l_{N-1,\alpha}, ..., l_{0,\alpha}},
    \label{waveguide ket}
\end{equation}
where $\ket{l_{n,\alpha}}$ is the number state from $-n \Delta t$ to $-(n+1) \Delta t$. In order to keep the size of this basis numerically accessible for simulations, we make two assumptions. First, $l_{n,\alpha} \in \{ 0,1 \}$ so that in each directional box of the waveguide there is a maximum of one photon. Second, we fix $\sum_{\alpha \in \{ L,R \}} \sum_{n=0}^{N-1} l_{n,\alpha} = M$, allowing for a maximum of $M$ excitations in the waveguide, where we can choose $M$ to maintain a basis size which is numerically accessible, typically this choice is $M = 1$ or $2$ (and we will use both later). These assumptions are good as long as $\Delta t$ is sufficiently small (equivalently, $N$ is sufficiently large) so that the dynamics of the interaction Hamiltonian are well resolved.

Unlike in the MPS formalism, once the photon has left the modelled waveguide segment, it is not accounted for in the system state. Therefore, before $U_{\rm W} (\Delta t)$ can be applied and the waveguide boxes move forward, the information in the final box must be accounted for. To do this, Eq.~\eqref{waveguide ket} can be separated into three components; one where the final box of each directional set of boxes is empty and two where each final box contains a photon but the other does not, so that
\begin{align}
    \ket{\psi_{\rm W} (t)} ={} & \ket{\psi_0 (t)} \ket{l_{0,L} = 0} \ket{l_{0,R} = 0} \\
    & {}+ \ket{\psi_{1,L} (t)} \ket{l_{0,L} = 1} \ket{l_{0,R} = 0} \nonumber \\
    & {}+ \ket{\psi_{1,R} (t)} \ket{l_{0,L} = 0} \ket{l_{0,R} = 1}. \nonumber
\end{align}

A simulated ``measurement'' is made on each of the final boxes with probability $\braket{\psi_{1,\alpha} (t) | \psi_{1,\alpha} (t)}$, similar to the check for a quantum jump in the QT formalism~\cite{Dalibard92,Carmichael92,Dum92a}. Here we make the approximation that the photon can only be measured in one of the final boxes each time step, which avoids the simultaneous detection of two photons. If a photon is determined to be present in the box propagating in direction $\alpha$, the system is projected into the $\ket{\psi_{1,\alpha}(t)}$ state and both final boxes are emptied. If no photon is present, the system is projected into the $\ket{\psi_0(t)}$ state with both final boxes already emptied. Then $U_W (\Delta t)$ can be applied to the system with periodic boundary conditions,
\begin{align}
    \ket{\psi_{\rm W} (t)} ={} & \prod_{\alpha \in \{ L,R \}} \ket{l_{N-1,\alpha}, ..., l_{0,\alpha}}, \\
    & \hspace{0.4cm} \Downarrow \nonumber \\
    \ket{\psi_{\rm W}' (t + \Delta t)} ={} &  \prod_{\alpha \in \{ L,R \}} \ket{0, l_{N-1,\alpha}, ..., l_{1,\alpha}}, \nonumber
\end{align}
without losing any information. This process does not conserve the norm of the system and so before the time step can be completed, the system must be renormalized. Thus, the ket vector for the waveguide after each time step is
\begin{equation}
    \ket{\psi_{\rm W} (t + \Delta t)} = \frac{\ket{\psi_{\rm W}' (t + \Delta t)}}{\braket{\psi_{\rm W}' (t + \Delta t) | \psi_{\rm W}' (t + \Delta t)}}.
\end{equation}

Due to the stochastic nature of this approach, each realization of the system will be a QT which needs to be averaged against a suitable number of trajectories to arrive at the ensemble average behaviour of the system. Also, by setting the new incoming boxes to be in the ground state, we are making the assumption that the incoming fields are empty, however this is not a strict restriction of the model.

This description of the waveguide will always be implemented as the final three steps of each time step: The final (outgoing) boxes will be measured to determine whether or not a photon is present and the state will be projected accordingly with the final boxes set to the vacuum state like an absorbing boundary condition. Then, the free evolution of the waveguide will be applied and all boxes stepped forward removing the now empty final box and introducing a new empty box at the start of the box chain. Lastly, the state must be renormalized before moving on to the next time step.

\subsection{Modelling the Schemes of Interest with Space Discretization}
\label{sec:SDW_Schemes}

Before applying the SDW model to the systems of interest presented in Sec.~\ref{sec:Hamiltonian}, we will explain how to apply this model to a general QO system of interest with the interaction Hamiltonian presented in the continuous frequency domain. The Hamiltonian for a waveguide coupled to some arbitrary QO system is
\begin{equation}
    H = H_{\rm S} + H_{\rm W} + H_{\rm I},
\end{equation}
where $H_{\rm S}$ is the Hamiltonian for the arbitrary QO system (including pumping) and $H_{\rm I}$ is the interaction Hamiltonian between the system and waveguide. 
In the continuous frequency domain, this is
\begin{equation}
\label{eq:HI}
    H_{\rm I} = \sum_{\alpha \in \{ L,R \}} \int_{-\infty}^{\infty} d\omega \left( \kappa_{\alpha}(\omega) a^{\dagger} b_{\alpha}(\omega) + {\rm H.c.} \right) ,
\end{equation}
where $a$ ($a^{\dagger}$) is arbitrarily chosen as the annihilation (creation) operator for the system and $\kappa_{\alpha}(\omega)$ is the frequency dependent coupling function between the QO system and the $\alpha$ directional frequency mode in the waveguide. 

Equation \eqref{eq:HI} is next transformed into the discrete frequency domain by converting the continuous integral to a discrete sum over the frequency modes and substituting the continuous operators to the discrete operators, $b_{k,\alpha}$. The result of this transformation is to introduce a factor of $\sqrt{2 \pi / L_0}$ to the Hamiltonian,
\begin{equation}
    H_{\rm I} = \sqrt{\frac{2 \pi}{L_0}} \sum_{\alpha \in \{ L,R \}} \sum_{k=0}^{N-1} \left( \kappa_{\alpha}(\omega_k) a^{\dagger} b_{k,\alpha} + {\rm H.c.} \right).
\end{equation}
Lastly, the interaction is transformed into the spatial frame by direct substitution of Eq.~\eqref{discrete-Ops} for $b_{k,\alpha}$, giving
\begin{equation}
    H_{\rm I} = \sum_{\alpha \in \{ L,R \}} \sum_{n = 0}^{N-1} \left( \lambda_{n,\alpha} a^{\dagger} B_{n,\alpha} + {\rm H.c.} \right),
\end{equation}
where
\begin{equation}
   \lambda_{n,\alpha} = \sqrt{\frac{2 \pi}{N L_0}} \sum_{k = 0}^{N-1} \kappa_{\alpha}(\omega_k) e^{(-1)^{m_{\alpha} +1} i \omega_k n \Delta t}.
\end{equation}
Then the system couples to the $n$'th waveguide box with a coupling rate of $\lambda_n$ for $n \in \{ 0, ..., N-1 \}$. This setup allows the system to couple to the waveguide at an arbitrary number of places, but in practice this is restricted to one or two choices for $n$ for this paper through the choice of $\kappa_{\alpha}(\omega_k)$.

Therefore, the complete ket vector for the model is
\begin{equation}
    \ket{\psi(t)} = \ket{\psi_{\rm S}(t)}  \ket{\psi_{\rm W}(t)},
\end{equation}
where $\ket{\psi_{\rm S}(t)}$ is the ket vector for the QO system of interest. The simulation over 
one time step ($\Delta t$) following a four step algorithm is:
\begin{enumerate}
    \item Evolve $\ket{\psi(t)}$ under the combined QO system and interaction Hamiltonians, $H_{\rm S} + H_{\rm I}$, by direct application of $e^{-i (H_{\rm S} + H_{\rm I}) \Delta t}$.
    \item Take a direct measurement on the final boxes of the waveguide and project the ket vector accordingly.
    \item Shift the waveguide boxes one step under the operator $U_{\rm W} (\Delta t)$.
    \item Renormalize the system, so that the next ket state (after the time step) is
    \begin{equation}
        \ket{\psi(t + \Delta t)} = \frac{ \ket{\psi_{\rm S}(t + \Delta t)}  \ket{\psi_{\rm W}' (t + \Delta t)} }{ \braket{\psi_{\rm W}' (t + \Delta t) | \psi_{\rm W}' (t + \Delta t)} }.
    \end{equation}
\end{enumerate}

In the following three parts, we will derive the interaction Hamiltonian for the three schemes of interest shown in Fig.~\ref{schematics2}, with the Hamiltonians described in Sec.~\ref{sec:Hamiltonian}.

\subsubsection{Scheme (i): Single Two Level System in an Infinite Waveguide}

The initial system of interest is a single TLS in an infinite waveguide as shown in Fig.~\ref{schematics2}(a). For this scheme, the continuous coupling function is $\kappa_{\alpha} (\omega_0) = \sqrt{\gamma_{\alpha} / 2\pi}$, where we allow for non-equal coupling to the left and right propagating modes of the waveguide. Then the spatial coupling function to the right is
\begin{equation}
    \lambda_{n,R} = \sqrt{\frac{2 \pi}{N L_0}} \sum_{k = 0}^{N-1} \sqrt{\frac{\gamma_{R}}{2 \pi}} e^{-i \omega_k n \Delta t} = \sqrt{ \frac{\gamma_{R}}{\Delta t} } \delta_{0,n},
\end{equation}
where we have used the identity 
\begin{equation}
    \delta_{n,m} = 1/N \sum_{k = 0}^{N-1} e^{i \frac{2 \pi k}{N} (n-m)},
\end{equation}
and similarly $\lambda_{n,L} = \sqrt{ \gamma_{L}/ \Delta t }\, \delta_{0,n}$. Although this derivation of the spatial coupling functions may seem somewhat circular, it makes a clear connection to the more common frequency domain representations of the interaction Hamiltonian (for waveguides) and follows a general approach to deriving these functions.

Since the dynamics of the photons in the waveguide are unimportant after leaving the TLS, the interaction with the waveguide can be described by a single box, and the interaction Hamiltonian is simply
\begin{equation}
\label{eq:92}
    H_{\rm I} = \sum_{\alpha \in \{ L,R \}} \sqrt{\frac{\gamma_{\alpha}} {\Delta t}} \left[ \sigma^+ B_{0,\alpha} + \rm{H.c.} \right].
\end{equation}

It is also worth highlighting  that an alternative approach to deriving the interaction Hamiltonian in the SDW model is to simply represent the waveguide by the total waveguide field at the location of the TLS ($x=0$), $\mathcal{E}_{L} (0) + \mathcal{E}_{R} (0)$. Then the interaction Hamiltonian is
\begin{equation}
    H_{\rm I} = \sum_{\alpha \in \{ L,R \}} \sqrt{\gamma_{\alpha}} \left[ \sigma^+ \mathcal{E}_{\alpha} (0) + \rm{H.c.} \right],
\end{equation}
where this presumes $\mathcal{E}_{\alpha} (0)$ is written in photon flux units and can thus be replaced with $\mathcal{E}_{\alpha} (0) = B_{0,\alpha} / \sqrt{\Delta t}$, since $B_{0,\alpha}$ is the spatial box at the location of the TLS. Thus we obtain the same result (cf.~Eq.~\eqref{eq:92}). Note that shifting the spatial box that the TLS is located will simply introduce $\exp(\pm i\omega x_{0})$ factors to the fields, and thus shift the spatial box that the TLS couples to.

\subsubsection{Scheme (ii): Single Two Level System in a Half Open Waveguide with a Time-Delayed Coherent Feedback}

\begin{figure}[tbh]
    \includegraphics[width=0.48\textwidth]{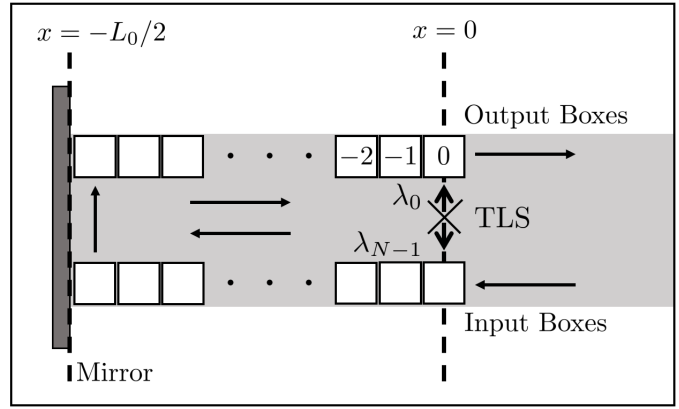}
    \caption{Schematic of a single TLS embedded in a half open waveguide introducing time-delayed feedback to the system in the SDW model.}
    \label{SDW_Schematic_FeedbackSystem}
\end{figure}

In order to include time-delayed feedback as shown in Fig.~\ref{schematics2}(b), we have to slightly change our approach to implementing the SDW model. Since the field that is emitted into the waveguide to the left is returned as the right propagating field from the mirror, we only need to use one set of boxes. These boxes enter, empty, travel to the left, propagate down the waveguide to the mirror, where they are reflected, and return to the TLS as the right propagating boxes. Once they arrive at the TLS, the box and TLS interact again and then the box leaves the system where it is measured for a photon. This is shown schematically in Fig.~\ref{SDW_Schematic_FeedbackSystem}.

The two coupling functions for this system are
\begin{align}
    \kappa_L (\omega_0) ={} & \sqrt{ \frac{\gamma_L}{2 \pi} }, \\ 
    \kappa_R (\omega) ={} & \sqrt{ \frac{\gamma_R}{2 \pi} } e^{i \phi} e^{i \omega \tau}, \nonumber 
\end{align}
where $\kappa_R (\omega)$ picks up the round trip phase of the photon in the interaction picture. Thus the coupling to the left is identical to the coupling without feedback, $\lambda_{n,L} = \sqrt{\gamma_L / \Delta t} \, \delta_{0,n}$, and the coupling to the right is
\begin{align}
    \lambda_{n,R} & {}= \sqrt{\frac{2 \pi}{N L_0}} \sum_{k = 0}^{N-1} \sqrt{\frac{\gamma_R}{2 \pi}} e^{i \phi} e^{-i \omega_k (n \Delta t - \tau)}, \\
    & {}= e^{i \phi} \sqrt{ \frac{\gamma_R}{\Delta t} } \delta_{n,-(N-1)}, \nonumber
\end{align}
modified by the presence of the mirror.

Of course, $n = -(N-1)$ does not occur for $n \in \{ 0, \hdots, N-1 \}$, so we reparamaterize and set the fictional $-(N-1)$ box to be $0$ and box $0$ to be $N-1$. Therefore, the interaction Hamiltonian for this scheme is
\begin{align}
    H_{\rm I} ={} & \sqrt{\frac{\gamma_L} {\Delta t}} \left[ \sigma^+ B_{N-1} + \rm{H.c.} \right] \\
    & {}+ e^{i \phi} \sqrt{\frac{\gamma_R} {\Delta t}} \left[ \sigma^+ B_{0} + \rm{H.c.} \right], \nonumber
\end{align}
where the directional subscript on $B_n$ has been dropped since there is only one set of boxes needed for this scheme.

\subsubsection{Scheme (iii): Two Waveguide-Coupled Two Level Systems Separated by a Time Delay}
\label{sec:SDW_2TLSs}
\begin{figure}[tbh]
    \includegraphics[width=0.48\textwidth]{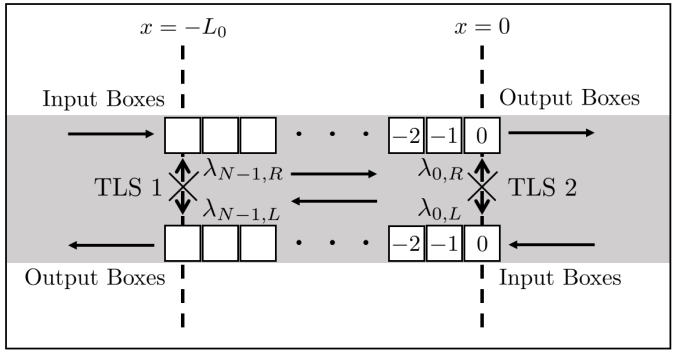}
    \caption{Schematic of two TLSs embedded in an open waveguide with non-negligible separation between them in the SDW model.}
    \label{SDW_Schematic_TwoTLS}
\end{figure}

The third scheme of interest is two spatially separated TLSs in an open waveguide as depicted in Fig.~\ref{schematics2}(c). This system is commonly investigated under the assumption that the spatial separation is negligible to the dynamics of the system in order to recover Markovian dynamics (thus neglecting time retardation), and when the non-Markovian effects are included, it is commonly in the single excitation regime \cite{PhysRevLett.124.043603}. We can relax this assumption with the SDW model and treat the non-Markovian effects from the separation of the TLSs with dynamics from up to four quanta in the entire system included (one in each TLS and two in the waveguide). The approach to modelling this system is similar to that of scheme (i), but now more than one box is needed to model the waveguide. Instead, $N$ boxes are introduced which span the distance between the two TLSs, shown schematically in Fig.~\ref{SDW_Schematic_TwoTLS}.

There are now four coupling functions which must be converted into their respective spatial coupling functions. These coupling functions are presented in Eq.~\eqref{twoTLS_interaction}, and following a similar approach to the previous two sections, the interaction Hamiltonian in the SDW model is
\begin{align}
    H_{\rm I} ={} & e^{i\phi} \sqrt{\frac{\gamma_{L1}} {\Delta t}} \left[ \sigma^+_1 B_{N-1,L} + \rm{H.c.} \right] \\
    & {}+ e^{i\phi} \sqrt{\frac{\gamma_{R1}} {\Delta t}} \left[ \sigma^+_1 B_{N-1,R} + \rm{H.c.} \right] \nonumber \\
    & {}+ \sqrt{\frac{\gamma_{L2}} {\Delta t}} \left[ \sigma^+_2 B_{N-1,L} + \rm{H.c.} \right] \nonumber \\
    & {}+ \sqrt{\frac{\gamma_{R2}} {\Delta t}} \left[ \sigma^+_2 B_{0,R} + \rm{H.c.} \right]. \nonumber
\end{align}

\subsection{Introducing Lindblad Output Channels}
\label{sec:Lindblad}

A central result of previous studies on coherent feedback systems is the ability to tune the phase of the returning feedback to enhance or suppress the output from the system. However, it is important to note that in physically realized systems, such as semiconductor quantum dots~\cite{PhysRevB.98.045309,
IlesSmith2017,
Kuhlmann2013,
PhysRevLett.104.017402,
PhysRevLett.118.253602,
RevModPhys.87.347,
PhysRevB.66.165312,
PhysRevLett.91.127401,
PhysRevB.63.155307,Trschmann2019}, feedback systems are ultimately less effective when one accounts for dissipation processes such as off-chip decay from the TLS and pure dephasing. To include these processes in the SDW model, our algorithm is amended to include Lindblad quantum jump operators from conventional QT theory~\cite{Dalibard92,Carmichael92,Dum92a}.

As an example, the previous schemes can be augmented by two quantum jump operators $C_0 = \sqrt{\gamma_0} \sigma^-$, representing off-chip decay from the TLS with rate $\gamma_0$, and $C_1 = \sqrt{\gamma'/2} \sigma_z$, representing pure dephasing in the TLS with rate $\gamma'$. It is important to note that the MPS approach neglects these terms and it is not clear how to include them in a numerically efficient way. In contrast, due to the already stochastic nature of the SDW model, including these output channels is quite natural and is one of the major advantages of exploiting QT theory to model waveguide QED.

In order to include these jumps, the algorithm's first step must be modified by evolving each QT under a non-Hermitian effective Hamiltonian,
\begin{equation}
    H_{\rm{eff}} = H_{\rm S} + H_{\rm I} - \frac{i}{2} \sum_{j = 0}^1 C^{\dagger}_j C_j.
\end{equation}
This evolution is further modified by stochastically introducing quantum jumps with a jump probability of
\begin{equation}
    \label{Lindblad_Prob}
    P(t) = \Delta t \sum_j \braket{\psi(t) | C^{\dagger}_j C_j | \psi(t)},
\end{equation}
for the time step beginning at $t$. If a jump is determined to occur, either $C_0$ or $C_1$ is chosen to be applied to the system according to their relative probabilities.
Note, an alternative approach to include the above processes is to add further streams of little boxes, one stream for each additional decay or dephasing channel. However, 
we find our presented QT formalism to be more intuitive. 

Since neither the evolution under $H_{\rm{eff}}$ or applying either quantum jump preserves the state norm, before moving on to the measurement of the final box in the second step of the algorithm, the state must be renormalized. Thus there must be two renormalizations during each step of the system evolution.

It is also important to note that, unlike in typical QT theory, if a quantum jump occurs, the waveguide Hamiltonian is still applied to the system, i.e., the boxes still shift. For a typical QT, if a jump occurs, the Hamiltonian is not applied to the system and instead the jump operator is applied. If this were to be followed for this model with feedback, then the feedback would return to the system at irregular times. Therefore, the waveguide Hamiltonian must be decoupled from the system and interaction Hamiltonians in order to maintain a consistent round trip time for the feedback.

\subsection{Implementation in {\sc Python}}
\label{SDW_Python}
One of the major benefits of the SDW model is the ease of implementing the model in the users preferred coding language, especially if the user is familiar with QTs in general. Similar to our MPS implementation, our SDW implementation uses {\sc Python 3.7}, which exploits its straightforward parallelization abilities and the sparse matrix capabilities of \pyth{scipy}.

To begin a simulation, the ket vector is initialized as an outer product of $\ket{\psi_{\rm S}}$ and $\ket{\psi_{\rm W}}$, which is represented as a vector of length $N_{\rm S} \times \sum_{j = 0}^{M} {2N \choose j}$; here $N_{\rm S}$ the size of the QO system basis and $\sum_{j = 0}^{M} {2N \choose j}$ the size of the waveguide basis. Note that $2N$ is used in the factorial because the limit of $M$ photons in the waveguide encompasses both directions of field propagation in the loop (the $2$ is dropped for the feedback scheme since there is only one row of boxes). To give an example, for a typical simulation, we would choose 20 boxes ($N=20$) and allow for two photons in the loop ($M=2$) which gives a vector length of 422 for the single TLS with feedback and 3284 for two TLSs in an open waveguide. The evolution under $e^{-i H_{\rm{eff}} \Delta t}$ is done similarly to our MPS implementation by utilizing the exponential function in \pyth{scipy linalg}, but we also convert it into a sparse matrix to save both memory and computation time since there are many levels of the ket vector which do not interact.

The simulation then runs by following the algorithm described in Sec.~\ref{sec:SDW_Schemes}, implementing each time step (with Lindblad output channels included) until the desired end time is reached. First, there is a check for whether a quantum jump from the Lindblad output channels occurs, with probability $P(t)$ (\eqref{Lindblad_Prob}). To do this check, a uniformly distributed random number, $\epsilon$, is generated and compared against $P(t)$. If $\epsilon < P(t)$ then a jump occurs, with the responsible jump operator chosen from their respective relative probabilities compared against a second uniformly distributed random number. Otherwise, if $\epsilon > P(t)$, then the system ket vector is evolved by direct multiplication of $e^{-i H_{\rm{eff}} \Delta t}$ and then renormalized. Note that this step can be done with a smaller time step, $\delta t < \Delta t$, multiple times per large time step in order to resolve fast system Hamiltonian dynamics. Other methods of evolution can be used such as various Runge-Kutte 
approaches, however these require a larger number of calculations per time step and can slow down the numerics significantly for small gains in accuracy.  Next, a direct measurement of the output boxes is taken to determine if a photon leaves the system from the waveguide. This is implemented similarly to the Lindblad jump operators with the probabilities now given by $\braket{\psi_{1,\alpha} (t) | \psi_{1,\alpha} (t)}$. Penultimately, the waveguide boxes are all shifted forward, and lastly, the ket vector is renormalized again.

The benefits of this implementation is that essentially each step of the code is obtained by simply writing down the explicit mathematical calculation that needs to be done without any outside functions. This makes the implementation quite easy and transparent for a simple system. The complexity arises as the system of interest becomes more intricate, for example through the inclusion of more Lindblad output channels or complex QO systems. Also, since the ket vector is known at each time step, any desired observables can be calculated either during or after the trajectory.

Of course, since each QT simulated is a single realization of the scheme of interest, in order to recover the ensemble average dynamics these trajectories must be averaged over a large number of realizations. Depending on the system dynamics or the desired precision of the observable, this can require anywhere from 500 to 10000 trajectories. Since each trajectory is inherently independent of the other trajectories, this implementation is a prime candidate for parallelization across multiple CPUs which we have done using the \pyth{mpi4py} package for {\sc Python}. This can also make use of the multithreading on a single high-performance workstation or the many nodes of a computing cluster.

\section{Results}
\label{sec:results}

In this section, we  present numerical results of the two methods discussed in detail above.
For convenience,
we will present the graphical results in normalized units,
in terms of $\gamma$,
defined from:
 $\tilde t = t \gamma$, $\tilde\tau = \tau \gamma$, $\tilde \Omega = \Omega / \gamma$ $\tilde\gamma_0=\gamma_0 / \gamma$ and $\tilde\gamma'=\gamma' / \gamma$.

\subsection{Single two level system in a waveguide with and without a time-delayed feedback: vacuum dynamics}

We will first explore the case of a single TLS in a waveguide, with and without feedback, beginning with the simple spontaneous emission dynamics in vacuum.

\begin{figure}[h]
    \centering
    \includegraphics[width=1 \columnwidth]{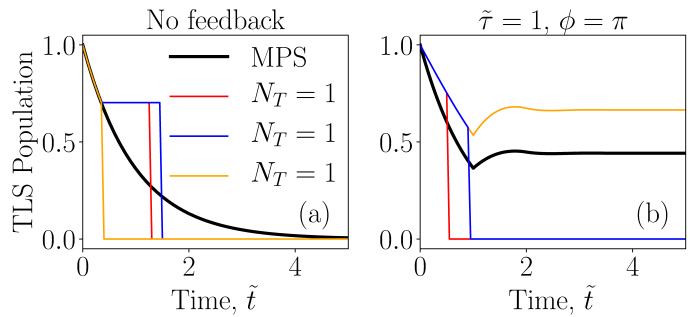}
    \caption{Vacuum decay of the TLS population for a single TLS in a waveguide with (a) no feedback, and (b) with feedback, using delay parameters $\tilde\tau=1$, $\phi=\pi$. The
    MPS result is shown in black
    and the SDW model with a single
    QT is shown in the colored 
    lines.
    The stochastic nature of the SDW model is clear, where for one trajectory the TLS population 
    decays to the ground state or is trapped after a single jump~\cite{crowder_quantum_2020}.
    Note that the SDW model results in {\em delayed conditioning} for single QT runs.
    The times shown are normalized to the decay rate, so the nominal decay rate of $\gamma$ would reach $1/e \approx 0.368$ at $\tilde t=1$, and thus for  $\tilde \tau=1$, this is also precisely when the feedback effects start to appear. }
    \label{multi1}
\end{figure}

As discussed earlier, since the SDW model computes  stochastic dynamics,
 expectation values 
can be obtained from an average over a finite
number of trajectories. 
To make this clear, at the few QT level, in  Fig.~\ref{multi1}, 
we show single trajectories calculated with this model
for a TLS in an infinite waveguide
and for a half open waveguide with a feedback delay (see sections \ref{subsec:scheme1} and \ref{subsec:scheme2}), with
 $\tilde \tau=1$
and $\phi=0$. We also  compare these  with the direct results given by the MPS, which shows a smooth decay in the
case of no feedback, and population trapping
for the case of feedback,
which recovers previous vacuum results that have been reported 
elsewhere~\cite{PhysRevA.92.053801,
PhysRevLett.110.013601,crowder_quantum_2020}.
Since we only show a single QT (which is deliberate), the results obviously do not overlap, since when one quantum jump happens, the TLS population simply decays to the ground or becomes trapped.
However, for a larger number of trajectories, we recover excellent agreement from both the
MPS  and SDW models as shown below.

Note that standard QTs for spontaneous emission begin as horizontal lines with no exponential decay.
In the present case, the exponential behaviour occurs
because there are 20 space bins interacting with the 
TLS before the interaction with the output bin begins; 
 thus, the QTs are effectively conditioned on photon counts made downstream from the TLS. 

\begin{figure}[h]
    \centering
    \includegraphics[width=1 \columnwidth]{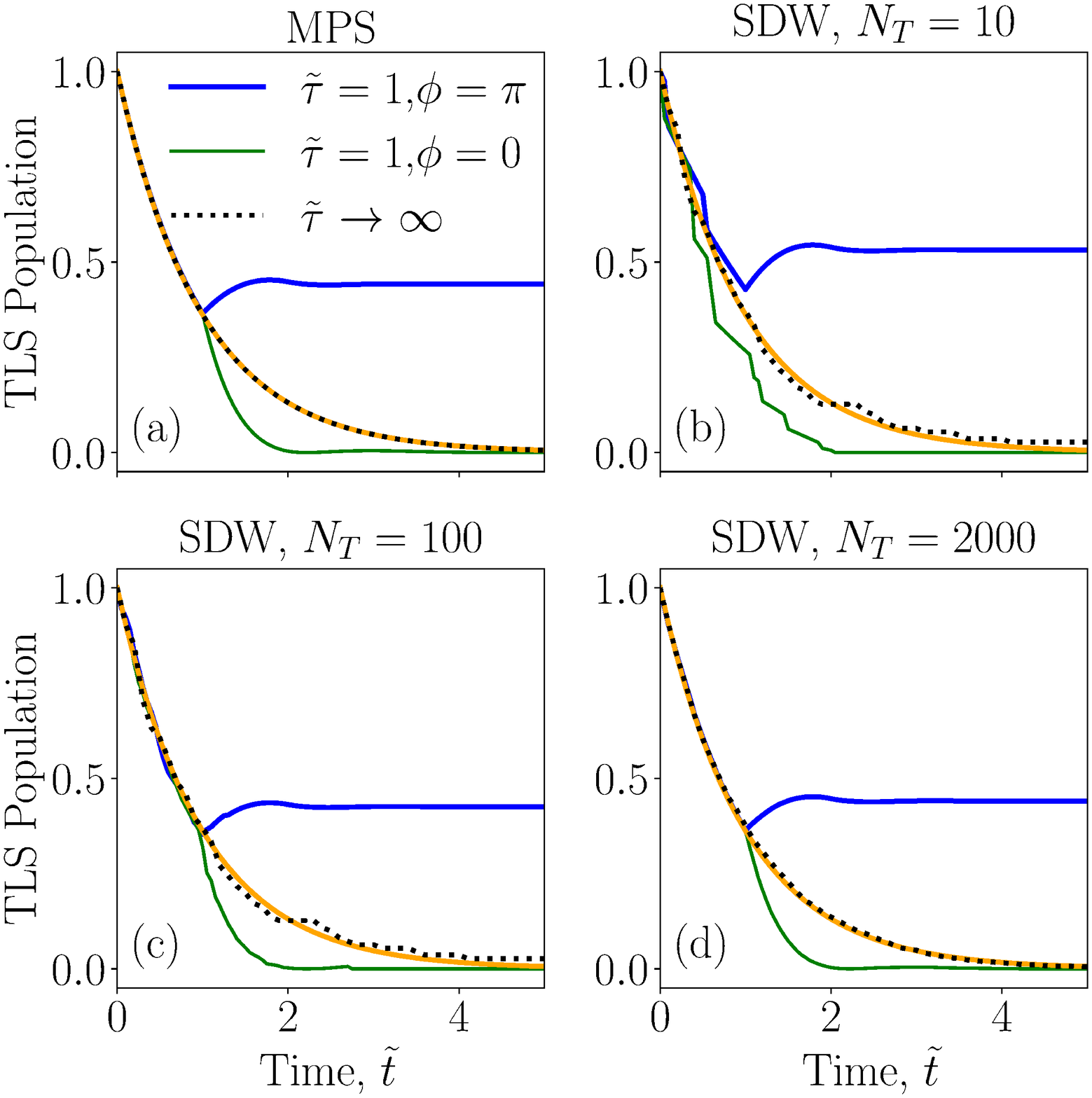}
    \caption{Decay of the TLS population for a single TLS in a waveguide with no feedback (black dashed), and with $\tilde \tau=1$ for the constructive case with $\phi=\pi$ (blue) and the destructive one (green). The case with no feedback is compared with the analytical solution (orange). In (a), the MPS method is used, and in (b-d) the SDW model is used averaged for 10, 100 and 2000 trajectories respectively. We see excellent agreement with both numbers
    after about $N_T>1000$.}
    \label{multi2}
    \vspace{0.2cm}
%
%
    \includegraphics[width=1 \columnwidth]{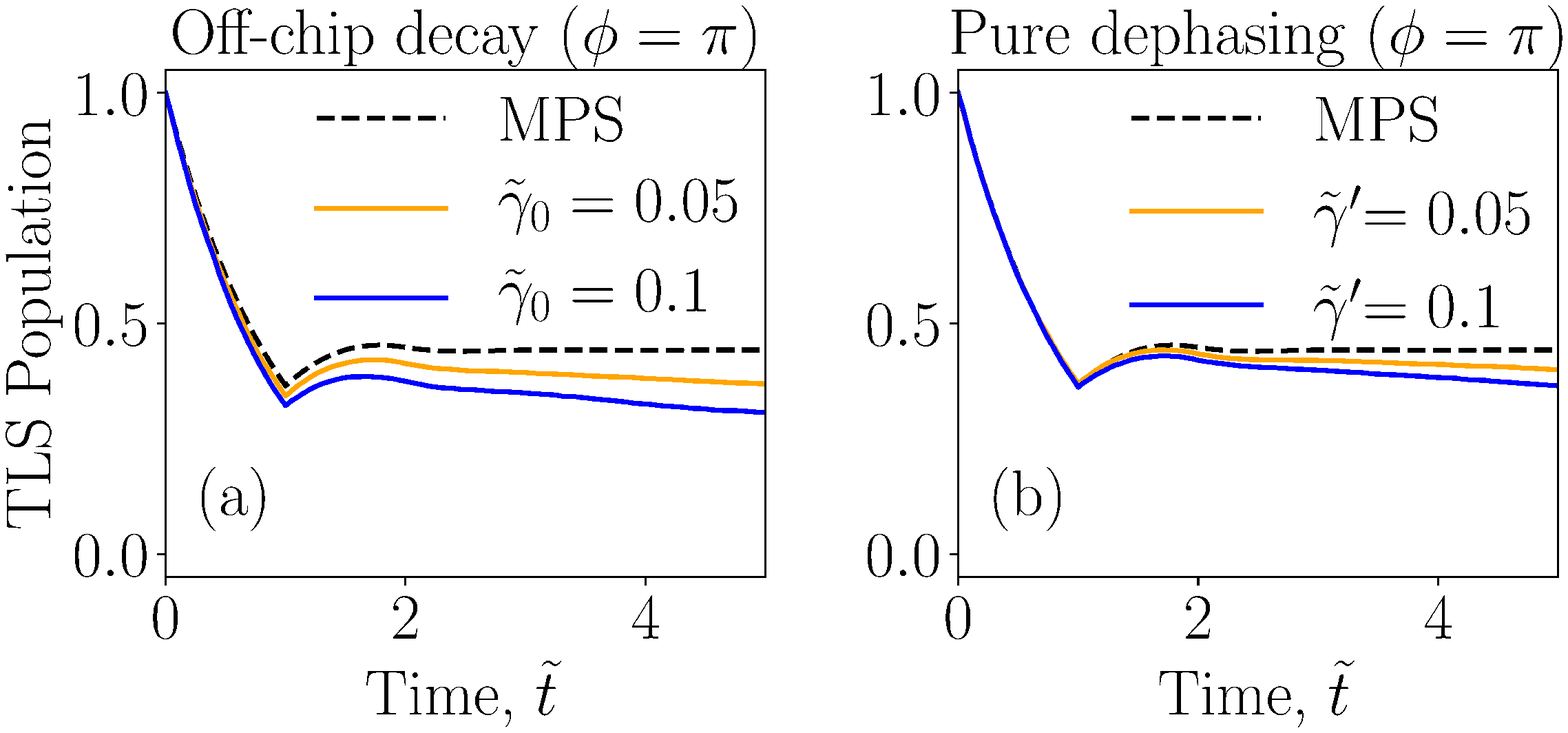}
    \caption{Decay of the TLS population for a single TLS in a waveguide with a constructive feedback ($\tilde \tau=1$, $\phi=\pi$). In (a), an off-chip decay ($\gamma_{0}$) is introduced in the SDW model. In (b), a pure dephasing process ($\gamma'$) is taken in account  for the SDW model. In both cases, we see that these additional dissipation processes prevents the case of perfect population trapping, and ultimately the population decays in the long time limit.
    }
    \label{multi3}
\end{figure}

In Fig.~\ref{multi2},
we next study the same TLS
decay for different feedback phases, yielding
 destructive interference with $\phi=0$ and constructive interference with $\phi=\pi$,
 and show how the number of
 QT averages affects the results when comparing with the MPSs.
 We now see how the phase
 can completely change the trapping scenario causing a faster decay with feedback when
 $\phi=0$.
 The MPS case is compared with 3 different cases for the SDW model, where 10, 100 and 2000 trajectories are considered. It can be seen that
 both methods agree extremely well
 in when $N_t=2000$, and we also show that both methods recover the simple analytical solution with no feedback, namely
 $n_a = \braket{\sigma^+\sigma^-}=\exp(-\gamma t)$.
 Note that there are other non-Markovian systems that lead to TLS population  trapping, such as
 fractional decay near the edge of a photonic bandgap~\cite{PhysRevA.50.1764,PhysRevA.47.3380}, 
 though in practise these would be very difficult
to realize~\cite{Kristensen2008}, most notably due to structural
disorder~\cite{PhysRevLett.94.033903}.

 Now that we have verified that both approaches can yield the same predictions
 for the vacuum dynamics, it is also important to compare the
 computational efficiencies, as well as the ease
 of numerical implementation (which we have discussed earlier).
 For these vacuum examples of a single TLS in a waveguide, with and without feedback, 
 the computational run times are compared in Tables~\ref{table1} and \ref{table2}. All the examples are run on the same computer workstation (125.6GB RAM, 3.70GHz, 16 cores).

%
\begin{table}[h]
\centering
\caption{Run times for TLS decay in an infinite waveguide (vacuum dynamics). $\Delta \tilde t = 0.05$, 10 boxes in the SDW code (non parallelized code), and a maximum bond dimension of 2 in the MPS code. 
}
\begin{tabular}{|p{2.0cm}|p{2.0cm}|p{2.0cm}| }
\hline
Model  & $N_T$ & Run Time (s) \\
\hline\hline
& 10 & 0.25 \\
SDW & 100 & 1.04 \\
 & 2000 &  14.22 \\
\hline
MPS &  &  0.11 \\
\hline
\end{tabular}
\label{table1}
%
\centering
\vspace{0.2cm}
\caption{Run times for TLS decay in semi-infinite waveguide (vacuum dynamics). $\Delta \tilde t = 0.05$, 20 boxes in the SDW code (non parallelized code), and a maximum bond dimension of 2 in the MPS code.}
\begin{tabular}{|p{1.6cm}|p{2.0cm}|p{2.0cm}|p{2.0cm}| }
\hline
Model  & \# of  & \multicolumn{2}{|c|}{Run Time (s)}\\
\cline{3-4}
& trajectories & $\phi=\pi$ & $\phi=0$ \\
\hline\hline
& 10 & 0.15 & 0.24\\
SDW & 100 & 0.69 & 1.76\\
& 2000 & 12.66 & 33.56 \\
\hline
MPS &  & 1.18 & 1.30 \\
\hline
\end{tabular}
\label{table2}
\end{table}

 In the infinite waveguide case
 (Table~\ref{table1}), the MPS code is faster than any of the cases given with the SDW.  Once the feedback is introduced into the MPS approach, as each time step involves more operations, the MPS code slows down. However, it is still comparable to around 100 trajectories. 

\begin{figure*}
    \centering
    \includegraphics[width=1 \textwidth]{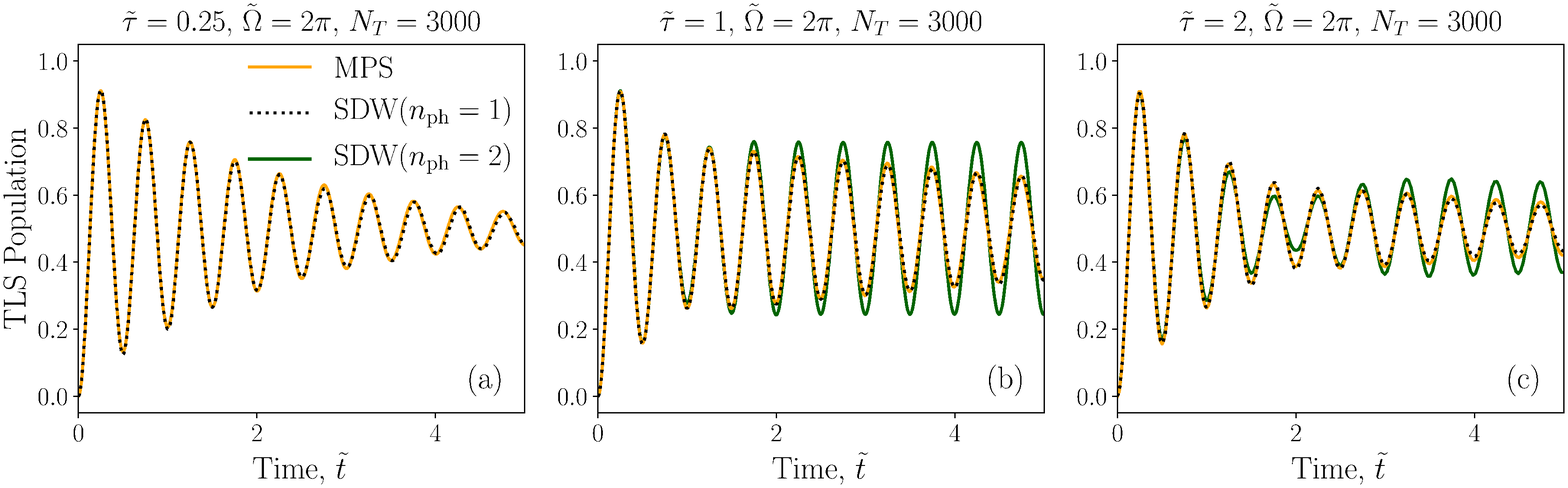}
    \caption{Single TLS driven by a CW pump field with  $\tilde\Omega=2\pi$. Comparison of the TLS population using MPSs (orange) and the SDW model for one photon in the waveguide (green) and 2 photons in the waveguide (dashed black). Different feedback lengths are presented: $\tilde\tau=0.25$ in (a), $\tilde\tau=1$ in (b), and $\tilde\tau=2$ in (c). All the SDW cases are run for 3000 trajectories.}
    \label{multi3b}
    \centering
    \includegraphics[width=0.8 \textwidth]{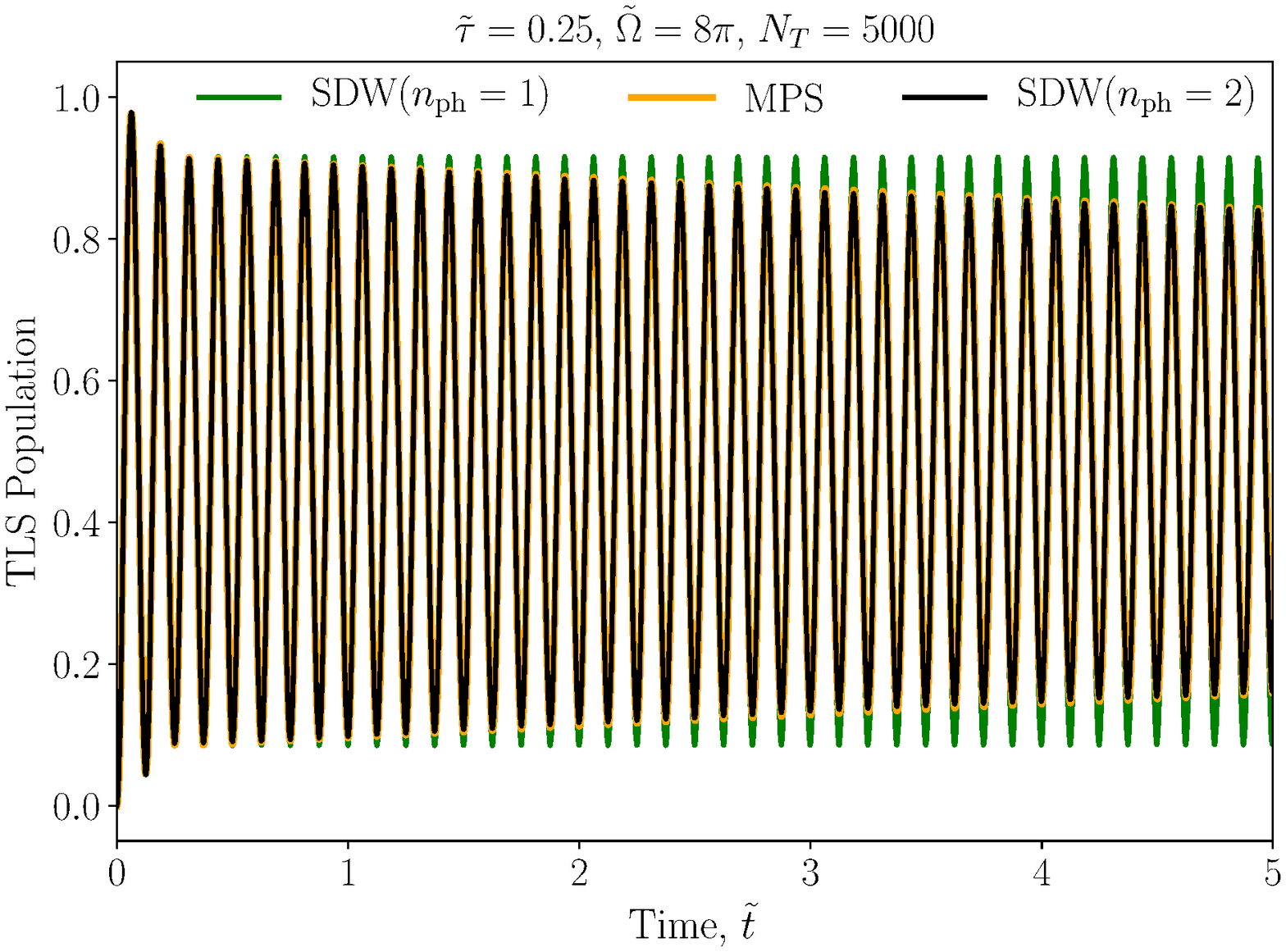}
    \caption{Single TLS driven by a strong CW pump field with  $\tilde\Omega=8\pi$. Comparison between the MPS (orange) and the SDW model for one photon in the waveguide (green) and 2 photons in the waveguide (black). The evolution of the TLS population shows a good agreement between the 2 photons SDW model and the MPS model. However, the 1 photon limit deviates from these two others showing a perfect trapping condition (within numerical precision)~\cite{Grimsmo2015,Droenner2019}.}
    \label{multi4}
\end{figure*}

While the MPS approach appears to be faster for the presented examples,
they are clearly both efficient, and it is important to note that the computational implementation
and intuitive understanding is much more complex. In addition,
the SDW model can easily add in additional dissipation processes
that are known to be important for connecting to real experiments, including
off-chip photon decay and pure dephasing -- the latter process is a well known feature with solid state quantum bits (quantum dots)~\cite{PhysRevB.98.045309,
IlesSmith2017,
Kuhlmann2013,
PhysRevLett.104.017402,
PhysRevLett.118.253602,
RevModPhys.87.347,
PhysRevB.66.165312,
PhysRevLett.91.127401,
PhysRevB.63.155307,Trschmann2019}.
As remarked earlier, implementing
such processes with MPSs is not well developed and non trivial. To demonstrate the role of these processes, Fig.~\ref{multi3} shows the behavior of the decay of a TLS in a waveguide with a coherent feedback when these two effects are considered, separately. Indeed, in both cases, we see that these additional dissipation channels break the regime of perfect TLS population trapping, and  it is essential to realize that these dephaning processes set a limit on how well one can exploit feedback in general. The run times increase around four times when these effects are added to the system, for the same parameters as the ones shown in Table~\ref{table2} with $N_T=2000$; this increase is more than reasonable given the complexity of the open system we are modelling, and the simulations are still efficient, even on a single computer.

It is also interesting to note that the role
of $\gamma_0$ and $\gamma'$ are qualitatively different in how they affect the trapping condition. This is because the $\gamma_0$ process affects both the population decay and the coherence, while the pure dephasing does not directly reduce the population, but instead dephases the coherence that is necessary for population trapping. Thus the effect of off-chip decay is more problematic, though both processes
lead to an overall decay of the trapped state.

\subsection{Single two level system in a waveguide with and without a time-delayed feedback: nonlinear dynamics with a coherent pump field}

Now that we have studied the vacuum decay dynamics of the TLS population, which can also easily be described
with classical linear response theories~\cite{PhysRevLett.98.083603}, the real power of our presented waveguide QED methods is in their ability to describe
nonlinear effects beyond a single quantum, namely unique 
quantum nonlinear effects that have no classical counterpart.
As an example,
one can explicitly include
one photon in the feedback loop, and the TLS or/and a side coupled cavity~\cite{crowder_quantum_2020}, which goes beyond the one quanta limit.
Thus we next add a coherent pump field to the system Hamiltonian
to access the quantum nonlinear regime.

Note,  one of the main advantages of using MPSs is that there is no restriction on the number of photons considered in the waveguide (subject to computational restrictions inherent in the method), while in the SDW model we are restricted to one or two photons in the waveguide for this study; the extension to include three or more photons is possible, 
but the computational overhead may be considerable.
However, it is very insightful to explicitly see the 
differences between  between the 1 photon and 2 photon results, and often 2 photons plus the TLS excitations
is enough for many few photon descriptions, even under extreme conditions (as we show below).
 
Figures~\ref{multi3b} and \ref{multi4} show coherent pump examples for the large drive strengths of $\tilde \Omega=2\pi$ and $\tilde \Omega=8\pi$, respectively, which easily break a weak excitation approximation (when the TLS is basically in the classical harmonic oscillator regime). 
 First, in Fig.~\ref{multi3b}, three different feedback lengths are considered for the same drive, yielding feedback times of $\tilde\tau=0.25, 1,2$. 
 It can be seen that, as the feedback time increases, the results with 1 photon deviate from the MPS model and the 2  photon SDW results, and thus becomes incorrect. However, the 2 photon case seems to agree well with the MPS model, showing that this approximation is very accurate here. We also highlight that  for  longer  feedback loops,  the SDW model becomes slower, especially in the case of the SDW with two photons. Physically, as the length of the feedback loop increases, the probability of having two photons in the waveguide increases as well, making the approach in which two photons are considered more accurate. 
 Table~\ref{table3} gives a summary of computational run times for the two models.

In  Fig.~\ref{multi4},
 it is important to note that the 1 photon limit case gives perfect population trapping~\cite{Grimsmo2015,PhysRevA.92.053801}, while there is no perfect trapping  in the 2 photons case and the MPS solution, and a decay can be seen. This is expected
 as it is practically impossible to phase match at two different frequencies, when quantum nonlinearies become important. This multi-photon influence on
 feedback-induced population trapping  is consistent with the results from  Grimsmo~\cite{Grimsmo2015} (whose results were limited to the early transient regime, showing only a few cycles). Note also, when running these codes for a non-trapping situation (e.g., with $\tilde\Omega=5$),  we found that the breakdown of the one photon case is 
 still important (though suppressed), and the 2 photon case again agrees quantitatively well with the MPS result.

\begin{table}[h]
\centering
\caption{Run times for a driven TLS with $\tilde\tau=2$ and $N_T=3000$. We use $\Delta \tilde t = 0.02$, 100 boxes in the SDW code (non parallelized code), and a maximum bond dimension of 32 in the MPS code. }
\begin{tabular}{|p{2.0cm}|p{2.0cm}|p{2.0cm}| }
\hline
Model  & \# Photons & Run Time (s) \\
\hline\hline
SDW & 1 & 63.10 \\
 & 2 &  693.89 \\
\hline
MPS &  &  43.43 \\
\hline
\end{tabular}
\label{table3}
\end{table}

In Fig.~\ref{multi4}, we now apply an even larger pump field ($\tilde\Omega=8\pi$) on the same system. We see again, that even in the very strong field regime,  the 2 photon SDW model is very accurate (agrees quantitatively well with the MPS results). However, we note the 2 photon SDW model needs more trajectories ($N_T=5000$)
to recover an accurate ensemble average, and a smaller time step in general, causing the SDW code to become somewhat slow and require more computational memory 
for accurate results; as an example,  with  a single multi-core workstation,
in this case the run time for the SDW is 2406s (for a time step $\delta \tilde t=0.002$) whereas the MPS code only takes 108s to run. Nevertheless, this is a very difficult nonlinear QO dynamic to model, and most other approaches to this problem would run into simulate computational problems or would not even be tractable (note we are also simulating for relatively long time scales with multiple oscillations). In addition, the longer feedback results are likely not as practical for applications, especially when one considers other realistic scattering processes (which are difficult for MPSs to handle).

\subsection{Two coupled two level systems in a waveguide with a finite delay time between them}

Next, we consider two TLSs in a waveguide  (see Sec.~\ref{subsec:scheme3}), with a finite separation between them. This example is a good test-best for beginning to exploit many-body interactions beyond the instantaneous coupling limit (an approximation that is frequently made when considering collective effects in the nonlinear regime). It is also a pedagogically important example, since
it is known to produce sub-radiant and super-radiant Dicke states \cite{Dinc2019,Zhang2019,PhysRevLett.124.043603},  bound states in the regime of ultrastrong waveguide QED~\cite{PhysRevA.102.023702,PhysRevA.100.013812}, and cause complex waveguide-mediated phase coupling \cite{Cheng2017}. 
For this system, we will again consider both vacuum dynamics and the case with strong optical pumping, as well as investigate the role of pure dephasing.

\begin{figure}[h]
    \centering
    \includegraphics[width=1 \columnwidth]{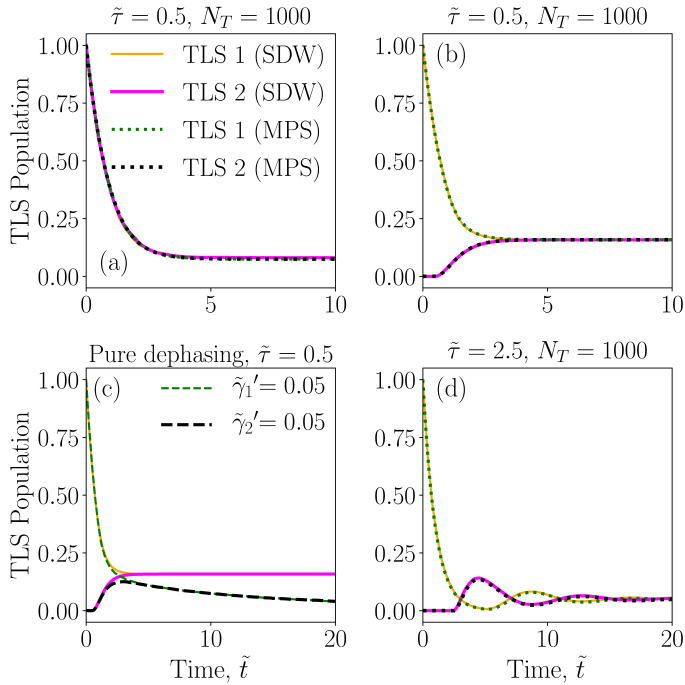}
    \caption{Decay of the TLSs populations for two TLSs in an infinite waveguide, with different spatial separations (delay times). A comparison between both methods is done in the case of (a) both TLSs on the excited state with a feedback of $\tilde\tau=0.5$,  and  (b) one TLS on the excited state and one on the ground state with a feedback of $\tilde\tau=0.5$; in (c) we show the same situation as (b) but now also with finite pure dephasing added to the TLSs;
    the effect of adding $\gamma_0$ (off-chip decay) has a very similar effect for the same decay rate so we do not show it. 
    In (d), one TLS is initialized on the excited state and one on the ground state with $\tilde\tau=2.5$. All the SDW cases are run for 1000 trajectories at a 2 photon truncation. 
    }
    \label{multi5}
\end{figure}

\begin{figure*}[ht]
    \centering
    \includegraphics[width=1 \textwidth]{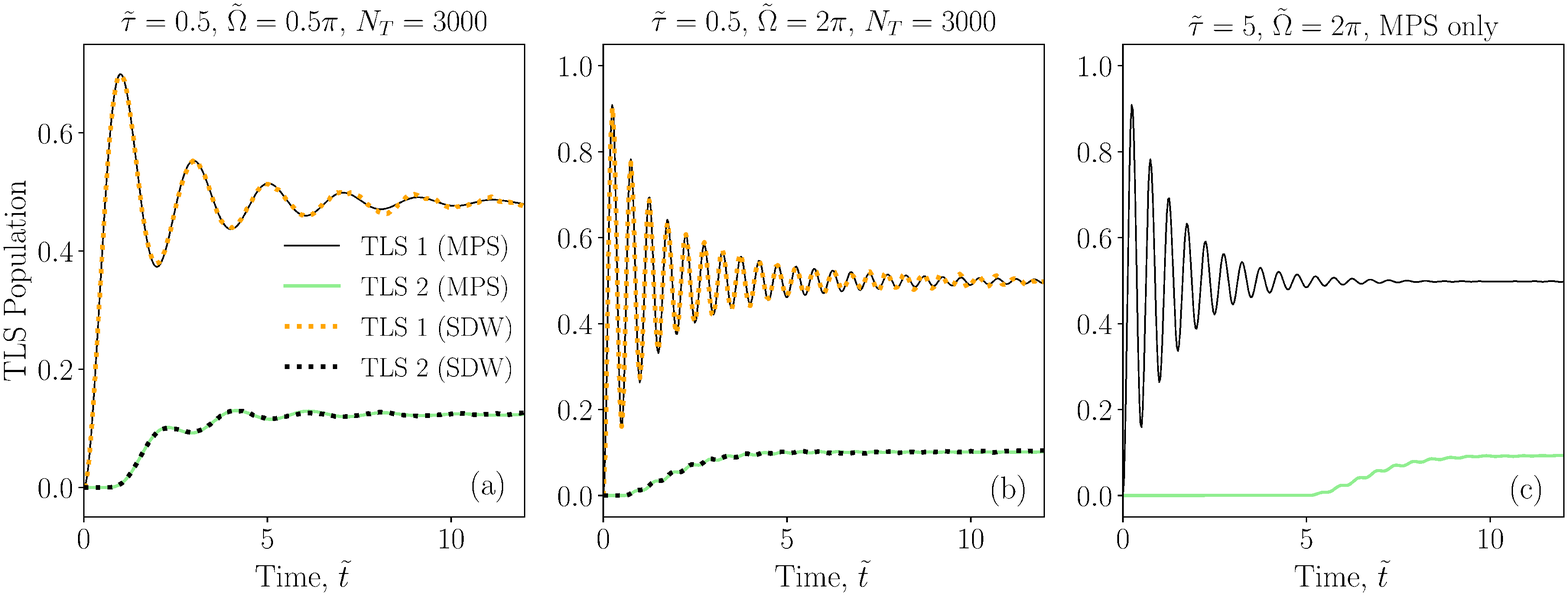}
    \caption{Evolution of the TLS populations for two TLSs in an infinite waveguide, when one of the TLS is driven by a pump field, with different pump strengths and spatial separations (delay times). In all cases, both TLSs start in the ground state. In (a), both the MPS and DWG methods are compared for  pump strength of  $\tilde\Omega=0.5\pi$, $\tilde\tau=0.5$, and 3000 trajectories for the SDW case. In (b), a stronger pump of $\tilde\Omega=2\pi$ for  $\tilde\tau=0.5$ is applied with both methods, and in (c), a long time delay $\tilde\tau=5$ with $\tilde\Omega=2\pi$ is solved using MPS.}
    \label{multi6}
\end{figure*}

In Fig.~\ref{multi5}, we first show results for  the vacuum decay case, assuming the same decay rate for both TLSs (this is not a model restriction in either model). We start with the two TLSs in the excited state (Fig.~\ref{multi5}(a)), and show how they decay equally, with a delay time of $\tilde \tau=0.5$. Then, we consider a different initial condition where one TLS is in the ground state and the other one is in the excited state (Fig.~\ref{multi5}(b)). It can be seen that both TLSs reach an equilibrium with a trapped TLS population, whose value decreases when the feedback time increases
 (cf.~Figs.~\ref{multi5} (b) and (d)). 
 Note the significant retardation oscillations that appear
 for the long delay time, initially causing a faster decay time. This shows that one can use the finite delay as a means to tune the emission dynamics of the distant TLS, similar to the effects of a distant mirror. Indeed, in the weak excitation regime, the other TLS acts a a resonant mirror (resonant scatterer), with a bandwidth that depends on the decay rate. Similar effects can be seen for TLSs embedded in cavities that are connected through a  waveguide~\cite{Yao2009}.
 All of these cases are calculated using MPSs and the SDW model, showing very good agreement for $N_T=1000$. Figure.~\ref{multi5}(c) shows the impact of a pure dephasing rate in the TLSs for the same conditions as in (b). This is performed using the SDW code and shows how, in the same way as in the 1 TLS case, the population trapping (and entanglement) is destroyed in the long time limit when this effect is considered.
The effect of adding an off-chip decay is very similar, so we do not bother showing it, and both effects cause a long time decay that depends on the additional decay rate.
 Further study in this coupling regime can be done, e.g.,   by changing the initial conditions and the phase between the TLSs, which would allow us to explore the impact on the known sub-radiant and super-radiant Dicke states \cite{GavinThesis}. 

In this non-Markovian example, the decay rate of one TLS can exceed the one given by the Dicke superradiance due to field emitted from the other TLS (see also Figs.~\ref{multi5}(d)). This can produce a constructive interference leading to a  ``super-superradiant'' state. In Ref.~\cite{PhysRevLett.124.043603} this is achieved in vacuum. We stress again that 
the dynamics in vacuum can easily be solved
exactly as has been demonstrated in a number
of works for coupling TLSs over macroscopic distances~\cite{Yao2009}, where the retardation dynamics are exactly accounted and shown to play a qualitatively important role on two TLS coupling. 
While both our approaches here can also recover the
super-superradiant state phenomena in vacuum, 
the real power of our QO waveguide approaches here is being apply to explore such regimes beyond the one quanta regime, and we will show examples of that below with CW pumping.

With regards to getting the same level of precision in both methods here, we can still use  $\Delta \tilde t=0.1$ in the MPS approach, but we need to go to $\Delta \tilde t=0.05$ in the SDW model, thus increasing the run times. Run times are compared in Table~\ref{table4}, where the parallel version of the SDW code is considered (on a single computer). The long feedback case significantly slows down the SDW code, showing in this case the greatest difference between run times. However, the long feedback case is used more as an academic study, since the coherent interactions typically become
less pronounced.

\begin{table}[h]
\centering
\caption{Run times for 
two TLSs in a infinite waveguide. We use $\Delta \tilde t = 0.05$ in the SDW model and $\Delta \tilde t = 0.1$ in the MPS model; we also use 10 and 50 boxes, respectively, in the SDW code (parallelized code with 2 photons, single computer), and a maximum bond dimension of 8 in the MPS code. Note that the $\tilde \tau = 0.5$ case is run for $\tilde t_{\rm max}=10$, and the $\tilde \tau = 2.5$ case is run for $\tilde t_{\rm max}=20$. Note also that for the vacuum dynamics, one could use a 1 photon SDW model which would yield significantly faster run times.}
\begin{tabular}{|p{1.6cm}|p{1.0cm}|p{5.0cm}| }
\hline
Model & $\tilde \tau$ & Run Time (s) \\
\hline\hline
SDW & 0.5 \phantom{\large{\cal B}} & 6.13 ($\Delta \tilde t=0.05$, $N_T=1000$)\\
MPS &  & 0.70 ($\Delta \tilde t=0.1$, bond = 8)\\
\hline
SDW & 2.5 \phantom{\large{\cal B}}  & 381.63 ($\Delta \tilde t=0.05$, $N_T=1000$) \\
MPS &  & 4.37 ($\Delta \tilde t=0.1$, bond = 8)\\
\hline
\end{tabular}
\label{table4}
\end{table}

We next consider two coupled TLS and a pumping field, 
with various results shown in Fig.~\ref{multi6}. In Fig.~\ref{multi6}(a), the case of a pump, $\tilde\Omega=0.5\pi$, with a delay time, $\tilde\tau=0.5$, is shown. This is calculated with both MPS and SDW codes. For the MPS, a bond dimension of 8 is required, taking 7.17s to run. For the SDW model, we need $N_T=3000$ taking 189s in its parallelized version (all results for a single computer). It can be seen how the pump in one TLS affects the TLS population of the other one, exciting it with a coherent damped coupling and eventually 
reaching a steady state.

We now consider a stronger driven case in Fig.~\ref{multi6}(b). It shows that, if the pump is too strong, an incoherent excitation of the second TLS appears, as the single TLS Rabi drive dominates the coherent oscillation of that population. It can be seen that the first TLS is apparently not affected by the second one, having a similar behaviour as in the case of having one driven TLS. The excitation of the second TLS is thus mainly through incoherent excitation. 

After confirming the excellent agreement with both methods, we next consider a much longer
delay time. In Fig.~\ref{multi6}) (c), the population results with $\tilde \tau = 5$ is shown. As we saw in the 1 TLS decay case, when the feedback increases substantially then the SDW code becomes much slower; also, here we  have to consider  the fact that there is also a significant pump field involved. For these reasons, this last example (c) is only run with the MPS approach. It is shown that  for a longer feedback, the pumping scenario is similar to Fig.~\ref{multi6}) (b), namely
the response of the second TLS to continuous driving
is through fluorescence from the first, 
 eventually reaching the same steady state. 

\subsection{Entanglement entropy for two 
coupled two level systems in a waveguide: role of retardation}

As a final application, we will study the
entanglement between the 2 TLSs, for different delay times.
The entanglement between the TLSs (joint system bin) and the waveguide can be measured through the entanglement entropy.

This is the Von Neumann entropy of the reduced density matrix \cite{Nielsen2009} (see also Eq~\eqref{vonneumann}). The Von Neumann entropy for a state $\rho$ is, 
\begin{equation}
    S(\rho)=-{\rm Tr}(\rho \log \rho),
\end{equation}
which can be rewritten in terms of the Schmidt coefficients,
\begin{equation}
    S(\rho)=-\sum_\alpha \Lambda_\alpha^2 \log_2 \Lambda_\alpha^2,
\end{equation}
where $\alpha$ indicates the position of the Schmidt coefficients in the diagonal matrix containing them.
%

\begin{figure}[b]
    \centering
    \includegraphics[width=1 \columnwidth]{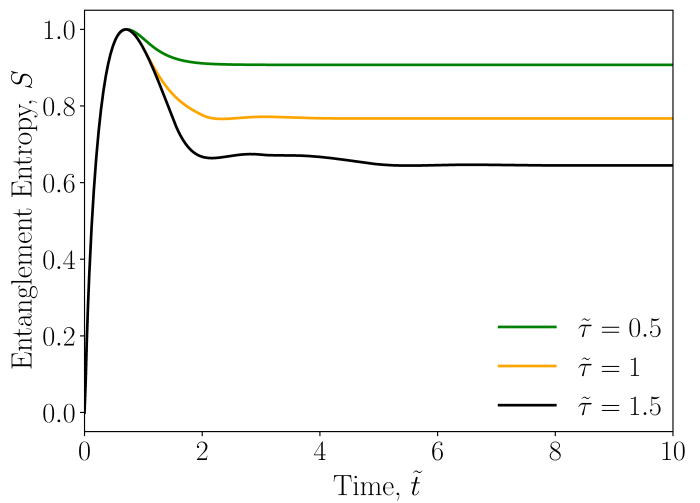}
    \caption{MPS calculation of the entanglement entropy between the TLSs bin and the waveguide (Eq.~\eqref{entsys}), for the decay 2 TLSs given different feedback: $\tilde\tau=0.5$ (green), $\tilde\tau=1$ (orange) and $\tilde\tau=1.5$ (black). In the three example cases, one TLS starts on the excited state and the other TLS starts on the ground state.
    }
    \label{entang1}
\end{figure}

Subsequently, the entanglement entropy between the TLSs bin and the waveguide  can be written as follows \cite{cajitas},
\begin{equation}
    \label{entsys}
    S(\rho_{\rm sys})=-\sum_\alpha \Lambda[S]^2_\alpha \log_2(\Lambda[S]^2_\alpha), 
\end{equation}
where $\rho_{\rm sys}$ represents the reduced density matrix of the TLSs bin, and $\Lambda[S]_\alpha$ are the Schmidt coefficients corresponding to the TLSs. 
Depending on the dimensions of our TLS(s) bin there is a different maximum value of the entanglement entropy as it counts the number of entangled qubits between the parts of the system, being the maximum~\cite{Nielsen2009}
$S_{\rm max}=k_{\rm qubits} \log_2 2$,
where $k_{\rm qubits}$ is the number of qubits. For example, in the case of 1 TLS, the maximum will be 1, and in the case of 2 TLS, the maximum will be 2.

Figure~\ref{entang1} shows the entanglement entropy between the TLSs bin
 and the waveguide for three different values of feedback ($\tilde\tau=0.5$, $\tilde\tau=1$ and $\tilde\tau=1.5$), where it can be seen that the longer the feedback the lower the entanglement after reaching a steady state. For these examples, we use the MPS approach only, though clearly we would obtain the same result with the SDW approach.

\section{Conclusions}
\label{sec:conclusions}

We have presented two different models
for solving quantum nonlinear light-matter in waveguide-QED systems, using MPSs and a SDW model. Both approaches are shown to efficiently  
describe the complicated non-Markovian cases of a
time-delayed coherent feedback and two spatially 
separated TLSs.
We applied these models to study
three different topical systems
in waveguide quantum circuits, 
including a TLS coupled to an infinite waveguide, a TLS coupled to a semi-infinite waveguide (with a time-delayed feedback), and two spatially separated TLS coupled to an infinite waveguide.
Both methods include waveguide photons that are quantized at the system level and, importantly, can explore both linear and nonlinear quantum regimes. While the MPS approach is intrinsically
non-Markovian, the SDW model solves
Markovian equations of motion and exploits QT theory which also provides 
physical insight into the underlying stochastic dynamics~\cite{Whalen_2017}.

After presenting the theory of MPSs and the SDW model,  results were shown and compared directly for the three QED-waveguide systems of interest. Numerically, 
we find excellent agreement between both methods
if the required number of waveguide photons is included in the SDW model, which works remarkably well with up to two photons in the loop, even under extreme pumping conditions. The SDW model also allows us to easily identify the differences between a one photon and a two photon approximation, yielding  information about the role of two photon interference effects.
 For the case of one TLS and a time-delayed feedback,
 we have studied both the vacuum dynamics and nonlinear dynamics, verifying that the one-photon-in-the  loop approximation breaks down in the presence of a strong pumping field. We also show how both methods can efficiently track the population trapped state, easily yielding coherent oscillations over a large number of  periods. In addition, it was shown how two
 spatially-separated TLSs can be efficiently modelled with both approaches, showing again the vacuum dynamics and nonlinear quantum dynamics with a coherent pump field. We investigated the  role of retardation, and briefly discussed how to quantify the entanglement entropy between the TLSs and the waveguide, with various delay times (spatial separations).

While we have shown that both approaches offer excellent complimentary information for modelling system-level waveguide QED,  each has
certain advantages and disadvantages
 for studying waveguide QED systems.
 The MPS model,  although significantly more complicated to implement,  offers faster run-time results in most of the cases studied in our paper, for the same level of precision. This becomes more noticeable when we increase the feedback length or/and if the system is driven by a very strong pump field. In these cases, the SDW model can run into computational memory problems; in contrast,
  in the MPS case, although the run times increase, there are no memory problems found
 for the examples presented. On the other hand, smaller delay times are more practical anyway.
The SDW model also  has some notable advantages over MPSs: (i) it can show results for different levels of approximation more clearly, such as results with one photon or two photons in the loop;
(ii) it is far easier to implement computationally, is perfectly parallelizable,  and uses well known techniques in quantum optics, such as QT theory; (iii) the equations are actually all Markovian, even though retardation effects are fully accounted for; (iv) the ease of adding in other dissipation channels such as off-chip decay and pure dephasing is fairly straightforward. In this latter case, we demonstrated the importance of these effects as an important limit to creating population trapped states and entangled qubits.  For connecting to real experiments, such as with semiconductor quantum dots, including such processes is critical.

Overall, our paper shows how one can implement both these two different methods to accurately model complicated waveguide QED systems, which can work together as  powerful and complementary models in quantum optics. Indeed, as we have demonstrated,
these methods can be used to improve our understanding and exploitation of  complex non-Markovian feedback systems. Both the SDW model and the MPS model  can also support the addition of more complicated circuits, including two TLSs  with a mirror-based coherent feedback, pulsed excitation, and input-output theory with input photon
states. 
 Although we find that the approximation to two photons is highly accurate for the results presented in this paper, the presence of more photons can become important in other cases, giving the opportunity to describe more complex systems in  future work, e.g. three quantum emitters (TLSs) in a waveguide~\cite{PhysRevResearch.2.013238} (where the side atoms can behave like mirrors in cavity QED \cite{Mirhosseini2019}), a higher number of qubits~\cite{Albrecht_2019,Finsterhlzl2020},  and 1D atomic arrays~\cite{Masson2019AtomicWaveguideQE,PhysRevLett.124.213601}.

\acknowledgements

This work was funded by the Natural Sciences and Engineering
Research Council of Canada, the Canadian Foundation for Innovation and Queen's University, Canada.
Howard Carmichael acknowledges the support of the New Zealand Tertiary Education Committee through the Dodd-Walls Centre for Photonic and Quantum Technologies.
We  thank
Nir Rotenberg for useful comments.

\bibliography{biblio}

\begin{thebibliography}{105}%
\makeatletter
\providecommand \@ifxundefined [1]{%
 \@ifx{#1\undefined}
}%
\providecommand \@ifnum [1]{%
 \ifnum #1\expandafter \@firstoftwo
 \else \expandafter \@secondoftwo
 \fi
}%
\providecommand \@ifx [1]{%
 \ifx #1\expandafter \@firstoftwo
 \else \expandafter \@secondoftwo
 \fi
}%
\providecommand \natexlab [1]{#1}%
\providecommand \enquote  [1]{``#1''}%
\providecommand \bibnamefont  [1]{#1}%
\providecommand \bibfnamefont [1]{#1}%
\providecommand \citenamefont [1]{#1}%
\providecommand \href@noop [0]{\@secondoftwo}%
\providecommand \href [0]{\begingroup \@sanitize@url \@href}%
\providecommand \@href[1]{\@@startlink{#1}\@@href}%
\providecommand \@@href[1]{\endgroup#1\@@endlink}%
\providecommand \@sanitize@url [0]{\catcode `\\12\catcode `\$12\catcode
  `\&12\catcode `\#12\catcode `\^12\catcode `\_12\catcode `\%12\relax}%
\providecommand \@@startlink[1]{}%
\providecommand \@@endlink[0]{}%
\providecommand \url  [0]{\begingroup\@sanitize@url \@url }%
\providecommand \@url [1]{\endgroup\@href {#1}{\urlprefix }}%
\providecommand \urlprefix  [0]{URL }%
\providecommand \Eprint [0]{\href }%
\providecommand \doibase [0]{https://doi.org/}%
\providecommand \selectlanguage [0]{\@gobble}%
\providecommand \bibinfo  [0]{\@secondoftwo}%
\providecommand \bibfield  [0]{\@secondoftwo}%
\providecommand \translation [1]{[#1]}%
\providecommand \BibitemOpen [0]{}%
\providecommand \bibitemStop [0]{}%
\providecommand \bibitemNoStop [0]{.\EOS\space}%
\providecommand \EOS [0]{\spacefactor3000\relax}%
\providecommand \BibitemShut  [1]{\csname bibitem#1\endcsname}%
\let\auto@bib@innerbib\@empty
\bibitem [{\citenamefont {Hughes}(2004)}]{Hughes2004}%
  \BibitemOpen
  \bibfield  {author} {\bibinfo {author} {\bibfnamefont {S.}~\bibnamefont
  {Hughes}},\ }\bibfield  {title} {\bibinfo {title} {Enhanced single-photon
  emission from quantum dots in photonic crystal waveguides and nanocavities},\
  }\href {https://doi.org/10.1364/ol.29.002659} {\bibfield  {journal} {\bibinfo
   {journal} {Optics Letters}\ }\textbf {\bibinfo {volume} {29}},\ \bibinfo
  {pages} {2659} (\bibinfo {year} {2004})}\BibitemShut {NoStop}%
\bibitem [{\citenamefont {Shen}\ and\ \citenamefont
  {Fan}(2007{\natexlab{a}})}]{PhysRevA.76.062709}%
  \BibitemOpen
  \bibfield  {author} {\bibinfo {author} {\bibfnamefont {J.-T.}\ \bibnamefont
  {Shen}}\ and\ \bibinfo {author} {\bibfnamefont {S.}~\bibnamefont {Fan}},\
  }\bibfield  {title} {\bibinfo {title} {Strongly correlated multiparticle
  transport in one dimension through a quantum impurity},\ }\href
  {https://doi.org/10.1103/PhysRevA.76.062709} {\bibfield  {journal} {\bibinfo
  {journal} {Phys. Rev. A}\ }\textbf {\bibinfo {volume} {76}},\ \bibinfo
  {pages} {062709} (\bibinfo {year} {2007}{\natexlab{a}})}\BibitemShut
  {NoStop}%
\bibitem [{\citenamefont {Shen}\ and\ \citenamefont
  {Fan}(2007{\natexlab{b}})}]{PhysRevLett.98.153003}%
  \BibitemOpen
  \bibfield  {author} {\bibinfo {author} {\bibfnamefont {J.-T.}\ \bibnamefont
  {Shen}}\ and\ \bibinfo {author} {\bibfnamefont {S.}~\bibnamefont {Fan}},\
  }\bibfield  {title} {\bibinfo {title} {Strongly correlated two-photon
  transport in a one-dimensional waveguide coupled to a two-level system},\
  }\href {https://doi.org/10.1103/PhysRevLett.98.153003} {\bibfield  {journal}
  {\bibinfo  {journal} {Phys. Rev. Lett.}\ }\textbf {\bibinfo {volume} {98}},\
  \bibinfo {pages} {153003} (\bibinfo {year} {2007}{\natexlab{b}})}\BibitemShut
  {NoStop}%
\bibitem [{\citenamefont {Zheng}\ \emph {et~al.}(2010)\citenamefont {Zheng},
  \citenamefont {Gauthier},\ and\ \citenamefont {Baranger}}]{Zheng2010}%
  \BibitemOpen
  \bibfield  {author} {\bibinfo {author} {\bibfnamefont {H.}~\bibnamefont
  {Zheng}}, \bibinfo {author} {\bibfnamefont {D.~J.}\ \bibnamefont
  {Gauthier}},\ and\ \bibinfo {author} {\bibfnamefont {H.~U.}\ \bibnamefont
  {Baranger}},\ }\bibfield  {title} {\bibinfo {title} {Waveguide {QED}:
  Many-body bound-state effects in coherent and fock-state scattering from a
  two-level system},\ }\href {https://doi.org/10.1103/physreva.82.063816}
  {\bibfield  {journal} {\bibinfo  {journal} {Phys. Rev. A}\ }\textbf {\bibinfo
  {volume} {82}},\ \bibinfo {pages} {063816} (\bibinfo {year}
  {2010})}\BibitemShut {NoStop}%
\bibitem [{\citenamefont {Witthaut}\ and\ \citenamefont
  {S{\o}rensen}(2010)}]{Witthaut_2010}%
  \BibitemOpen
  \bibfield  {author} {\bibinfo {author} {\bibfnamefont {D.}~\bibnamefont
  {Witthaut}}\ and\ \bibinfo {author} {\bibfnamefont {A.~S.}\ \bibnamefont
  {S{\o}rensen}},\ }\bibfield  {title} {\bibinfo {title} {Photon scattering by
  a three-level emitter in a one-dimensional waveguide},\ }\href
  {https://doi.org/10.1088/1367-2630/12/4/043052} {\bibfield  {journal}
  {\bibinfo  {journal} {New Journal of Physics}\ }\textbf {\bibinfo {volume}
  {12}},\ \bibinfo {pages} {043052} (\bibinfo {year} {2010})}\BibitemShut
  {NoStop}%
\bibitem [{\citenamefont {Longo}\ \emph {et~al.}(2011)\citenamefont {Longo},
  \citenamefont {Schmitteckert},\ and\ \citenamefont
  {Busch}}]{PhysRevA.83.063828}%
  \BibitemOpen
  \bibfield  {author} {\bibinfo {author} {\bibfnamefont {P.}~\bibnamefont
  {Longo}}, \bibinfo {author} {\bibfnamefont {P.}~\bibnamefont
  {Schmitteckert}},\ and\ \bibinfo {author} {\bibfnamefont {K.}~\bibnamefont
  {Busch}},\ }\bibfield  {title} {\bibinfo {title} {Few-photon transport in
  low-dimensional systems},\ }\href
  {https://doi.org/10.1103/PhysRevA.83.063828} {\bibfield  {journal} {\bibinfo
  {journal} {Phys. Rev. A}\ }\textbf {\bibinfo {volume} {83}},\ \bibinfo
  {pages} {063828} (\bibinfo {year} {2011})}\BibitemShut {NoStop}%
\bibitem [{\citenamefont {Roy}(2011)}]{PhysRevLett.106.053601}%
  \BibitemOpen
  \bibfield  {author} {\bibinfo {author} {\bibfnamefont {D.}~\bibnamefont
  {Roy}},\ }\bibfield  {title} {\bibinfo {title} {Two-photon scattering by a
  driven three-level emitter in a one-dimensional waveguide and
  electromagnetically induced transparency},\ }\href
  {https://doi.org/10.1103/PhysRevLett.106.053601} {\bibfield  {journal}
  {\bibinfo  {journal} {Phys. Rev. Lett.}\ }\textbf {\bibinfo {volume} {106}},\
  \bibinfo {pages} {053601} (\bibinfo {year} {2011})}\BibitemShut {NoStop}%
\bibitem [{\citenamefont {Sanchez-Burillo}\ \emph {et~al.}(2014)\citenamefont
  {Sanchez-Burillo}, \citenamefont {Zueco}, \citenamefont {Garcia-Ripoll},\
  and\ \citenamefont {Martin-Moreno}}]{PhysRevLett.113.263604}%
  \BibitemOpen
  \bibfield  {author} {\bibinfo {author} {\bibfnamefont {E.}~\bibnamefont
  {Sanchez-Burillo}}, \bibinfo {author} {\bibfnamefont {D.}~\bibnamefont
  {Zueco}}, \bibinfo {author} {\bibfnamefont {J.~J.}\ \bibnamefont
  {Garcia-Ripoll}},\ and\ \bibinfo {author} {\bibfnamefont {L.}~\bibnamefont
  {Martin-Moreno}},\ }\bibfield  {title} {\bibinfo {title} {Scattering in the
  ultrastrong regime: Nonlinear optics with one photon},\ }\href
  {https://doi.org/10.1103/PhysRevLett.113.263604} {\bibfield  {journal}
  {\bibinfo  {journal} {Phys. Rev. Lett.}\ }\textbf {\bibinfo {volume} {113}},\
  \bibinfo {pages} {263604} (\bibinfo {year} {2014})}\BibitemShut {NoStop}%
\bibitem [{\citenamefont {Calaj\'o}\ \emph {et~al.}(2016)\citenamefont
  {Calaj\'o}, \citenamefont {Ciccarello}, \citenamefont {Chang},\ and\
  \citenamefont {Rabl}}]{Calaj2016}%
  \BibitemOpen
  \bibfield  {author} {\bibinfo {author} {\bibfnamefont {G.}~\bibnamefont
  {Calaj\'o}}, \bibinfo {author} {\bibfnamefont {F.}~\bibnamefont
  {Ciccarello}}, \bibinfo {author} {\bibfnamefont {D.}~\bibnamefont {Chang}},\
  and\ \bibinfo {author} {\bibfnamefont {P.}~\bibnamefont {Rabl}},\ }\bibfield
  {title} {\bibinfo {title} {Atom-field dressed states in slow-light waveguide
  qed},\ }\href {https://doi.org/10.1103/PhysRevA.93.033833} {\bibfield
  {journal} {\bibinfo  {journal} {Phys. Rev. A}\ }\textbf {\bibinfo {volume}
  {93}},\ \bibinfo {pages} {033833} (\bibinfo {year} {2016})}\BibitemShut
  {NoStop}%
\bibitem [{\citenamefont {Pichler}\ and\ \citenamefont
  {Zoller}(2016)}]{cajitas}%
  \BibitemOpen
  \bibfield  {author} {\bibinfo {author} {\bibfnamefont {H.}~\bibnamefont
  {Pichler}}\ and\ \bibinfo {author} {\bibfnamefont {P.}~\bibnamefont
  {Zoller}},\ }\bibfield  {title} {\bibinfo {title} {Photonic circuits with
  time delays and quantum feedback},\ }\href
  {https://doi.org/10.1103/PhysRevLett.116.093601} {\bibfield  {journal}
  {\bibinfo  {journal} {Phys. Rev. Lett.}\ }\textbf {\bibinfo {volume} {116}},\
  \bibinfo {pages} {093601} (\bibinfo {year} {2016})}\BibitemShut {NoStop}%
\bibitem [{\citenamefont {Manga~Rao}\ and\ \citenamefont
  {Hughes}(2007)}]{PhysRevB.75.205437}%
  \BibitemOpen
  \bibfield  {author} {\bibinfo {author} {\bibfnamefont {V.~S.~C.}\
  \bibnamefont {Manga~Rao}}\ and\ \bibinfo {author} {\bibfnamefont
  {S.}~\bibnamefont {Hughes}},\ }\bibfield  {title} {\bibinfo {title} {Single
  quantum-dot purcell factor and $\ensuremath{\beta}$ factor in a photonic
  crystal waveguide},\ }\href {https://doi.org/10.1103/PhysRevB.75.205437}
  {\bibfield  {journal} {\bibinfo  {journal} {Phys. Rev. B}\ }\textbf {\bibinfo
  {volume} {75}},\ \bibinfo {pages} {205437} (\bibinfo {year}
  {2007})}\BibitemShut {NoStop}%
\bibitem [{\citenamefont {Lund-Hansen}\ \emph {et~al.}(2008)\citenamefont
  {Lund-Hansen}, \citenamefont {Stobbe}, \citenamefont {Julsgaard},
  \citenamefont {Thyrrestrup}, \citenamefont {S\"unner}, \citenamefont {Kamp},
  \citenamefont {Forchel},\ and\ \citenamefont
  {Lodahl}}]{PhysRevLett.101.113903}%
  \BibitemOpen
  \bibfield  {author} {\bibinfo {author} {\bibfnamefont {T.}~\bibnamefont
  {Lund-Hansen}}, \bibinfo {author} {\bibfnamefont {S.}~\bibnamefont {Stobbe}},
  \bibinfo {author} {\bibfnamefont {B.}~\bibnamefont {Julsgaard}}, \bibinfo
  {author} {\bibfnamefont {H.}~\bibnamefont {Thyrrestrup}}, \bibinfo {author}
  {\bibfnamefont {T.}~\bibnamefont {S\"unner}}, \bibinfo {author}
  {\bibfnamefont {M.}~\bibnamefont {Kamp}}, \bibinfo {author} {\bibfnamefont
  {A.}~\bibnamefont {Forchel}},\ and\ \bibinfo {author} {\bibfnamefont
  {P.}~\bibnamefont {Lodahl}},\ }\bibfield  {title} {\bibinfo {title}
  {Experimental realization of highly efficient broadband coupling of single
  quantum dots to a photonic crystal waveguide},\ }\href
  {https://doi.org/10.1103/PhysRevLett.101.113903} {\bibfield  {journal}
  {\bibinfo  {journal} {Phys. Rev. Lett.}\ }\textbf {\bibinfo {volume} {101}},\
  \bibinfo {pages} {113903} (\bibinfo {year} {2008})}\BibitemShut {NoStop}%
\bibitem [{\citenamefont {Laucht}\ \emph {et~al.}(2012)\citenamefont {Laucht},
  \citenamefont {P\"utz}, \citenamefont {G\"unthner}, \citenamefont {Hauke},
  \citenamefont {Saive}, \citenamefont {Fr\'ed\'erick}, \citenamefont
  {Bichler}, \citenamefont {Amann}, \citenamefont {Holleitner}, \citenamefont
  {Kaniber},\ and\ \citenamefont {Finley}}]{PhysRevX.2.011014}%
  \BibitemOpen
  \bibfield  {author} {\bibinfo {author} {\bibfnamefont {A.}~\bibnamefont
  {Laucht}}, \bibinfo {author} {\bibfnamefont {S.}~\bibnamefont {P\"utz}},
  \bibinfo {author} {\bibfnamefont {T.}~\bibnamefont {G\"unthner}}, \bibinfo
  {author} {\bibfnamefont {N.}~\bibnamefont {Hauke}}, \bibinfo {author}
  {\bibfnamefont {R.}~\bibnamefont {Saive}}, \bibinfo {author} {\bibfnamefont
  {S.}~\bibnamefont {Fr\'ed\'erick}}, \bibinfo {author} {\bibfnamefont
  {M.}~\bibnamefont {Bichler}}, \bibinfo {author} {\bibfnamefont {M.-C.}\
  \bibnamefont {Amann}}, \bibinfo {author} {\bibfnamefont {A.~W.}\ \bibnamefont
  {Holleitner}}, \bibinfo {author} {\bibfnamefont {M.}~\bibnamefont
  {Kaniber}},\ and\ \bibinfo {author} {\bibfnamefont {J.~J.}\ \bibnamefont
  {Finley}},\ }\bibfield  {title} {\bibinfo {title} {A waveguide-coupled
  on-chip single-photon source},\ }\href
  {https://doi.org/10.1103/PhysRevX.2.011014} {\bibfield  {journal} {\bibinfo
  {journal} {Phys. Rev. X}\ }\textbf {\bibinfo {volume} {2}},\ \bibinfo {pages}
  {011014} (\bibinfo {year} {2012})}\BibitemShut {NoStop}%
\bibitem [{\citenamefont {Gardiner}\ and\ \citenamefont
  {Zoller}(2010)}]{gardiner_zoller_2010}%
  \BibitemOpen
  \bibfield  {author} {\bibinfo {author} {\bibfnamefont {C.~W.}\ \bibnamefont
  {Gardiner}}\ and\ \bibinfo {author} {\bibfnamefont {P.}~\bibnamefont
  {Zoller}},\ }\href@noop {} {\emph {\bibinfo {title} {Quantum noise: a
  handbook of Markovian and non-Markovian quantum stochastic methods with
  applications to quantum optics}}}\ (\bibinfo  {publisher} {Springer},\
  \bibinfo {address} {Berlin},\ \bibinfo {year} {2010})\BibitemShut {NoStop}%
\bibitem [{\citenamefont {Droenner}\ \emph {et~al.}(2019)\citenamefont
  {Droenner}, \citenamefont {Naumann}, \citenamefont {Schöll}, \citenamefont
  {Knorr},\ and\ \citenamefont {Carmele}}]{Droenner2019}%
  \BibitemOpen
  \bibfield  {author} {\bibinfo {author} {\bibfnamefont {L.}~\bibnamefont
  {Droenner}}, \bibinfo {author} {\bibfnamefont {N.~L.}\ \bibnamefont
  {Naumann}}, \bibinfo {author} {\bibfnamefont {E.}~\bibnamefont {Schöll}},
  \bibinfo {author} {\bibfnamefont {A.}~\bibnamefont {Knorr}},\ and\ \bibinfo
  {author} {\bibfnamefont {A.}~\bibnamefont {Carmele}},\ }\bibfield  {title}
  {{\selectlanguage {en}\bibinfo {title} {Quantum {Pyragas} control:
  {Selective} control of individual photon probabilities}},\ }\href
  {https://doi.org/10.1103/PhysRevA.99.023840} {\bibfield  {journal} {\bibinfo
  {journal} {Physical Review A}\ }\textbf {\bibinfo {volume} {99}},\ \bibinfo
  {pages} {023840} (\bibinfo {year} {2019})}\BibitemShut {NoStop}%
\bibitem [{\citenamefont {Crowder}\ \emph {et~al.}(2020)\citenamefont
  {Crowder}, \citenamefont {Carmichael},\ and\ \citenamefont
  {Hughes}}]{crowder_quantum_2020}%
  \BibitemOpen
  \bibfield  {author} {\bibinfo {author} {\bibfnamefont {G.}~\bibnamefont
  {Crowder}}, \bibinfo {author} {\bibfnamefont {H.}~\bibnamefont
  {Carmichael}},\ and\ \bibinfo {author} {\bibfnamefont {S.}~\bibnamefont
  {Hughes}},\ }\bibfield  {title} {{\selectlanguage {en}\bibinfo {title}
  {Quantum trajectory theory of few-photon cavity-{QED} systems with a
  time-delayed coherent feedback}},\ }\href
  {https://doi.org/10.1103/PhysRevA.101.023807} {\bibfield  {journal} {\bibinfo
   {journal} {Physical Review A}\ }\textbf {\bibinfo {volume} {101}},\ \bibinfo
  {pages} {023807} (\bibinfo {year} {2020})}\BibitemShut {NoStop}%
\bibitem [{\citenamefont {Dorner}\ and\ \citenamefont
  {Zoller}(2002)}]{Dorner2002}%
  \BibitemOpen
  \bibfield  {author} {\bibinfo {author} {\bibfnamefont {U.}~\bibnamefont
  {Dorner}}\ and\ \bibinfo {author} {\bibfnamefont {P.}~\bibnamefont
  {Zoller}},\ }\bibfield  {title} {{\selectlanguage {en}\bibinfo {title}
  {Laser-driven atoms in half-cavities}},\ }\href
  {https://doi.org/10.1103/PhysRevA.66.023816} {\bibfield  {journal} {\bibinfo
  {journal} {Physical Review A}\ }\textbf {\bibinfo {volume} {66}},\ \bibinfo
  {pages} {023816} (\bibinfo {year} {2002})}\BibitemShut {NoStop}%
\bibitem [{\citenamefont {Tufarelli}\ \emph {et~al.}(2013)\citenamefont
  {Tufarelli}, \citenamefont {Ciccarello},\ and\ \citenamefont
  {Kim}}]{Tufarelli2013}%
  \BibitemOpen
  \bibfield  {author} {\bibinfo {author} {\bibfnamefont {T.}~\bibnamefont
  {Tufarelli}}, \bibinfo {author} {\bibfnamefont {F.}~\bibnamefont
  {Ciccarello}},\ and\ \bibinfo {author} {\bibfnamefont {M.~S.}\ \bibnamefont
  {Kim}},\ }\bibfield  {title} {{\selectlanguage {en}\bibinfo {title} {Dynamics
  of spontaneous emission in a single-end photonic waveguide}},\ }\href
  {https://doi.org/10.1103/PhysRevA.87.013820} {\bibfield  {journal} {\bibinfo
  {journal} {Physical Review A}\ }\textbf {\bibinfo {volume} {87}},\ \bibinfo
  {pages} {013820} (\bibinfo {year} {2013})}\BibitemShut {NoStop}%
\bibitem [{\citenamefont {Carmele}\ \emph {et~al.}(2013)\citenamefont
  {Carmele}, \citenamefont {Kabuss}, \citenamefont {Schulze}, \citenamefont
  {Reitzenstein},\ and\ \citenamefont {Knorr}}]{PhysRevLett.110.013601}%
  \BibitemOpen
  \bibfield  {author} {\bibinfo {author} {\bibfnamefont {A.}~\bibnamefont
  {Carmele}}, \bibinfo {author} {\bibfnamefont {J.}~\bibnamefont {Kabuss}},
  \bibinfo {author} {\bibfnamefont {F.}~\bibnamefont {Schulze}}, \bibinfo
  {author} {\bibfnamefont {S.}~\bibnamefont {Reitzenstein}},\ and\ \bibinfo
  {author} {\bibfnamefont {A.}~\bibnamefont {Knorr}},\ }\bibfield  {title}
  {\bibinfo {title} {Single photon delayed feedback: A way to stabilize
  intrinsic quantum cavity electrodynamics},\ }\href
  {https://doi.org/10.1103/PhysRevLett.110.013601} {\bibfield  {journal}
  {\bibinfo  {journal} {Phys. Rev. Lett.}\ }\textbf {\bibinfo {volume} {110}},\
  \bibinfo {pages} {013601} (\bibinfo {year} {2013})}\BibitemShut {NoStop}%
\bibitem [{\citenamefont {Német}\ \emph {et~al.}(2019)\citenamefont {Német},
  \citenamefont {Carmele}, \citenamefont {Parkins},\ and\ \citenamefont
  {Knorr}}]{Nemet2019}%
  \BibitemOpen
  \bibfield  {author} {\bibinfo {author} {\bibfnamefont {N.}~\bibnamefont
  {Német}}, \bibinfo {author} {\bibfnamefont {A.}~\bibnamefont {Carmele}},
  \bibinfo {author} {\bibfnamefont {S.}~\bibnamefont {Parkins}},\ and\ \bibinfo
  {author} {\bibfnamefont {A.}~\bibnamefont {Knorr}},\ }\bibfield  {title}
  {\bibinfo {title} {Comparison between continuous- and discrete-mode coherent
  feedback for the {Jaynes}-{Cummings} model},\ }\href
  {https://doi.org/10.1103/PhysRevA.100.023805} {\bibfield  {journal} {\bibinfo
   {journal} {Physical Review A}\ }\textbf {\bibinfo {volume} {100}},\ \bibinfo
  {pages} {023805} (\bibinfo {year} {2019})}\BibitemShut {NoStop}%
\bibitem [{\citenamefont {Grimsmo}(2015)}]{Grimsmo2015}%
  \BibitemOpen
  \bibfield  {author} {\bibinfo {author} {\bibfnamefont {A.~L.}\ \bibnamefont
  {Grimsmo}},\ }\bibfield  {title} {\bibinfo {title} {Time-delayed quantum
  feedback control},\ }\href {https://doi.org/10.1103/PhysRevLett.115.060402}
  {\bibfield  {journal} {\bibinfo  {journal} {Phys. Rev. Lett.}\ }\textbf
  {\bibinfo {volume} {115}},\ \bibinfo {pages} {060402} (\bibinfo {year}
  {2015})}\BibitemShut {NoStop}%
\bibitem [{\citenamefont {Whalen}\ \emph {et~al.}(2017)\citenamefont {Whalen},
  \citenamefont {Grimsmo},\ and\ \citenamefont {Carmichael}}]{Whalen_2017}%
  \BibitemOpen
  \bibfield  {author} {\bibinfo {author} {\bibfnamefont {S.~J.}\ \bibnamefont
  {Whalen}}, \bibinfo {author} {\bibfnamefont {A.~L.}\ \bibnamefont
  {Grimsmo}},\ and\ \bibinfo {author} {\bibfnamefont {H.~J.}\ \bibnamefont
  {Carmichael}},\ }\bibfield  {title} {\bibinfo {title} {Open quantum systems
  with delayed coherent feedback},\ }\href
  {https://doi.org/10.1088/2058-9565/aa8331} {\bibfield  {journal} {\bibinfo
  {journal} {Quantum Science and Technology}\ }\textbf {\bibinfo {volume}
  {2}},\ \bibinfo {pages} {044008} (\bibinfo {year} {2017})}\BibitemShut
  {NoStop}%
\bibitem [{\citenamefont {Chalabi}\ and\ \citenamefont
  {Waks}(2018)}]{PhysRevA.98.063832}%
  \BibitemOpen
  \bibfield  {author} {\bibinfo {author} {\bibfnamefont {H.}~\bibnamefont
  {Chalabi}}\ and\ \bibinfo {author} {\bibfnamefont {E.}~\bibnamefont {Waks}},\
  }\bibfield  {title} {\bibinfo {title} {Interaction of photons with a coupled
  atom-cavity system through a bidirectional time-delayed feedback},\ }\href
  {https://doi.org/10.1103/PhysRevA.98.063832} {\bibfield  {journal} {\bibinfo
  {journal} {Phys. Rev. A}\ }\textbf {\bibinfo {volume} {98}},\ \bibinfo
  {pages} {063832} (\bibinfo {year} {2018})}\BibitemShut {NoStop}%
\bibitem [{\citenamefont {Kubanek}\ \emph {et~al.}(2009)\citenamefont
  {Kubanek}, \citenamefont {Koch}, \citenamefont {Sames}, \citenamefont
  {Ourjoumtsev}, \citenamefont {Pinkse}, \citenamefont {Murr},\ and\
  \citenamefont {Rempe}}]{Kubanek2009}%
  \BibitemOpen
  \bibfield  {author} {\bibinfo {author} {\bibfnamefont {A.}~\bibnamefont
  {Kubanek}}, \bibinfo {author} {\bibfnamefont {M.}~\bibnamefont {Koch}},
  \bibinfo {author} {\bibfnamefont {C.}~\bibnamefont {Sames}}, \bibinfo
  {author} {\bibfnamefont {A.}~\bibnamefont {Ourjoumtsev}}, \bibinfo {author}
  {\bibfnamefont {P.~W.~H.}\ \bibnamefont {Pinkse}}, \bibinfo {author}
  {\bibfnamefont {K.}~\bibnamefont {Murr}},\ and\ \bibinfo {author}
  {\bibfnamefont {G.}~\bibnamefont {Rempe}},\ }\bibfield  {title} {\bibinfo
  {title} {Photon-by-photon feedback control of a single-atom trajectory},\
  }\href {https://doi.org/10.1038/nature08563} {\bibfield  {journal} {\bibinfo
  {journal} {Nature}\ }\textbf {\bibinfo {volume} {462}},\ \bibinfo {pages}
  {898} (\bibinfo {year} {2009})}\BibitemShut {NoStop}%
\bibitem [{\citenamefont {Gillett}\ \emph {et~al.}(2010)\citenamefont
  {Gillett}, \citenamefont {Dalton}, \citenamefont {Lanyon}, \citenamefont
  {Almeida}, \citenamefont {Barbieri}, \citenamefont {Pryde}, \citenamefont
  {O'Brien}, \citenamefont {Resch}, \citenamefont {Bartlett},\ and\
  \citenamefont {White}}]{Gillett}%
  \BibitemOpen
  \bibfield  {author} {\bibinfo {author} {\bibfnamefont {G.~G.}\ \bibnamefont
  {Gillett}}, \bibinfo {author} {\bibfnamefont {R.~B.}\ \bibnamefont {Dalton}},
  \bibinfo {author} {\bibfnamefont {B.~P.}\ \bibnamefont {Lanyon}}, \bibinfo
  {author} {\bibfnamefont {M.~P.}\ \bibnamefont {Almeida}}, \bibinfo {author}
  {\bibfnamefont {M.}~\bibnamefont {Barbieri}}, \bibinfo {author}
  {\bibfnamefont {G.~J.}\ \bibnamefont {Pryde}}, \bibinfo {author}
  {\bibfnamefont {J.~L.}\ \bibnamefont {O'Brien}}, \bibinfo {author}
  {\bibfnamefont {K.~J.}\ \bibnamefont {Resch}}, \bibinfo {author}
  {\bibfnamefont {S.~D.}\ \bibnamefont {Bartlett}},\ and\ \bibinfo {author}
  {\bibfnamefont {A.~G.}\ \bibnamefont {White}},\ }\bibfield  {title} {\bibinfo
  {title} {Experimental feedback control of quantum systems using weak
  measurements},\ }\href {https://doi.org/10.1103/PhysRevLett.104.080503}
  {\bibfield  {journal} {\bibinfo  {journal} {Phys. Rev. Lett.}\ }\textbf
  {\bibinfo {volume} {104}},\ \bibinfo {pages} {080503} (\bibinfo {year}
  {2010})}\BibitemShut {NoStop}%
\bibitem [{\citenamefont {Brandes}(2010)}]{Brandes}%
  \BibitemOpen
  \bibfield  {author} {\bibinfo {author} {\bibfnamefont {T.}~\bibnamefont
  {Brandes}},\ }\bibfield  {title} {\bibinfo {title} {Feedback control of
  quantum transport},\ }\href {https://doi.org/10.1103/PhysRevLett.105.060602}
  {\bibfield  {journal} {\bibinfo  {journal} {Phys. Rev. Lett.}\ }\textbf
  {\bibinfo {volume} {105}},\ \bibinfo {pages} {060602} (\bibinfo {year}
  {2010})}\BibitemShut {NoStop}%
\bibitem [{\citenamefont {Balouchi}\ and\ \citenamefont
  {Jacobs}(2017)}]{Balouchi2017}%
  \BibitemOpen
  \bibfield  {author} {\bibinfo {author} {\bibfnamefont {A.}~\bibnamefont
  {Balouchi}}\ and\ \bibinfo {author} {\bibfnamefont {K.}~\bibnamefont
  {Jacobs}},\ }\bibfield  {title} {{\selectlanguage {en}\bibinfo {title}
  {Coherent versus measurement-based feedback for controlling a single
  qubit}},\ }\href {https://doi.org/10.1088/2058-9565/aa5409} {\bibfield
  {journal} {\bibinfo  {journal} {Quantum Science and Technology}\ }\textbf
  {\bibinfo {volume} {2}},\ \bibinfo {pages} {025001} (\bibinfo {year}
  {2017})}\BibitemShut {NoStop}%
\bibitem [{\citenamefont {Calajó}\ \emph {et~al.}(2019)\citenamefont
  {Calajó}, \citenamefont {Fang}, \citenamefont {Baranger},\ and\
  \citenamefont {Ciccarello}}]{Calajo2019}%
  \BibitemOpen
  \bibfield  {author} {\bibinfo {author} {\bibfnamefont {G.}~\bibnamefont
  {Calajó}}, \bibinfo {author} {\bibfnamefont {Y.-L.~L.}\ \bibnamefont
  {Fang}}, \bibinfo {author} {\bibfnamefont {H.~U.}\ \bibnamefont {Baranger}},\
  and\ \bibinfo {author} {\bibfnamefont {F.}~\bibnamefont {Ciccarello}},\
  }\bibfield  {title} {\bibinfo {title} {Exciting a {Bound} {State} in the
  {Continuum} through {Multiphoton} {Scattering} {Plus} {Delayed} {Quantum}
  {Feedback}},\ }\href {https://doi.org/10.1103/PhysRevLett.122.073601}
  {\bibfield  {journal} {\bibinfo  {journal} {Physical Review Letters}\
  }\textbf {\bibinfo {volume} {122}},\ \bibinfo {pages} {073601} (\bibinfo
  {year} {2019})}\BibitemShut {NoStop}%
\bibitem [{\citenamefont {Yao}\ and\ \citenamefont
  {Hughes}(2009{\natexlab{a}})}]{Yao2009}%
  \BibitemOpen
  \bibfield  {author} {\bibinfo {author} {\bibfnamefont {P.}~\bibnamefont
  {Yao}}\ and\ \bibinfo {author} {\bibfnamefont {S.}~\bibnamefont {Hughes}},\
  }\bibfield  {title} {\bibinfo {title} {Macroscopic entanglement and violation
  of bell's inequalities between two spatially separated quantum dots in a
  planar photonic crystal system},\ }\href
  {https://doi.org/10.1364/oe.17.011505} {\bibfield  {journal} {\bibinfo
  {journal} {Optics Express}\ }\textbf {\bibinfo {volume} {17}},\ \bibinfo
  {pages} {11505} (\bibinfo {year} {2009}{\natexlab{a}})}\BibitemShut {NoStop}%
\bibitem [{\citenamefont {Hein}\ \emph {et~al.}(2016)\citenamefont {Hein},
  \citenamefont {Carmele},\ and\ \citenamefont {Knorr}}]{Hein2016}%
  \BibitemOpen
  \bibfield  {author} {\bibinfo {author} {\bibfnamefont {S.~M.}\ \bibnamefont
  {Hein}}, \bibinfo {author} {\bibfnamefont {A.}~\bibnamefont {Carmele}},\ and\
  \bibinfo {author} {\bibfnamefont {A.}~\bibnamefont {Knorr}},\ }\bibfield
  {title} {\bibinfo {title} {Creation and control of entanglement by
  time-delayed quantum-coherent feedback},\ }in\ \href
  {https://doi.org/10.1117/12.2207671} {\emph {\bibinfo {booktitle} {Physics
  and Simulation of Optoelectronic Devices {XXIV}}}},\ \bibinfo {editor}
  {edited by\ \bibinfo {editor} {\bibfnamefont {B.}~\bibnamefont {Witzigmann}},
  \bibinfo {editor} {\bibfnamefont {M.}~\bibnamefont {Osi{\'{n}}ski}},\ and\
  \bibinfo {editor} {\bibfnamefont {Y.}~\bibnamefont {Arakawa}}}\ (\bibinfo
  {publisher} {{SPIE}},\ \bibinfo {year} {2016})\BibitemShut {NoStop}%
\bibitem [{\citenamefont {Buckley}\ \emph {et~al.}(2012)\citenamefont
  {Buckley}, \citenamefont {Rivoire},\ and\ \citenamefont
  {Vu{\v{c}}kovi{\'{c}}}}]{Buckley_2012}%
  \BibitemOpen
  \bibfield  {author} {\bibinfo {author} {\bibfnamefont {S.}~\bibnamefont
  {Buckley}}, \bibinfo {author} {\bibfnamefont {K.}~\bibnamefont {Rivoire}},\
  and\ \bibinfo {author} {\bibfnamefont {J.}~\bibnamefont
  {Vu{\v{c}}kovi{\'{c}}}},\ }\bibfield  {title} {\bibinfo {title} {Engineered
  quantum dot single-photon sources},\ }\href
  {https://doi.org/10.1088/0034-4885/75/12/126503} {\bibfield  {journal}
  {\bibinfo  {journal} {Reports on Progress in Physics}\ }\textbf {\bibinfo
  {volume} {75}},\ \bibinfo {pages} {126503} (\bibinfo {year}
  {2012})}\BibitemShut {NoStop}%
\bibitem [{\citenamefont {Matthiesen}\ \emph {et~al.}(2012)\citenamefont
  {Matthiesen}, \citenamefont {Vamivakas},\ and\ \citenamefont
  {Atat\"ure}}]{Matthiese}%
  \BibitemOpen
  \bibfield  {author} {\bibinfo {author} {\bibfnamefont {C.}~\bibnamefont
  {Matthiesen}}, \bibinfo {author} {\bibfnamefont {A.~N.}\ \bibnamefont
  {Vamivakas}},\ and\ \bibinfo {author} {\bibfnamefont {M.}~\bibnamefont
  {Atat\"ure}},\ }\bibfield  {title} {\bibinfo {title} {Subnatural linewidth
  single photons from a quantum dot},\ }\href
  {https://doi.org/10.1103/PhysRevLett.108.093602} {\bibfield  {journal}
  {\bibinfo  {journal} {Phys. Rev. Lett.}\ }\textbf {\bibinfo {volume} {108}},\
  \bibinfo {pages} {093602} (\bibinfo {year} {2012})}\BibitemShut {NoStop}%
\bibitem [{\citenamefont {Heinze}\ \emph {et~al.}(2015)\citenamefont {Heinze},
  \citenamefont {Breddermann}, \citenamefont {Zrenner},\ and\ \citenamefont
  {Schumacher}}]{Heinze2015}%
  \BibitemOpen
  \bibfield  {author} {\bibinfo {author} {\bibfnamefont {D.}~\bibnamefont
  {Heinze}}, \bibinfo {author} {\bibfnamefont {D.}~\bibnamefont {Breddermann}},
  \bibinfo {author} {\bibfnamefont {A.}~\bibnamefont {Zrenner}},\ and\ \bibinfo
  {author} {\bibfnamefont {S.}~\bibnamefont {Schumacher}},\ }\bibfield  {title}
  {\bibinfo {title} {A quantum dot single-photon source with on-the-fly
  all-optical polarization control and timed emission},\ }\href
  {https://doi.org/10.1038/ncomms9473} {\bibfield  {journal} {\bibinfo
  {journal} {Nature Communications}\ }\textbf {\bibinfo {volume} {6}} (\bibinfo
  {year} {2015})}\BibitemShut {NoStop}%
\bibitem [{\citenamefont {T\"{u}rschmann}\ \emph {et~al.}(2019)\citenamefont
  {T\"{u}rschmann}, \citenamefont {Jeannic}, \citenamefont {Simonsen},
  \citenamefont {Haakh}, \citenamefont {G\"{o}tzinger}, \citenamefont
  {Sandoghdar}, \citenamefont {Lodahl},\ and\ \citenamefont
  {Rotenberg}}]{Trschmann2019}%
  \BibitemOpen
  \bibfield  {author} {\bibinfo {author} {\bibfnamefont {P.}~\bibnamefont
  {T\"{u}rschmann}}, \bibinfo {author} {\bibfnamefont {H.~L.}\ \bibnamefont
  {Jeannic}}, \bibinfo {author} {\bibfnamefont {S.~F.}\ \bibnamefont
  {Simonsen}}, \bibinfo {author} {\bibfnamefont {H.~R.}\ \bibnamefont {Haakh}},
  \bibinfo {author} {\bibfnamefont {S.}~\bibnamefont {G\"{o}tzinger}}, \bibinfo
  {author} {\bibfnamefont {V.}~\bibnamefont {Sandoghdar}}, \bibinfo {author}
  {\bibfnamefont {P.}~\bibnamefont {Lodahl}},\ and\ \bibinfo {author}
  {\bibfnamefont {N.}~\bibnamefont {Rotenberg}},\ }\bibfield  {title} {\bibinfo
  {title} {Coherent nonlinear optics of quantum emitters in nanophotonic
  waveguides},\ }\href {https://doi.org/10.1515/nanoph-2019-0126} {\bibfield
  {journal} {\bibinfo  {journal} {Nanophotonics}\ }\textbf {\bibinfo {volume}
  {8}},\ \bibinfo {pages} {1641} (\bibinfo {year} {2019})}\BibitemShut
  {NoStop}%
\bibitem [{\citenamefont {Gu}\ \emph {et~al.}(2017)\citenamefont {Gu},
  \citenamefont {Kockum}, \citenamefont {Miranowicz}, \citenamefont {xi~Liu},\
  and\ \citenamefont {Nori}}]{Gu2017}%
  \BibitemOpen
  \bibfield  {author} {\bibinfo {author} {\bibfnamefont {X.}~\bibnamefont
  {Gu}}, \bibinfo {author} {\bibfnamefont {A.~F.}\ \bibnamefont {Kockum}},
  \bibinfo {author} {\bibfnamefont {A.}~\bibnamefont {Miranowicz}}, \bibinfo
  {author} {\bibfnamefont {Y.}~\bibnamefont {xi~Liu}},\ and\ \bibinfo {author}
  {\bibfnamefont {F.}~\bibnamefont {Nori}},\ }\bibfield  {title} {\bibinfo
  {title} {Microwave photonics with superconducting quantum circuits},\ }\href
  {https://doi.org/10.1016/j.physrep.2017.10.002} {\bibfield  {journal}
  {\bibinfo  {journal} {Physics Reports}\ }\textbf {\bibinfo {volume}
  {718-719}},\ \bibinfo {pages} {1} (\bibinfo {year} {2017})}\BibitemShut
  {NoStop}%
\bibitem [{\citenamefont {Kockum}\ \emph {et~al.}(2018)\citenamefont {Kockum},
  \citenamefont {Johansson},\ and\ \citenamefont
  {Nori}}]{PhysRevLett.120.140404}%
  \BibitemOpen
  \bibfield  {author} {\bibinfo {author} {\bibfnamefont {A.~F.}\ \bibnamefont
  {Kockum}}, \bibinfo {author} {\bibfnamefont {G.}~\bibnamefont {Johansson}},\
  and\ \bibinfo {author} {\bibfnamefont {F.}~\bibnamefont {Nori}},\ }\bibfield
  {title} {\bibinfo {title} {Decoherence-free interaction between giant atoms
  in waveguide quantum electrodynamics},\ }\href
  {https://doi.org/10.1103/PhysRevLett.120.140404} {\bibfield  {journal}
  {\bibinfo  {journal} {Phys. Rev. Lett.}\ }\textbf {\bibinfo {volume} {120}},\
  \bibinfo {pages} {140404} (\bibinfo {year} {2018})}\BibitemShut {NoStop}%
\bibitem [{\citenamefont {Kannan}\ \emph {et~al.}(2020)\citenamefont {Kannan},
  \citenamefont {Ruckriegel}, \citenamefont {Campbell}, \citenamefont {Kockum},
  \citenamefont {Braum\"{u}ller}, \citenamefont {Kim}, \citenamefont
  {Kjaergaard}, \citenamefont {Krantz}, \citenamefont {Melville}, \citenamefont
  {Niedzielski}, \citenamefont {Veps\"{a}l\"{a}inen}, \citenamefont {Winik},
  \citenamefont {Yoder}, \citenamefont {Nori}, \citenamefont {Orlando},
  \citenamefont {Gustavsson},\ and\ \citenamefont {Oliver}}]{Kannan2020}%
  \BibitemOpen
  \bibfield  {author} {\bibinfo {author} {\bibfnamefont {B.}~\bibnamefont
  {Kannan}}, \bibinfo {author} {\bibfnamefont {M.~J.}\ \bibnamefont
  {Ruckriegel}}, \bibinfo {author} {\bibfnamefont {D.~L.}\ \bibnamefont
  {Campbell}}, \bibinfo {author} {\bibfnamefont {A.~F.}\ \bibnamefont
  {Kockum}}, \bibinfo {author} {\bibfnamefont {J.}~\bibnamefont
  {Braum\"{u}ller}}, \bibinfo {author} {\bibfnamefont {D.~K.}\ \bibnamefont
  {Kim}}, \bibinfo {author} {\bibfnamefont {M.}~\bibnamefont {Kjaergaard}},
  \bibinfo {author} {\bibfnamefont {P.}~\bibnamefont {Krantz}}, \bibinfo
  {author} {\bibfnamefont {A.}~\bibnamefont {Melville}}, \bibinfo {author}
  {\bibfnamefont {B.~M.}\ \bibnamefont {Niedzielski}}, \bibinfo {author}
  {\bibfnamefont {A.}~\bibnamefont {Veps\"{a}l\"{a}inen}}, \bibinfo {author}
  {\bibfnamefont {R.}~\bibnamefont {Winik}}, \bibinfo {author} {\bibfnamefont
  {J.~L.}\ \bibnamefont {Yoder}}, \bibinfo {author} {\bibfnamefont
  {F.}~\bibnamefont {Nori}}, \bibinfo {author} {\bibfnamefont {T.~P.}\
  \bibnamefont {Orlando}}, \bibinfo {author} {\bibfnamefont {S.}~\bibnamefont
  {Gustavsson}},\ and\ \bibinfo {author} {\bibfnamefont {W.~D.}\ \bibnamefont
  {Oliver}},\ }\bibfield  {title} {\bibinfo {title} {Waveguide quantum
  electrodynamics with superconducting artificial giant atoms},\ }\href
  {https://doi.org/10.1038/s41586-020-2529-9} {\bibfield  {journal} {\bibinfo
  {journal} {Nature}\ }\textbf {\bibinfo {volume} {583}},\ \bibinfo {pages}
  {775} (\bibinfo {year} {2020})}\BibitemShut {NoStop}%
\bibitem [{\citenamefont {Yang}\ \emph {et~al.}(2018)\citenamefont {Yang},
  \citenamefont {Binder}, \citenamefont {Narasimhachar},\ and\ \citenamefont
  {Gu}}]{yang_matrix_2018}%
  \BibitemOpen
  \bibfield  {author} {\bibinfo {author} {\bibfnamefont {C.}~\bibnamefont
  {Yang}}, \bibinfo {author} {\bibfnamefont {F.~C.}\ \bibnamefont {Binder}},
  \bibinfo {author} {\bibfnamefont {V.}~\bibnamefont {Narasimhachar}},\ and\
  \bibinfo {author} {\bibfnamefont {M.}~\bibnamefont {Gu}},\ }\bibfield
  {title} {\bibinfo {title} {Matrix product states for quantum stochastic
  modeling},\ }\href {https://doi.org/10.1103/PhysRevLett.121.260602}
  {\bibfield  {journal} {\bibinfo  {journal} {Physical Review Letters}\
  }\textbf {\bibinfo {volume} {121}},\ \bibinfo {pages} {260602} (\bibinfo
  {year} {2018})}\BibitemShut {NoStop}%
\bibitem [{\citenamefont {Vanderstraeten}(2017)}]{vanderstraeten_tensor_2017}%
  \BibitemOpen
  \bibfield  {author} {\bibinfo {author} {\bibfnamefont {L.}~\bibnamefont
  {Vanderstraeten}},\ }\href@noop {} {\emph {\bibinfo {title} {Tensor Network
  States and Effective Particles for Low Dimensional Quantum Spin Systems}}}\
  (\bibinfo  {publisher} {Springer, Cham},\ \bibinfo {year} {2017})\BibitemShut
  {NoStop}%
\bibitem [{\citenamefont {Naumann}\ \emph {et~al.}(2017)\citenamefont
  {Naumann}, \citenamefont {Hein}, \citenamefont {Kraft}, \citenamefont
  {Knorr},\ and\ \citenamefont {Carmele}}]{Naumann2017}%
  \BibitemOpen
  \bibfield  {author} {\bibinfo {author} {\bibfnamefont {N.~L.}\ \bibnamefont
  {Naumann}}, \bibinfo {author} {\bibfnamefont {S.~M.}\ \bibnamefont {Hein}},
  \bibinfo {author} {\bibfnamefont {M.}~\bibnamefont {Kraft}}, \bibinfo
  {author} {\bibfnamefont {A.}~\bibnamefont {Knorr}},\ and\ \bibinfo {author}
  {\bibfnamefont {A.}~\bibnamefont {Carmele}},\ }\bibfield  {title} {\bibinfo
  {title} {Feedback control of photon statistics},\ }in\ \href
  {https://doi.org/10.1117/12.2251952} {\emph {\bibinfo {booktitle} {Physics
  and {Simulation} of {Optoelectronic} {Devices} {XXV}}}},\ Vol.\ \bibinfo
  {volume} {10098}\ (\bibinfo  {publisher} {International Society for Optics
  and Photonics},\ \bibinfo {year} {2017})\ p.\ \bibinfo {pages}
  {100980N}\BibitemShut {NoStop}%
\bibitem [{\citenamefont {Ciccarello}(2017)}]{Ciccarello2017}%
  \BibitemOpen
  \bibfield  {author} {\bibinfo {author} {\bibfnamefont {F.}~\bibnamefont
  {Ciccarello}},\ }\bibfield  {title} {{\selectlanguage {en}\bibinfo {title}
  {Collision models in quantum optics}},\ }\href
  {https://doi.org/10.1515/qmetro-2017-0007} {\bibfield  {journal} {\bibinfo
  {journal} {QMTR}\ }\textbf {\bibinfo {volume} {4}},\ \bibinfo {pages} {53}
  (\bibinfo {year} {2017})}\BibitemShut {NoStop}%
\bibitem [{\citenamefont {Whalen}(2019)}]{PhysRevA.100.052113}%
  \BibitemOpen
  \bibfield  {author} {\bibinfo {author} {\bibfnamefont {S.~J.}\ \bibnamefont
  {Whalen}},\ }\bibfield  {title} {\bibinfo {title} {Collision model for
  non-markovian quantum trajectories},\ }\href
  {https://doi.org/10.1103/PhysRevA.100.052113} {\bibfield  {journal} {\bibinfo
   {journal} {Phys. Rev. A}\ }\textbf {\bibinfo {volume} {100}},\ \bibinfo
  {pages} {052113} (\bibinfo {year} {2019})}\BibitemShut {NoStop}%
\bibitem [{\citenamefont {Crowder}(2020)}]{GavinThesis}%
  \BibitemOpen
  \bibfield  {author} {\bibinfo {author} {\bibfnamefont {G.}~\bibnamefont
  {Crowder}},\ }\emph {\bibinfo {title} {Quantum Trajectory Theory of Open
  Cavity-QED Systems with a Time Delayed Coherent Optical Feedback}},\ \href
  {http://hdl.handle.net/1974/27840} {Master's thesis},\ \bibinfo  {school}
  {Queen's University at Kingston} (\bibinfo {year} {2020})\BibitemShut
  {NoStop}%
\bibitem [{\citenamefont {Cilluffo}\ \emph {et~al.}(2020)\citenamefont
  {Cilluffo}, \citenamefont {Carollo}, \citenamefont {Lorenzo}, \citenamefont
  {Gross}, \citenamefont {Palma},\ and\ \citenamefont
  {Ciccarello}}]{Cilluffo2020}%
  \BibitemOpen
  \bibfield  {author} {\bibinfo {author} {\bibfnamefont {D.}~\bibnamefont
  {Cilluffo}}, \bibinfo {author} {\bibfnamefont {A.}~\bibnamefont {Carollo}},
  \bibinfo {author} {\bibfnamefont {S.}~\bibnamefont {Lorenzo}}, \bibinfo
  {author} {\bibfnamefont {J.~A.}\ \bibnamefont {Gross}}, \bibinfo {author}
  {\bibfnamefont {G.~M.}\ \bibnamefont {Palma}},\ and\ \bibinfo {author}
  {\bibfnamefont {F.}~\bibnamefont {Ciccarello}},\ }\bibfield  {title}
  {{\selectlanguage {en}\bibinfo {title} {Collisional picture of quantum optics
  with giant emitters}},\ }\href {http://arxiv.org/abs/2006.08631} {\bibfield
  {journal} {\bibinfo  {journal} {arXiv:2006.08631 [quant-ph]}\ } (\bibinfo
  {year} {2020})}\BibitemShut {NoStop}%
\bibitem [{\citenamefont {Orús}(2014)}]{orus_practical_2014}%
  \BibitemOpen
  \bibfield  {author} {\bibinfo {author} {\bibfnamefont {R.}~\bibnamefont
  {Orús}},\ }\bibfield  {title} {\bibinfo {title} {A practical introduction to
  tensor networks: Matrix product states and projected entangled pair states},\
  }\href {https://doi.org/10.1016/j.aop.2014.06.013} {\bibfield  {journal}
  {\bibinfo  {journal} {Annals of Physics}\ }\textbf {\bibinfo {volume}
  {349}},\ \bibinfo {pages} {117} (\bibinfo {year} {2014})}\BibitemShut
  {NoStop}%
\bibitem [{\citenamefont {Droenner}(2019)}]{droenner_out--equilibrium_2019}%
  \BibitemOpen
  \bibfield  {author} {\bibinfo {author} {\bibfnamefont {L.~J.}\ \bibnamefont
  {Droenner}},\ }\emph {\bibinfo {title} {Out-of-equilibrium dynamics of open
  quantum many-body systems}},\ \href
  {https://doi.org/10.14279/depositonce-8154} {\bibinfo {type} {Doctoral
  thesis}},\ \bibinfo  {school} {Technischen Universität Berlin} (\bibinfo
  {year} {2019})\BibitemShut {NoStop}%
\bibitem [{\citenamefont {Eduardo}(2017)}]{eduardo_one-dimensional_2017}%
  \BibitemOpen
  \bibfield  {author} {\bibinfo {author} {\bibfnamefont {S.~B.}\ \bibnamefont
  {Eduardo}},\ }\href
  {https://books.google.ca/books?hl=en&lr=&id=afxFDwAAQBAJ&oi=fnd&pg=PA211&dq=One-dimensional+few-+photon+scattering:+Numerical+and+analytical+techniques&ots=wYrgHWkcA9&sig=A4376ptXGtTapRbs2lhTwSNPpYs#v=onepage&q=One-dimensional%20few-%20photon%20scattering%3A%20Numerical%20and%20analytical%20techniques&f=false}
  {\emph {\bibinfo {title} {One-dimensional few-photon scattering}}}\ (\bibinfo
   {publisher} {Prensas de la Universidad de Zaragoza},\ \bibinfo {address}
  {Zaragoza},\ \bibinfo {year} {2017})\BibitemShut {NoStop}%
\bibitem [{\citenamefont {Brun}(2002)}]{Brun2002}%
  \BibitemOpen
  \bibfield  {author} {\bibinfo {author} {\bibfnamefont {T.~A.}\ \bibnamefont
  {Brun}},\ }\bibfield  {title} {\bibinfo {title} {A simple model of quantum
  trajectories},\ }\href {https://doi.org/10.1119/1.1475328} {\bibfield
  {journal} {\bibinfo  {journal} {American Journal of Physics}\ }\textbf
  {\bibinfo {volume} {70}},\ \bibinfo {pages} {719} (\bibinfo {year}
  {2002})}\BibitemShut {NoStop}%
\bibitem [{\citenamefont {Kretschmer}\ \emph {et~al.}(2016)\citenamefont
  {Kretschmer}, \citenamefont {Luoma},\ and\ \citenamefont
  {Strunz}}]{Kretschmer2016}%
  \BibitemOpen
  \bibfield  {author} {\bibinfo {author} {\bibfnamefont {S.}~\bibnamefont
  {Kretschmer}}, \bibinfo {author} {\bibfnamefont {K.}~\bibnamefont {Luoma}},\
  and\ \bibinfo {author} {\bibfnamefont {W.~T.}\ \bibnamefont {Strunz}},\
  }\bibfield  {title} {\bibinfo {title} {Collision model for non-{Markovian}
  quantum dynamics},\ }\href {https://doi.org/10.1103/PhysRevA.94.012106}
  {\bibfield  {journal} {\bibinfo  {journal} {Physical Review A}\ }\textbf
  {\bibinfo {volume} {94}},\ \bibinfo {pages} {012106} (\bibinfo {year}
  {2016})}\BibitemShut {NoStop}%
\bibitem [{\citenamefont {Gustin}\ and\ \citenamefont
  {Hughes}(2018)}]{PhysRevB.98.045309}%
  \BibitemOpen
  \bibfield  {author} {\bibinfo {author} {\bibfnamefont {C.}~\bibnamefont
  {Gustin}}\ and\ \bibinfo {author} {\bibfnamefont {S.}~\bibnamefont
  {Hughes}},\ }\bibfield  {title} {\bibinfo {title} {Pulsed excitation dynamics
  in quantum-dot--cavity systems: Limits to optimizing the fidelity of
  on-demand single-photon sources},\ }\href
  {https://doi.org/10.1103/PhysRevB.98.045309} {\bibfield  {journal} {\bibinfo
  {journal} {Phys. Rev. B}\ }\textbf {\bibinfo {volume} {98}},\ \bibinfo
  {pages} {045309} (\bibinfo {year} {2018})}\BibitemShut {NoStop}%
\bibitem [{\citenamefont {Iles-Smith}\ \emph {et~al.}(2017)\citenamefont
  {Iles-Smith}, \citenamefont {McCutcheon}, \citenamefont {Nazir},\ and\
  \citenamefont {M{\o}rk}}]{IlesSmith2017}%
  \BibitemOpen
  \bibfield  {author} {\bibinfo {author} {\bibfnamefont {J.}~\bibnamefont
  {Iles-Smith}}, \bibinfo {author} {\bibfnamefont {D.~P.~S.}\ \bibnamefont
  {McCutcheon}}, \bibinfo {author} {\bibfnamefont {A.}~\bibnamefont {Nazir}},\
  and\ \bibinfo {author} {\bibfnamefont {J.}~\bibnamefont {M{\o}rk}},\
  }\bibfield  {title} {\bibinfo {title} {Phonon scattering inhibits
  simultaneous near-unity efficiency and indistinguishability in semiconductor
  single-photon sources},\ }\href {https://doi.org/10.1038/nphoton.2017.101}
  {\bibfield  {journal} {\bibinfo  {journal} {Nature Photonics}\ }\textbf
  {\bibinfo {volume} {11}},\ \bibinfo {pages} {521} (\bibinfo {year}
  {2017})}\BibitemShut {NoStop}%
\bibitem [{\citenamefont {Kuhlmann}\ \emph {et~al.}(2013)\citenamefont
  {Kuhlmann}, \citenamefont {Houel}, \citenamefont {Ludwig}, \citenamefont
  {Greuter}, \citenamefont {Reuter}, \citenamefont {Wieck}, \citenamefont
  {Poggio},\ and\ \citenamefont {Warburton}}]{Kuhlmann2013}%
  \BibitemOpen
  \bibfield  {author} {\bibinfo {author} {\bibfnamefont {A.~V.}\ \bibnamefont
  {Kuhlmann}}, \bibinfo {author} {\bibfnamefont {J.}~\bibnamefont {Houel}},
  \bibinfo {author} {\bibfnamefont {A.}~\bibnamefont {Ludwig}}, \bibinfo
  {author} {\bibfnamefont {L.}~\bibnamefont {Greuter}}, \bibinfo {author}
  {\bibfnamefont {D.}~\bibnamefont {Reuter}}, \bibinfo {author} {\bibfnamefont
  {A.~D.}\ \bibnamefont {Wieck}}, \bibinfo {author} {\bibfnamefont
  {M.}~\bibnamefont {Poggio}},\ and\ \bibinfo {author} {\bibfnamefont {R.~J.}\
  \bibnamefont {Warburton}},\ }\bibfield  {title} {\bibinfo {title} {Charge
  noise and spin noise in a semiconductor quantum device},\ }\href
  {https://doi.org/10.1038/nphys2688} {\bibfield  {journal} {\bibinfo
  {journal} {Nature Physics}\ }\textbf {\bibinfo {volume} {9}},\ \bibinfo
  {pages} {570} (\bibinfo {year} {2013})}\BibitemShut {NoStop}%
\bibitem [{\citenamefont {Ramsay}\ \emph {et~al.}(2010)\citenamefont {Ramsay},
  \citenamefont {Gopal}, \citenamefont {Gauger}, \citenamefont {Nazir},
  \citenamefont {Lovett}, \citenamefont {Fox},\ and\ \citenamefont
  {Skolnick}}]{PhysRevLett.104.017402}%
  \BibitemOpen
  \bibfield  {author} {\bibinfo {author} {\bibfnamefont {A.~J.}\ \bibnamefont
  {Ramsay}}, \bibinfo {author} {\bibfnamefont {A.~V.}\ \bibnamefont {Gopal}},
  \bibinfo {author} {\bibfnamefont {E.~M.}\ \bibnamefont {Gauger}}, \bibinfo
  {author} {\bibfnamefont {A.}~\bibnamefont {Nazir}}, \bibinfo {author}
  {\bibfnamefont {B.~W.}\ \bibnamefont {Lovett}}, \bibinfo {author}
  {\bibfnamefont {A.~M.}\ \bibnamefont {Fox}},\ and\ \bibinfo {author}
  {\bibfnamefont {M.~S.}\ \bibnamefont {Skolnick}},\ }\bibfield  {title}
  {\bibinfo {title} {Damping of exciton rabi rotations by acoustic phonons in
  optically excited $\mathrm{InGaAs}/\mathrm{GaAs}$ quantum dots},\ }\href
  {https://doi.org/10.1103/PhysRevLett.104.017402} {\bibfield  {journal}
  {\bibinfo  {journal} {Phys. Rev. Lett.}\ }\textbf {\bibinfo {volume} {104}},\
  \bibinfo {pages} {017402} (\bibinfo {year} {2010})}\BibitemShut {NoStop}%
\bibitem [{\citenamefont {Grange}\ \emph {et~al.}(2017)\citenamefont {Grange},
  \citenamefont {Somaschi}, \citenamefont {Ant\'on}, \citenamefont {De~Santis},
  \citenamefont {Coppola}, \citenamefont {Giesz}, \citenamefont
  {Lema\^{\i}tre}, \citenamefont {Sagnes}, \citenamefont {Auff\`eves},\ and\
  \citenamefont {Senellart}}]{PhysRevLett.118.253602}%
  \BibitemOpen
  \bibfield  {author} {\bibinfo {author} {\bibfnamefont {T.}~\bibnamefont
  {Grange}}, \bibinfo {author} {\bibfnamefont {N.}~\bibnamefont {Somaschi}},
  \bibinfo {author} {\bibfnamefont {C.}~\bibnamefont {Ant\'on}}, \bibinfo
  {author} {\bibfnamefont {L.}~\bibnamefont {De~Santis}}, \bibinfo {author}
  {\bibfnamefont {G.}~\bibnamefont {Coppola}}, \bibinfo {author} {\bibfnamefont
  {V.}~\bibnamefont {Giesz}}, \bibinfo {author} {\bibfnamefont
  {A.}~\bibnamefont {Lema\^{\i}tre}}, \bibinfo {author} {\bibfnamefont
  {I.}~\bibnamefont {Sagnes}}, \bibinfo {author} {\bibfnamefont
  {A.}~\bibnamefont {Auff\`eves}},\ and\ \bibinfo {author} {\bibfnamefont
  {P.}~\bibnamefont {Senellart}},\ }\bibfield  {title} {\bibinfo {title}
  {Reducing phonon-induced decoherence in solid-state single-photon sources
  with cavity quantum electrodynamics},\ }\href
  {https://doi.org/10.1103/PhysRevLett.118.253602} {\bibfield  {journal}
  {\bibinfo  {journal} {Phys. Rev. Lett.}\ }\textbf {\bibinfo {volume} {118}},\
  \bibinfo {pages} {253602} (\bibinfo {year} {2017})}\BibitemShut {NoStop}%
\bibitem [{\citenamefont {Lodahl}\ \emph {et~al.}(2015)\citenamefont {Lodahl},
  \citenamefont {Mahmoodian},\ and\ \citenamefont
  {Stobbe}}]{RevModPhys.87.347}%
  \BibitemOpen
  \bibfield  {author} {\bibinfo {author} {\bibfnamefont {P.}~\bibnamefont
  {Lodahl}}, \bibinfo {author} {\bibfnamefont {S.}~\bibnamefont {Mahmoodian}},\
  and\ \bibinfo {author} {\bibfnamefont {S.}~\bibnamefont {Stobbe}},\
  }\bibfield  {title} {\bibinfo {title} {Interfacing single photons and single
  quantum dots with photonic nanostructures},\ }\href
  {https://doi.org/10.1103/RevModPhys.87.347} {\bibfield  {journal} {\bibinfo
  {journal} {Rev. Mod. Phys.}\ }\textbf {\bibinfo {volume} {87}},\ \bibinfo
  {pages} {347} (\bibinfo {year} {2015})}\BibitemShut {NoStop}%
\bibitem [{\citenamefont {Vagov}\ \emph {et~al.}(2002)\citenamefont {Vagov},
  \citenamefont {Axt},\ and\ \citenamefont {Kuhn}}]{PhysRevB.66.165312}%
  \BibitemOpen
  \bibfield  {author} {\bibinfo {author} {\bibfnamefont {A.}~\bibnamefont
  {Vagov}}, \bibinfo {author} {\bibfnamefont {V.~M.}\ \bibnamefont {Axt}},\
  and\ \bibinfo {author} {\bibfnamefont {T.}~\bibnamefont {Kuhn}},\ }\bibfield
  {title} {\bibinfo {title} {Electron-phonon dynamics in optically excited
  quantum dots: Exact solution for multiple ultrashort laser pulses},\ }\href
  {https://doi.org/10.1103/PhysRevB.66.165312} {\bibfield  {journal} {\bibinfo
  {journal} {Phys. Rev. B}\ }\textbf {\bibinfo {volume} {66}},\ \bibinfo
  {pages} {165312} (\bibinfo {year} {2002})}\BibitemShut {NoStop}%
\bibitem [{\citenamefont {F\"orstner}\ \emph {et~al.}(2003)\citenamefont
  {F\"orstner}, \citenamefont {Weber}, \citenamefont {Danckwerts},\ and\
  \citenamefont {Knorr}}]{PhysRevLett.91.127401}%
  \BibitemOpen
  \bibfield  {author} {\bibinfo {author} {\bibfnamefont {J.}~\bibnamefont
  {F\"orstner}}, \bibinfo {author} {\bibfnamefont {C.}~\bibnamefont {Weber}},
  \bibinfo {author} {\bibfnamefont {J.}~\bibnamefont {Danckwerts}},\ and\
  \bibinfo {author} {\bibfnamefont {A.}~\bibnamefont {Knorr}},\ }\bibfield
  {title} {\bibinfo {title} {Phonon-assisted damping of rabi oscillations in
  semiconductor quantum dots},\ }\href
  {https://doi.org/10.1103/PhysRevLett.91.127401} {\bibfield  {journal}
  {\bibinfo  {journal} {Phys. Rev. Lett.}\ }\textbf {\bibinfo {volume} {91}},\
  \bibinfo {pages} {127401} (\bibinfo {year} {2003})}\BibitemShut {NoStop}%
\bibitem [{\citenamefont {Besombes}\ \emph {et~al.}(2001)\citenamefont
  {Besombes}, \citenamefont {Kheng}, \citenamefont {Marsal},\ and\
  \citenamefont {Mariette}}]{PhysRevB.63.155307}%
  \BibitemOpen
  \bibfield  {author} {\bibinfo {author} {\bibfnamefont {L.}~\bibnamefont
  {Besombes}}, \bibinfo {author} {\bibfnamefont {K.}~\bibnamefont {Kheng}},
  \bibinfo {author} {\bibfnamefont {L.}~\bibnamefont {Marsal}},\ and\ \bibinfo
  {author} {\bibfnamefont {H.}~\bibnamefont {Mariette}},\ }\bibfield  {title}
  {\bibinfo {title} {Acoustic phonon broadening mechanism in single quantum dot
  emission},\ }\href {https://doi.org/10.1103/PhysRevB.63.155307} {\bibfield
  {journal} {\bibinfo  {journal} {Phys. Rev. B}\ }\textbf {\bibinfo {volume}
  {63}},\ \bibinfo {pages} {155307} (\bibinfo {year} {2001})}\BibitemShut
  {NoStop}%
\bibitem [{\citenamefont {le~Feber}\ \emph {et~al.}(2015)\citenamefont
  {le~Feber}, \citenamefont {Rotenberg},\ and\ \citenamefont
  {Kuipers}}]{leFeber2015}%
  \BibitemOpen
  \bibfield  {author} {\bibinfo {author} {\bibfnamefont {B.}~\bibnamefont
  {le~Feber}}, \bibinfo {author} {\bibfnamefont {N.}~\bibnamefont
  {Rotenberg}},\ and\ \bibinfo {author} {\bibfnamefont {L.}~\bibnamefont
  {Kuipers}},\ }\bibfield  {title} {\bibinfo {title} {Nanophotonic control of
  circular dipole emission},\ }\bibfield  {journal} {\bibinfo  {journal}
  {Nature Communications}\ }\textbf {\bibinfo {volume} {6}},\ \href
  {https://doi.org/10.1038/ncomms7695} {10.1038/ncomms7695} (\bibinfo {year}
  {2015})\BibitemShut {NoStop}%
\bibitem [{\citenamefont {Young}\ \emph {et~al.}(2015)\citenamefont {Young},
  \citenamefont {Thijssen}, \citenamefont {Beggs}, \citenamefont
  {Androvitsaneas}, \citenamefont {Kuipers}, \citenamefont {Rarity},
  \citenamefont {Hughes},\ and\ \citenamefont
  {Oulton}}]{PhysRevLett.115.153901}%
  \BibitemOpen
  \bibfield  {author} {\bibinfo {author} {\bibfnamefont {A.~B.}\ \bibnamefont
  {Young}}, \bibinfo {author} {\bibfnamefont {A.~C.~T.}\ \bibnamefont
  {Thijssen}}, \bibinfo {author} {\bibfnamefont {D.~M.}\ \bibnamefont {Beggs}},
  \bibinfo {author} {\bibfnamefont {P.}~\bibnamefont {Androvitsaneas}},
  \bibinfo {author} {\bibfnamefont {L.}~\bibnamefont {Kuipers}}, \bibinfo
  {author} {\bibfnamefont {J.~G.}\ \bibnamefont {Rarity}}, \bibinfo {author}
  {\bibfnamefont {S.}~\bibnamefont {Hughes}},\ and\ \bibinfo {author}
  {\bibfnamefont {R.}~\bibnamefont {Oulton}},\ }\bibfield  {title} {\bibinfo
  {title} {Polarization engineering in photonic crystal waveguides for
  spin-photon entanglers},\ }\href
  {https://doi.org/10.1103/PhysRevLett.115.153901} {\bibfield  {journal}
  {\bibinfo  {journal} {Phys. Rev. Lett.}\ }\textbf {\bibinfo {volume} {115}},\
  \bibinfo {pages} {153901} (\bibinfo {year} {2015})}\BibitemShut {NoStop}%
\bibitem [{\citenamefont {S\"{o}llner}\ \emph {et~al.}(2015)\citenamefont
  {S\"{o}llner}, \citenamefont {Mahmoodian}, \citenamefont {Hansen},
  \citenamefont {Midolo}, \citenamefont {Javadi}, \citenamefont
  {Kir{\v{s}}ansk{\.{e}}}, \citenamefont {Pregnolato}, \citenamefont {El-Ella},
  \citenamefont {Lee}, \citenamefont {Song}, \citenamefont {Stobbe},\ and\
  \citenamefont {Lodahl}}]{Sllner2015}%
  \BibitemOpen
  \bibfield  {author} {\bibinfo {author} {\bibfnamefont {I.}~\bibnamefont
  {S\"{o}llner}}, \bibinfo {author} {\bibfnamefont {S.}~\bibnamefont
  {Mahmoodian}}, \bibinfo {author} {\bibfnamefont {S.~L.}\ \bibnamefont
  {Hansen}}, \bibinfo {author} {\bibfnamefont {L.}~\bibnamefont {Midolo}},
  \bibinfo {author} {\bibfnamefont {A.}~\bibnamefont {Javadi}}, \bibinfo
  {author} {\bibfnamefont {G.}~\bibnamefont {Kir{\v{s}}ansk{\.{e}}}}, \bibinfo
  {author} {\bibfnamefont {T.}~\bibnamefont {Pregnolato}}, \bibinfo {author}
  {\bibfnamefont {H.}~\bibnamefont {El-Ella}}, \bibinfo {author} {\bibfnamefont
  {E.~H.}\ \bibnamefont {Lee}}, \bibinfo {author} {\bibfnamefont {J.~D.}\
  \bibnamefont {Song}}, \bibinfo {author} {\bibfnamefont {S.}~\bibnamefont
  {Stobbe}},\ and\ \bibinfo {author} {\bibfnamefont {P.}~\bibnamefont
  {Lodahl}},\ }\bibfield  {title} {\bibinfo {title} {Deterministic
  photon{\textendash}emitter coupling in chiral photonic circuits},\ }\href
  {https://doi.org/10.1038/nnano.2015.159} {\bibfield  {journal} {\bibinfo
  {journal} {Nature Nanotechnology}\ }\textbf {\bibinfo {volume} {10}},\
  \bibinfo {pages} {775} (\bibinfo {year} {2015})}\BibitemShut {NoStop}%
\bibitem [{\citenamefont {Barik}\ \emph {et~al.}(2018)\citenamefont {Barik},
  \citenamefont {Karasahin}, \citenamefont {Flower}, \citenamefont {Cai},
  \citenamefont {Miyake}, \citenamefont {DeGottardi}, \citenamefont {Hafezi},\
  and\ \citenamefont {Waks}}]{Barik2018}%
  \BibitemOpen
  \bibfield  {author} {\bibinfo {author} {\bibfnamefont {S.}~\bibnamefont
  {Barik}}, \bibinfo {author} {\bibfnamefont {A.}~\bibnamefont {Karasahin}},
  \bibinfo {author} {\bibfnamefont {C.}~\bibnamefont {Flower}}, \bibinfo
  {author} {\bibfnamefont {T.}~\bibnamefont {Cai}}, \bibinfo {author}
  {\bibfnamefont {H.}~\bibnamefont {Miyake}}, \bibinfo {author} {\bibfnamefont
  {W.}~\bibnamefont {DeGottardi}}, \bibinfo {author} {\bibfnamefont
  {M.}~\bibnamefont {Hafezi}},\ and\ \bibinfo {author} {\bibfnamefont
  {E.}~\bibnamefont {Waks}},\ }\bibfield  {title} {\bibinfo {title} {A
  topological quantum optics interface},\ }\href
  {https://doi.org/10.1126/science.aaq0327} {\bibfield  {journal} {\bibinfo
  {journal} {Science}\ }\textbf {\bibinfo {volume} {359}},\ \bibinfo {pages}
  {666} (\bibinfo {year} {2018})}\BibitemShut {NoStop}%
\bibitem [{\citenamefont {Lodahl}\ \emph {et~al.}(2017)\citenamefont {Lodahl},
  \citenamefont {Mahmoodian}, \citenamefont {Stobbe}, \citenamefont
  {Rauschenbeutel}, \citenamefont {Schneeweiss}, \citenamefont {Volz},
  \citenamefont {Pichler},\ and\ \citenamefont {Zoller}}]{Lodahl2017}%
  \BibitemOpen
  \bibfield  {author} {\bibinfo {author} {\bibfnamefont {P.}~\bibnamefont
  {Lodahl}}, \bibinfo {author} {\bibfnamefont {S.}~\bibnamefont {Mahmoodian}},
  \bibinfo {author} {\bibfnamefont {S.}~\bibnamefont {Stobbe}}, \bibinfo
  {author} {\bibfnamefont {A.}~\bibnamefont {Rauschenbeutel}}, \bibinfo
  {author} {\bibfnamefont {P.}~\bibnamefont {Schneeweiss}}, \bibinfo {author}
  {\bibfnamefont {J.}~\bibnamefont {Volz}}, \bibinfo {author} {\bibfnamefont
  {H.}~\bibnamefont {Pichler}},\ and\ \bibinfo {author} {\bibfnamefont
  {P.}~\bibnamefont {Zoller}},\ }\bibfield  {title} {\bibinfo {title} {Chiral
  quantum optics},\ }\href {https://doi.org/10.1038/nature21037} {\bibfield
  {journal} {\bibinfo  {journal} {Nature}\ }\textbf {\bibinfo {volume} {541}},\
  \bibinfo {pages} {473} (\bibinfo {year} {2017})}\BibitemShut {NoStop}%
\bibitem [{\citenamefont {Bliokh}\ and\ \citenamefont
  {Nori}(2012)}]{PhysRevA.85.061801}%
  \BibitemOpen
  \bibfield  {author} {\bibinfo {author} {\bibfnamefont {K.~Y.}\ \bibnamefont
  {Bliokh}}\ and\ \bibinfo {author} {\bibfnamefont {F.}~\bibnamefont {Nori}},\
  }\bibfield  {title} {\bibinfo {title} {Transverse spin of a surface
  polariton},\ }\href {https://doi.org/10.1103/PhysRevA.85.061801} {\bibfield
  {journal} {\bibinfo  {journal} {Phys. Rev. A}\ }\textbf {\bibinfo {volume}
  {85}},\ \bibinfo {pages} {061801} (\bibinfo {year} {2012})}\BibitemShut
  {NoStop}%
\bibitem [{\citenamefont {Coles}\ \emph {et~al.}(2016)\citenamefont {Coles},
  \citenamefont {Price}, \citenamefont {Dixon}, \citenamefont {Royall},
  \citenamefont {Clarke}, \citenamefont {Kok}, \citenamefont {Skolnick},
  \citenamefont {Fox},\ and\ \citenamefont {Makhonin}}]{Coles2016}%
  \BibitemOpen
  \bibfield  {author} {\bibinfo {author} {\bibfnamefont {R.~J.}\ \bibnamefont
  {Coles}}, \bibinfo {author} {\bibfnamefont {D.~M.}\ \bibnamefont {Price}},
  \bibinfo {author} {\bibfnamefont {J.~E.}\ \bibnamefont {Dixon}}, \bibinfo
  {author} {\bibfnamefont {B.}~\bibnamefont {Royall}}, \bibinfo {author}
  {\bibfnamefont {E.}~\bibnamefont {Clarke}}, \bibinfo {author} {\bibfnamefont
  {P.}~\bibnamefont {Kok}}, \bibinfo {author} {\bibfnamefont {M.~S.}\
  \bibnamefont {Skolnick}}, \bibinfo {author} {\bibfnamefont {A.~M.}\
  \bibnamefont {Fox}},\ and\ \bibinfo {author} {\bibfnamefont {M.~N.}\
  \bibnamefont {Makhonin}},\ }\bibfield  {title} {\bibinfo {title} {Chirality
  of nanophotonic waveguide with embedded quantum emitter for unidirectional
  spin transfer},\ }\bibfield  {journal} {\bibinfo  {journal} {Nature
  Communications}\ }\textbf {\bibinfo {volume} {7}},\ \href
  {https://doi.org/10.1038/ncomms11183} {10.1038/ncomms11183} (\bibinfo {year}
  {2016})\BibitemShut {NoStop}%
\bibitem [{\citenamefont {Martin-Cano}\ \emph {et~al.}(2019)\citenamefont
  {Martin-Cano}, \citenamefont {Haakh},\ and\ \citenamefont
  {Rotenberg}}]{MartinCano2019}%
  \BibitemOpen
  \bibfield  {author} {\bibinfo {author} {\bibfnamefont {D.}~\bibnamefont
  {Martin-Cano}}, \bibinfo {author} {\bibfnamefont {H.~R.}\ \bibnamefont
  {Haakh}},\ and\ \bibinfo {author} {\bibfnamefont {N.}~\bibnamefont
  {Rotenberg}},\ }\bibfield  {title} {\bibinfo {title} {Chiral emission into
  nanophotonic resonators},\ }\href
  {https://doi.org/10.1021/acsphotonics.8b01555} {\bibfield  {journal}
  {\bibinfo  {journal} {{ACS} Photonics}\ }\textbf {\bibinfo {volume} {6}},\
  \bibinfo {pages} {961} (\bibinfo {year} {2019})}\BibitemShut {NoStop}%
\bibitem [{\citenamefont {Orazbayev}\ \emph {et~al.}(2018)\citenamefont
  {Orazbayev}, \citenamefont {Kaina},\ and\ \citenamefont
  {Fleury}}]{PhysRevApplied.10.054069}%
  \BibitemOpen
  \bibfield  {author} {\bibinfo {author} {\bibfnamefont {B.}~\bibnamefont
  {Orazbayev}}, \bibinfo {author} {\bibfnamefont {N.}~\bibnamefont {Kaina}},\
  and\ \bibinfo {author} {\bibfnamefont {R.}~\bibnamefont {Fleury}},\
  }\bibfield  {title} {\bibinfo {title} {Chiral waveguides for robust
  waveguiding at the deep subwavelength scale},\ }\href
  {https://doi.org/10.1103/PhysRevApplied.10.054069} {\bibfield  {journal}
  {\bibinfo  {journal} {Phys. Rev. Applied}\ }\textbf {\bibinfo {volume}
  {10}},\ \bibinfo {pages} {054069} (\bibinfo {year} {2018})}\BibitemShut
  {NoStop}%
\bibitem [{\citenamefont {Petersen}\ \emph {et~al.}(2014)\citenamefont
  {Petersen}, \citenamefont {Volz},\ and\ \citenamefont
  {Rauschenbeutel}}]{Petersen2014}%
  \BibitemOpen
  \bibfield  {author} {\bibinfo {author} {\bibfnamefont {J.}~\bibnamefont
  {Petersen}}, \bibinfo {author} {\bibfnamefont {J.}~\bibnamefont {Volz}},\
  and\ \bibinfo {author} {\bibfnamefont {A.}~\bibnamefont {Rauschenbeutel}},\
  }\bibfield  {title} {\bibinfo {title} {Chiral nanophotonic waveguide
  interface based on spin-orbit interaction of light},\ }\href
  {https://doi.org/10.1126/science.1257671} {\bibfield  {journal} {\bibinfo
  {journal} {Science}\ }\textbf {\bibinfo {volume} {346}},\ \bibinfo {pages}
  {67} (\bibinfo {year} {2014})}\BibitemShut {NoStop}%
\bibitem [{\citenamefont {Yao}\ and\ \citenamefont
  {Hughes}(2009{\natexlab{b}})}]{PhysRevB.80.165128}%
  \BibitemOpen
  \bibfield  {author} {\bibinfo {author} {\bibfnamefont {P.}~\bibnamefont
  {Yao}}\ and\ \bibinfo {author} {\bibfnamefont {S.}~\bibnamefont {Hughes}},\
  }\bibfield  {title} {\bibinfo {title} {Controlled cavity qed and
  single-photon emission using a photonic-crystal waveguide cavity system},\
  }\href {https://doi.org/10.1103/PhysRevB.80.165128} {\bibfield  {journal}
  {\bibinfo  {journal} {Phys. Rev. B}\ }\textbf {\bibinfo {volume} {80}},\
  \bibinfo {pages} {165128} (\bibinfo {year} {2009}{\natexlab{b}})}\BibitemShut
  {NoStop}%
\bibitem [{\citenamefont {Pichler}\ \emph {et~al.}(2017)\citenamefont
  {Pichler}, \citenamefont {Choi}, \citenamefont {Zoller},\ and\ \citenamefont
  {Lukin}}]{Pichler11362}%
  \BibitemOpen
  \bibfield  {author} {\bibinfo {author} {\bibfnamefont {H.}~\bibnamefont
  {Pichler}}, \bibinfo {author} {\bibfnamefont {S.}~\bibnamefont {Choi}},
  \bibinfo {author} {\bibfnamefont {P.}~\bibnamefont {Zoller}},\ and\ \bibinfo
  {author} {\bibfnamefont {M.~D.}\ \bibnamefont {Lukin}},\ }\bibfield  {title}
  {\bibinfo {title} {Universal photonic quantum computation via time-delayed
  feedback},\ }\href {https://doi.org/10.1073/pnas.1711003114} {\bibfield
  {journal} {\bibinfo  {journal} {PNAS}\ }\textbf {\bibinfo {volume} {114}},\
  \bibinfo {pages} {11362} (\bibinfo {year} {2017})}\BibitemShut {NoStop}%
\bibitem [{\citenamefont {McCulloch}(2007)}]{mcculloch_density-matrix_2007}%
  \BibitemOpen
  \bibfield  {author} {\bibinfo {author} {\bibfnamefont {I.~P.}\ \bibnamefont
  {McCulloch}},\ }\bibfield  {title} {\bibinfo {title} {From density-matrix
  renormalization group to matrix product states},\ }\href
  {https://doi.org/10.1088/1742-5468/2007/10/p10014} {\bibfield  {journal}
  {\bibinfo  {journal} {Journal of Statistical Mechanics}\ }\textbf {\bibinfo
  {volume} {2007}}\bibinfo  {number} { (10)},\ \bibinfo {pages}
  {P10014}}\BibitemShut {NoStop}%
\bibitem [{\citenamefont {Woolfe}(2015)}]{woolfe_matrix_2015}%
  \BibitemOpen
\bibfield  {number} {  }\bibfield  {author} {\bibinfo {author} {\bibfnamefont
  {K.}~\bibnamefont {Woolfe}},\ }{\selectlanguage {en}\emph {\bibinfo {title}
  {Matrix product operator simulations of quantum algorithms}}},\ \href
  {http://minerva-access.unimelb.edu.au/handle/11343/55335} {Ph.D. thesis},\
  \bibinfo  {school} {University of Melbourne} (\bibinfo {year}
  {2015})\BibitemShut {NoStop}%
\bibitem [{\citenamefont {Schollwöck}(2011)}]{schollwock_density-matrix_2011}%
  \BibitemOpen
  \bibfield  {author} {\bibinfo {author} {\bibfnamefont {U.}~\bibnamefont
  {Schollwöck}},\ }\bibfield  {title} {\bibinfo {title} {The density-matrix
  renormalization group in the age of matrix product states},\ }\href
  {https://doi.org/10.1016/j.aop.2010.09.012} {\bibfield  {journal} {\bibinfo
  {journal} {Annals of Physics}\ }\textbf {\bibinfo {volume} {326}},\ \bibinfo
  {pages} {96} (\bibinfo {year} {2011})}\BibitemShut {NoStop}%
\bibitem [{\citenamefont {Iblisdir}\ \emph {et~al.}(2007)\citenamefont
  {Iblisdir}, \citenamefont {Orús},\ and\ \citenamefont
  {Latorre}}]{iblisdir_matrix_2007}%
  \BibitemOpen
  \bibfield  {author} {\bibinfo {author} {\bibfnamefont {S.}~\bibnamefont
  {Iblisdir}}, \bibinfo {author} {\bibfnamefont {R.}~\bibnamefont {Orús}},\
  and\ \bibinfo {author} {\bibfnamefont {J.~I.}\ \bibnamefont {Latorre}},\
  }\bibfield  {title} {\bibinfo {title} {Matrix product states algorithms and
  continuous systems},\ }\href {https://doi.org/10.1103/PhysRevB.75.104305}
  {\bibfield  {journal} {\bibinfo  {journal} {Physical Review B}\ }\textbf
  {\bibinfo {volume} {75}},\ \bibinfo {pages} {104305} (\bibinfo {year}
  {2007})}\BibitemShut {NoStop}%
\bibitem [{\citenamefont {Pavarini}\ \emph {et~al.}(2013)\citenamefont
  {Pavarini}, \citenamefont {Koch},\ and\ \citenamefont
  {Schollwöck}}]{pavarini_emergent_2013}%
  \BibitemOpen
  \bibfield  {author} {\bibinfo {author} {\bibfnamefont {E.}~\bibnamefont
  {Pavarini}}, \bibinfo {author} {\bibfnamefont {E.}~\bibnamefont {Koch}},\
  and\ \bibinfo {author} {\bibfnamefont {U.}~\bibnamefont {Schollwöck}},\
  }\href@noop {} {{\selectlanguage {eng}\emph {\bibinfo {title} {Emergent
  phenomena in correlated matter: lecture notes of the {Autumn} {School}
  {Correlated} {Electrons} 2013 at {Forschungszentrum} {Jülich}, 23 – 27
  {September} 2013}}}},\ edited by\ \bibinfo {editor} {\bibnamefont {{Institute
  for Advanced Simulation}}},\ \bibinfo {series} {Schriften des
  {Forschungszentrums} {Jülich} {Reihe} {Modeling} and {Simulation}}\
  No.~\bibinfo {number} {3}\ (\bibinfo  {publisher} {Forschungszentrum
  Jülich},\ \bibinfo {address} {Jülich},\ \bibinfo {year} {2013})\BibitemShut
  {NoStop}%
\bibitem [{\citenamefont {Vanderstraeten}\ \emph {et~al.}(2019)\citenamefont
  {Vanderstraeten}, \citenamefont {Haegeman},\ and\ \citenamefont
  {Verstraete}}]{vanderstraeten_tangent-space_2019}%
  \BibitemOpen
  \bibfield  {author} {\bibinfo {author} {\bibfnamefont {L.}~\bibnamefont
  {Vanderstraeten}}, \bibinfo {author} {\bibfnamefont {J.}~\bibnamefont
  {Haegeman}},\ and\ \bibinfo {author} {\bibfnamefont {F.}~\bibnamefont
  {Verstraete}},\ }\bibfield  {title} {\bibinfo {title} {Tangent-space methods
  for uniform matrix product states},\ }\href
  {https://doi.org/10.21468/SciPostPhysLectNotes.7} {\bibfield  {journal}
  {\bibinfo  {journal} {SciPost Physics Lecture Notes}\ ,\ \bibinfo {pages}
  {7}} (\bibinfo {year} {2019})}\BibitemShut {NoStop}%
\bibitem [{\citenamefont {Hubig}\ \emph {et~al.}(2017)\citenamefont {Hubig},
  \citenamefont {McCulloch},\ and\ \citenamefont {Schollw\"ock}}]{genericmpo}%
  \BibitemOpen
  \bibfield  {author} {\bibinfo {author} {\bibfnamefont {C.}~\bibnamefont
  {Hubig}}, \bibinfo {author} {\bibfnamefont {I.~P.}\ \bibnamefont
  {McCulloch}},\ and\ \bibinfo {author} {\bibfnamefont {U.}~\bibnamefont
  {Schollw\"ock}},\ }\bibfield  {title} {\bibinfo {title} {Generic construction
  of efficient matrix product operators},\ }\href
  {https://doi.org/10.1103/PhysRevB.95.035129} {\bibfield  {journal} {\bibinfo
  {journal} {Phys. Rev. B}\ }\textbf {\bibinfo {volume} {95}},\ \bibinfo
  {pages} {035129} (\bibinfo {year} {2017})}\BibitemShut {NoStop}%
\bibitem [{\citenamefont {Suba{\c{s}}{\i}}\ \emph {et~al.}(2019)\citenamefont
  {Suba{\c{s}}{\i}}, \citenamefont {Cincio},\ and\ \citenamefont
  {Coles}}]{Suba2019}%
  \BibitemOpen
  \bibfield  {author} {\bibinfo {author} {\bibfnamefont {Y.}~\bibnamefont
  {Suba{\c{s}}{\i}}}, \bibinfo {author} {\bibfnamefont {L.}~\bibnamefont
  {Cincio}},\ and\ \bibinfo {author} {\bibfnamefont {P.~J.}\ \bibnamefont
  {Coles}},\ }\bibfield  {title} {\bibinfo {title} {Entanglement spectroscopy
  with a depth-two quantum circuit},\ }\href
  {https://doi.org/10.1088/1751-8121/aaf54d} {\bibfield  {journal} {\bibinfo
  {journal} {Journal of Physics A}\ }\textbf {\bibinfo {volume} {52}},\
  \bibinfo {pages} {044001} (\bibinfo {year} {2019})}\BibitemShut {NoStop}%
\bibitem [{\citenamefont {Nielsen}\ \emph {et~al.}(2003)\citenamefont
  {Nielsen}, \citenamefont {Dawson}, \citenamefont {Dodd}, \citenamefont
  {Gilchrist}, \citenamefont {Mortimer}, \citenamefont {Osborne}, \citenamefont
  {Bremner}, \citenamefont {Harrow},\ and\ \citenamefont
  {Hines}}]{nielsen_quantum_2003}%
  \BibitemOpen
  \bibfield  {author} {\bibinfo {author} {\bibfnamefont {M.~A.}\ \bibnamefont
  {Nielsen}}, \bibinfo {author} {\bibfnamefont {C.~M.}\ \bibnamefont {Dawson}},
  \bibinfo {author} {\bibfnamefont {J.~L.}\ \bibnamefont {Dodd}}, \bibinfo
  {author} {\bibfnamefont {A.}~\bibnamefont {Gilchrist}}, \bibinfo {author}
  {\bibfnamefont {D.}~\bibnamefont {Mortimer}}, \bibinfo {author}
  {\bibfnamefont {T.~J.}\ \bibnamefont {Osborne}}, \bibinfo {author}
  {\bibfnamefont {M.~J.}\ \bibnamefont {Bremner}}, \bibinfo {author}
  {\bibfnamefont {A.~W.}\ \bibnamefont {Harrow}},\ and\ \bibinfo {author}
  {\bibfnamefont {A.}~\bibnamefont {Hines}},\ }\bibfield  {title}
  {{\selectlanguage {en}\bibinfo {title} {Quantum dynamics as a physical
  resource}},\ }\href {https://doi.org/10.1103/PhysRevA.67.052301} {\bibfield
  {journal} {\bibinfo  {journal} {Physical Review A}\ }\textbf {\bibinfo
  {volume} {67}},\ \bibinfo {pages} {052301} (\bibinfo {year}
  {2003})}\BibitemShut {NoStop}%
\bibitem [{\citenamefont {Band}\ and\ \citenamefont
  {Avishai}(2013)}]{band_quantum_2013}%
  \BibitemOpen
  \bibfield  {author} {\bibinfo {author} {\bibfnamefont {Y.~B.}\ \bibnamefont
  {Band}}\ and\ \bibinfo {author} {\bibfnamefont {Y.}~\bibnamefont {Avishai}},\
  }\href@noop {} {\emph {\bibinfo {title} {Quantum mechanics with applications
  to nanotechnology and information science}}},\ \bibinfo {edition} {1st}\ ed.\
  (\bibinfo  {publisher} {Academic Press},\ \bibinfo {address} {Amsterdam ; New
  York},\ \bibinfo {year} {2013})\BibitemShut {NoStop}%
\bibitem [{\citenamefont {S{\'{a}}nchez-Burillo}\ \emph
  {et~al.}(2015)\citenamefont {S{\'{a}}nchez-Burillo}, \citenamefont
  {Garc{\'{\i}}a-Ripoll}, \citenamefont {Mart{\'{\i}}n-Moreno},\ and\
  \citenamefont {Zueco}}]{SnchezBurillo2015}%
  \BibitemOpen
  \bibfield  {author} {\bibinfo {author} {\bibfnamefont {E.}~\bibnamefont
  {S{\'{a}}nchez-Burillo}}, \bibinfo {author} {\bibfnamefont {J.}~\bibnamefont
  {Garc{\'{\i}}a-Ripoll}}, \bibinfo {author} {\bibfnamefont {L.}~\bibnamefont
  {Mart{\'{\i}}n-Moreno}},\ and\ \bibinfo {author} {\bibfnamefont
  {D.}~\bibnamefont {Zueco}},\ }\bibfield  {title} {\bibinfo {title} {Nonlinear
  quantum optics in the (ultra)strong light{\textendash}matter coupling},\
  }\href {https://doi.org/10.1039/c4fd00206g} {\bibfield  {journal} {\bibinfo
  {journal} {Faraday Discussions}\ }\textbf {\bibinfo {volume} {178}},\
  \bibinfo {pages} {335} (\bibinfo {year} {2015})}\BibitemShut {NoStop}%
\bibitem [{\citenamefont {Xu}\ and\ \citenamefont {Fan}(2018)}]{Xu2018}%
  \BibitemOpen
  \bibfield  {author} {\bibinfo {author} {\bibfnamefont {S.}~\bibnamefont
  {Xu}}\ and\ \bibinfo {author} {\bibfnamefont {S.}~\bibnamefont {Fan}},\
  }\bibfield  {title} {\bibinfo {title} {Generate tensor network state by
  sequential single-photon scattering in waveguide {QED} systems},\ }\href
  {https://doi.org/10.1063/1.5044248} {\bibfield  {journal} {\bibinfo
  {journal} {{APL} Photonics}\ }\textbf {\bibinfo {volume} {3}},\ \bibinfo
  {pages} {116102} (\bibinfo {year} {2018})}\BibitemShut {NoStop}%
\bibitem [{\citenamefont {Pfeifer}\ \emph {et~al.}(2015)\citenamefont
  {Pfeifer}, \citenamefont {Evenbly}, \citenamefont {Singh},\ and\
  \citenamefont {Vidal}}]{pfeifer_ncon:_2015}%
  \BibitemOpen
  \bibfield  {author} {\bibinfo {author} {\bibfnamefont {R.~N.~C.}\
  \bibnamefont {Pfeifer}}, \bibinfo {author} {\bibfnamefont {G.}~\bibnamefont
  {Evenbly}}, \bibinfo {author} {\bibfnamefont {S.}~\bibnamefont {Singh}},\
  and\ \bibinfo {author} {\bibfnamefont {G.}~\bibnamefont {Vidal}},\ }\bibfield
   {title} {\bibinfo {title} {{NCON}: {A} tensor network contractor for
  {MATLAB}},\ }\href {http://arxiv.org/abs/1402.0939} {\bibfield  {journal}
  {\bibinfo  {journal} {arXiv:1402.0939}\ } (\bibinfo {year}
  {2015})}\BibitemShut {NoStop}%
\bibitem [{\citenamefont {Dalibard}\ \emph {et~al.}(1992)\citenamefont
  {Dalibard}, \citenamefont {Castin},\ and\ \citenamefont
  {M{\o}lmer}}]{Dalibard92}%
  \BibitemOpen
  \bibfield  {author} {\bibinfo {author} {\bibfnamefont {J.}~\bibnamefont
  {Dalibard}}, \bibinfo {author} {\bibfnamefont {Y.}~\bibnamefont {Castin}},\
  and\ \bibinfo {author} {\bibfnamefont {K.}~\bibnamefont {M{\o}lmer}},\
  }\bibfield  {title} {\bibinfo {title} {Wave-function approach to dissipative
  processes in quantum optics},\ }\href
  {https://doi.org/10.1103/PhysRevLett.68.580} {\bibfield  {journal} {\bibinfo
  {journal} {Phys. Rev. Lett.}\ }\textbf {\bibinfo {volume} {68}},\ \bibinfo
  {pages} {580} (\bibinfo {year} {1992})}\BibitemShut {NoStop}%
\bibitem [{\citenamefont {Tian}\ and\ \citenamefont
  {Carmichael}(1992)}]{Carmichael92}%
  \BibitemOpen
  \bibfield  {author} {\bibinfo {author} {\bibfnamefont {L.}~\bibnamefont
  {Tian}}\ and\ \bibinfo {author} {\bibfnamefont {H.~J.}\ \bibnamefont
  {Carmichael}},\ }\bibfield  {title} {\bibinfo {title} {Quantum trajectory
  simulations of two-state behavior in an optical cavity containing one atom},\
  }\href {https://doi.org/10.1103/PhysRevA.46.R6801} {\bibfield  {journal}
  {\bibinfo  {journal} {Phys. Rev. A}\ }\textbf {\bibinfo {volume} {46}},\
  \bibinfo {pages} {R6801} (\bibinfo {year} {1992})}\BibitemShut {NoStop}%
\bibitem [{\citenamefont {Dum}\ \emph {et~al.}(1992)\citenamefont {Dum},
  \citenamefont {Zoller},\ and\ \citenamefont {Ritsch}}]{Dum92a}%
  \BibitemOpen
  \bibfield  {author} {\bibinfo {author} {\bibfnamefont {R.}~\bibnamefont
  {Dum}}, \bibinfo {author} {\bibfnamefont {P.}~\bibnamefont {Zoller}},\ and\
  \bibinfo {author} {\bibfnamefont {H.}~\bibnamefont {Ritsch}},\ }\bibfield
  {title} {\bibinfo {title} {Monte carlo simulation of the atomic master
  equation for spontaneous emission},\ }\href
  {https://doi.org/10.1103/PhysRevA.45.4879} {\bibfield  {journal} {\bibinfo
  {journal} {Phys. Rev. A}\ }\textbf {\bibinfo {volume} {45}},\ \bibinfo
  {pages} {4879} (\bibinfo {year} {1992})}\BibitemShut {NoStop}%
\bibitem [{\citenamefont {Sinha}\ \emph {et~al.}(2020)\citenamefont {Sinha},
  \citenamefont {Meystre}, \citenamefont {Goldschmidt}, \citenamefont {Fatemi},
  \citenamefont {Rolston},\ and\ \citenamefont
  {Solano}}]{PhysRevLett.124.043603}%
  \BibitemOpen
  \bibfield  {author} {\bibinfo {author} {\bibfnamefont {K.}~\bibnamefont
  {Sinha}}, \bibinfo {author} {\bibfnamefont {P.}~\bibnamefont {Meystre}},
  \bibinfo {author} {\bibfnamefont {E.~A.}\ \bibnamefont {Goldschmidt}},
  \bibinfo {author} {\bibfnamefont {F.~K.}\ \bibnamefont {Fatemi}}, \bibinfo
  {author} {\bibfnamefont {S.~L.}\ \bibnamefont {Rolston}},\ and\ \bibinfo
  {author} {\bibfnamefont {P.}~\bibnamefont {Solano}},\ }\bibfield  {title}
  {\bibinfo {title} {Non-markovian collective emission from macroscopically
  separated emitters},\ }\href {https://doi.org/10.1103/PhysRevLett.124.043603}
  {\bibfield  {journal} {\bibinfo  {journal} {Phys. Rev. Lett.}\ }\textbf
  {\bibinfo {volume} {124}},\ \bibinfo {pages} {043603} (\bibinfo {year}
  {2020})}\BibitemShut {NoStop}%
\bibitem [{\citenamefont {Kabuss}\ \emph {et~al.}(2015)\citenamefont {Kabuss},
  \citenamefont {Krimer}, \citenamefont {Rotter}, \citenamefont {Stannigel},
  \citenamefont {Knorr},\ and\ \citenamefont {Carmele}}]{PhysRevA.92.053801}%
  \BibitemOpen
  \bibfield  {author} {\bibinfo {author} {\bibfnamefont {J.}~\bibnamefont
  {Kabuss}}, \bibinfo {author} {\bibfnamefont {D.~O.}\ \bibnamefont {Krimer}},
  \bibinfo {author} {\bibfnamefont {S.}~\bibnamefont {Rotter}}, \bibinfo
  {author} {\bibfnamefont {K.}~\bibnamefont {Stannigel}}, \bibinfo {author}
  {\bibfnamefont {A.}~\bibnamefont {Knorr}},\ and\ \bibinfo {author}
  {\bibfnamefont {A.}~\bibnamefont {Carmele}},\ }\bibfield  {title} {\bibinfo
  {title} {Analytical study of quantum-feedback-enhanced rabi oscillations},\
  }\href {https://doi.org/10.1103/PhysRevA.92.053801} {\bibfield  {journal}
  {\bibinfo  {journal} {Phys. Rev. A}\ }\textbf {\bibinfo {volume} {92}},\
  \bibinfo {pages} {053801} (\bibinfo {year} {2015})}\BibitemShut {NoStop}%
\bibitem [{\citenamefont {John}\ and\ \citenamefont
  {Quang}(1994)}]{PhysRevA.50.1764}%
  \BibitemOpen
  \bibfield  {author} {\bibinfo {author} {\bibfnamefont {S.}~\bibnamefont
  {John}}\ and\ \bibinfo {author} {\bibfnamefont {T.}~\bibnamefont {Quang}},\
  }\bibfield  {title} {\bibinfo {title} {Spontaneous emission near the edge of
  a photonic band gap},\ }\href {https://doi.org/10.1103/PhysRevA.50.1764}
  {\bibfield  {journal} {\bibinfo  {journal} {Phys. Rev. A}\ }\textbf {\bibinfo
  {volume} {50}},\ \bibinfo {pages} {1764} (\bibinfo {year}
  {1994})}\BibitemShut {NoStop}%
\bibitem [{\citenamefont {Nabiev}\ \emph {et~al.}(1993)\citenamefont {Nabiev},
  \citenamefont {Yeh},\ and\ \citenamefont
  {Sanchez-Mondragon}}]{PhysRevA.47.3380}%
  \BibitemOpen
  \bibfield  {author} {\bibinfo {author} {\bibfnamefont {R.~F.}\ \bibnamefont
  {Nabiev}}, \bibinfo {author} {\bibfnamefont {P.}~\bibnamefont {Yeh}},\ and\
  \bibinfo {author} {\bibfnamefont {J.~J.}\ \bibnamefont {Sanchez-Mondragon}},\
  }\bibfield  {title} {\bibinfo {title} {Dynamics of the spontaneous emission
  of an atom into the photon-density-of-states gap: Solvable
  quantum-electrodynamical model},\ }\href
  {https://doi.org/10.1103/PhysRevA.47.3380} {\bibfield  {journal} {\bibinfo
  {journal} {Phys. Rev. A}\ }\textbf {\bibinfo {volume} {47}},\ \bibinfo
  {pages} {3380} (\bibinfo {year} {1993})}\BibitemShut {NoStop}%
\bibitem [{\citenamefont {Kristensen}\ \emph {et~al.}(2008)\citenamefont
  {Kristensen}, \citenamefont {Koenderink}, \citenamefont {Lodahl},
  \citenamefont {Tromborg},\ and\ \citenamefont {M{\o}rk}}]{Kristensen2008}%
  \BibitemOpen
  \bibfield  {author} {\bibinfo {author} {\bibfnamefont {P.}~\bibnamefont
  {Kristensen}}, \bibinfo {author} {\bibfnamefont {A.~F.}\ \bibnamefont
  {Koenderink}}, \bibinfo {author} {\bibfnamefont {P.}~\bibnamefont {Lodahl}},
  \bibinfo {author} {\bibfnamefont {B.}~\bibnamefont {Tromborg}},\ and\
  \bibinfo {author} {\bibfnamefont {J.}~\bibnamefont {M{\o}rk}},\ }\bibfield
  {title} {\bibinfo {title} {Fractional decay of quantum dots in real photonic
  crystals},\ }\href {https://doi.org/10.1364/ol.33.001557} {\bibfield
  {journal} {\bibinfo  {journal} {Optics Letters}\ }\textbf {\bibinfo {volume}
  {33}},\ \bibinfo {pages} {1557} (\bibinfo {year} {2008})}\BibitemShut
  {NoStop}%
\bibitem [{\citenamefont {Hughes}\ \emph {et~al.}(2005)\citenamefont {Hughes},
  \citenamefont {Ramunno}, \citenamefont {Young},\ and\ \citenamefont
  {Sipe}}]{PhysRevLett.94.033903}%
  \BibitemOpen
  \bibfield  {author} {\bibinfo {author} {\bibfnamefont {S.}~\bibnamefont
  {Hughes}}, \bibinfo {author} {\bibfnamefont {L.}~\bibnamefont {Ramunno}},
  \bibinfo {author} {\bibfnamefont {J.~F.}\ \bibnamefont {Young}},\ and\
  \bibinfo {author} {\bibfnamefont {J.~E.}\ \bibnamefont {Sipe}},\ }\bibfield
  {title} {\bibinfo {title} {Extrinsic optical scattering loss in photonic
  crystal waveguides: Role of fabrication disorder and photon group velocity},\
  }\href {https://doi.org/10.1103/PhysRevLett.94.033903} {\bibfield  {journal}
  {\bibinfo  {journal} {Phys. Rev. Lett.}\ }\textbf {\bibinfo {volume} {94}},\
  \bibinfo {pages} {033903} (\bibinfo {year} {2005})}\BibitemShut {NoStop}%
\bibitem [{\citenamefont {Hughes}(2007)}]{PhysRevLett.98.083603}%
  \BibitemOpen
  \bibfield  {author} {\bibinfo {author} {\bibfnamefont {S.}~\bibnamefont
  {Hughes}},\ }\bibfield  {title} {\bibinfo {title} {Coupled-cavity qed using
  planar photonic crystals},\ }\href
  {https://doi.org/10.1103/PhysRevLett.98.083603} {\bibfield  {journal}
  {\bibinfo  {journal} {Phys. Rev. Lett.}\ }\textbf {\bibinfo {volume} {98}},\
  \bibinfo {pages} {083603} (\bibinfo {year} {2007})}\BibitemShut {NoStop}%
\bibitem [{\citenamefont {Dinc}\ and\ \citenamefont
  {Bra{\'{n}}czyk}(2019)}]{Dinc2019}%
  \BibitemOpen
  \bibfield  {author} {\bibinfo {author} {\bibfnamefont {F.}~\bibnamefont
  {Dinc}}\ and\ \bibinfo {author} {\bibfnamefont {A.~M.}\ \bibnamefont
  {Bra{\'{n}}czyk}},\ }\bibfield  {title} {\bibinfo {title} {Non-markovian
  super-superradiance in a linear chain of up to 100 qubits},\ }\href
  {https://doi.org/10.1103/physrevresearch.1.032042} {\bibfield  {journal}
  {\bibinfo  {journal} {Phys. Rev. Research}\ }\textbf {\bibinfo {volume}
  {1}},\ \bibinfo {pages} {032042} (\bibinfo {year} {2019})}\BibitemShut
  {NoStop}%
\bibitem [{\citenamefont {Zhang}\ and\ \citenamefont
  {M{\o}lmer}(2019)}]{Zhang2019}%
  \BibitemOpen
  \bibfield  {author} {\bibinfo {author} {\bibfnamefont {Y.-X.}\ \bibnamefont
  {Zhang}}\ and\ \bibinfo {author} {\bibfnamefont {K.}~\bibnamefont
  {M{\o}lmer}},\ }\bibfield  {title} {\bibinfo {title} {Theory of subradiant
  states of a one-dimensional two-level atom chain},\ }\href
  {https://doi.org/10.1103/physrevlett.122.203605} {\bibfield  {journal}
  {\bibinfo  {journal} {Physical Review Letters}\ }\textbf {\bibinfo {volume}
  {122}},\ \bibinfo {pages} {203605} (\bibinfo {year} {2019})}\BibitemShut
  {NoStop}%
\bibitem [{\citenamefont {Rom\'an-Roche}\ \emph {et~al.}(2020)\citenamefont
  {Rom\'an-Roche}, \citenamefont {S\'anchez-Burillo},\ and\ \citenamefont
  {Zueco}}]{PhysRevA.102.023702}%
  \BibitemOpen
  \bibfield  {author} {\bibinfo {author} {\bibfnamefont {J.}~\bibnamefont
  {Rom\'an-Roche}}, \bibinfo {author} {\bibfnamefont {E.}~\bibnamefont
  {S\'anchez-Burillo}},\ and\ \bibinfo {author} {\bibfnamefont
  {D.}~\bibnamefont {Zueco}},\ }\bibfield  {title} {\bibinfo {title} {Bound
  states in ultrastrong waveguide {QED}},\ }\href
  {https://doi.org/10.1103/PhysRevA.102.023702} {\bibfield  {journal} {\bibinfo
   {journal} {Phys. Rev. A}\ }\textbf {\bibinfo {volume} {102}},\ \bibinfo
  {pages} {023702} (\bibinfo {year} {2020})}\BibitemShut {NoStop}%
\bibitem [{\citenamefont {Mukhopadhyay}\ and\ \citenamefont
  {Agarwal}(2019)}]{PhysRevA.100.013812}%
  \BibitemOpen
  \bibfield  {author} {\bibinfo {author} {\bibfnamefont {D.}~\bibnamefont
  {Mukhopadhyay}}\ and\ \bibinfo {author} {\bibfnamefont {G.~S.}\ \bibnamefont
  {Agarwal}},\ }\bibfield  {title} {\bibinfo {title} {Multiple fano
  interferences due to waveguide-mediated phase coupling between atoms},\
  }\href {https://doi.org/10.1103/PhysRevA.100.013812} {\bibfield  {journal}
  {\bibinfo  {journal} {Phys. Rev. A}\ }\textbf {\bibinfo {volume} {100}},\
  \bibinfo {pages} {013812} (\bibinfo {year} {2019})}\BibitemShut {NoStop}%
\bibitem [{\citenamefont {Cheng}\ \emph {et~al.}(2017)\citenamefont {Cheng},
  \citenamefont {Xu},\ and\ \citenamefont {Agarwal}}]{Cheng2017}%
  \BibitemOpen
  \bibfield  {author} {\bibinfo {author} {\bibfnamefont {M.-T.}\ \bibnamefont
  {Cheng}}, \bibinfo {author} {\bibfnamefont {J.}~\bibnamefont {Xu}},\ and\
  \bibinfo {author} {\bibfnamefont {G.~S.}\ \bibnamefont {Agarwal}},\
  }\bibfield  {title} {\bibinfo {title} {Waveguide transport mediated by strong
  coupling with atoms},\ }\href {https://doi.org/10.1103/physreva.95.053807}
  {\bibfield  {journal} {\bibinfo  {journal} {Physical Review A}\ }\textbf
  {\bibinfo {volume} {95}},\ \bibinfo {pages} {053807} (\bibinfo {year}
  {2017})}\BibitemShut {NoStop}%
\bibitem [{\citenamefont {Nielsen}\ and\ \citenamefont
  {Chuang}(2009)}]{Nielsen2009}%
  \BibitemOpen
  \bibfield  {author} {\bibinfo {author} {\bibfnamefont {M.~A.}\ \bibnamefont
  {Nielsen}}\ and\ \bibinfo {author} {\bibfnamefont {I.~L.}\ \bibnamefont
  {Chuang}},\ }\href {https://doi.org/10.1017/cbo9780511976667} {\emph
  {\bibinfo {title} {Quantum Computation and Quantum Information}}}\ (\bibinfo
  {publisher} {Cambridge University Press},\ \bibinfo {year}
  {2009})\BibitemShut {NoStop}%
\bibitem [{\citenamefont {Carmele}\ \emph {et~al.}(2020)\citenamefont
  {Carmele}, \citenamefont {Nemet}, \citenamefont {Canela},\ and\ \citenamefont
  {Parkins}}]{PhysRevResearch.2.013238}%
  \BibitemOpen
  \bibfield  {author} {\bibinfo {author} {\bibfnamefont {A.}~\bibnamefont
  {Carmele}}, \bibinfo {author} {\bibfnamefont {N.}~\bibnamefont {Nemet}},
  \bibinfo {author} {\bibfnamefont {V.}~\bibnamefont {Canela}},\ and\ \bibinfo
  {author} {\bibfnamefont {S.}~\bibnamefont {Parkins}},\ }\bibfield  {title}
  {\bibinfo {title} {Pronounced non-markovian features in multiply excited,
  multiple emitter waveguide qed: Retardation induced anomalous population
  trapping},\ }\href {https://doi.org/10.1103/PhysRevResearch.2.013238}
  {\bibfield  {journal} {\bibinfo  {journal} {Phys. Rev. Research}\ }\textbf
  {\bibinfo {volume} {2}},\ \bibinfo {pages} {013238} (\bibinfo {year}
  {2020})}\BibitemShut {NoStop}%
\bibitem [{\citenamefont {Mirhosseini}\ \emph {et~al.}(2019)\citenamefont
  {Mirhosseini}, \citenamefont {Kim}, \citenamefont {Zhang}, \citenamefont
  {Sipahigil}, \citenamefont {Dieterle}, \citenamefont {Keller}, \citenamefont
  {Asenjo-Garcia}, \citenamefont {Chang},\ and\ \citenamefont
  {Painter}}]{Mirhosseini2019}%
  \BibitemOpen
  \bibfield  {author} {\bibinfo {author} {\bibfnamefont {M.}~\bibnamefont
  {Mirhosseini}}, \bibinfo {author} {\bibfnamefont {E.}~\bibnamefont {Kim}},
  \bibinfo {author} {\bibfnamefont {X.}~\bibnamefont {Zhang}}, \bibinfo
  {author} {\bibfnamefont {A.}~\bibnamefont {Sipahigil}}, \bibinfo {author}
  {\bibfnamefont {P.~B.}\ \bibnamefont {Dieterle}}, \bibinfo {author}
  {\bibfnamefont {A.~J.}\ \bibnamefont {Keller}}, \bibinfo {author}
  {\bibfnamefont {A.}~\bibnamefont {Asenjo-Garcia}}, \bibinfo {author}
  {\bibfnamefont {D.~E.}\ \bibnamefont {Chang}},\ and\ \bibinfo {author}
  {\bibfnamefont {O.}~\bibnamefont {Painter}},\ }\bibfield  {title} {\bibinfo
  {title} {Cavity quantum electrodynamics with atom-like mirrors},\ }\href
  {https://doi.org/10.1038/s41586-019-1196-1} {\bibfield  {journal} {\bibinfo
  {journal} {Nature}\ }\textbf {\bibinfo {volume} {569}},\ \bibinfo {pages}
  {692} (\bibinfo {year} {2019})}\BibitemShut {NoStop}%
\bibitem [{\citenamefont {Albrecht}\ \emph {et~al.}(2019)\citenamefont
  {Albrecht}, \citenamefont {Henriet}, \citenamefont {Asenjo-Garcia},
  \citenamefont {Dieterle}, \citenamefont {Painter},\ and\ \citenamefont
  {Chang}}]{Albrecht_2019}%
  \BibitemOpen
  \bibfield  {author} {\bibinfo {author} {\bibfnamefont {A.}~\bibnamefont
  {Albrecht}}, \bibinfo {author} {\bibfnamefont {L.}~\bibnamefont {Henriet}},
  \bibinfo {author} {\bibfnamefont {A.}~\bibnamefont {Asenjo-Garcia}}, \bibinfo
  {author} {\bibfnamefont {P.~B.}\ \bibnamefont {Dieterle}}, \bibinfo {author}
  {\bibfnamefont {O.}~\bibnamefont {Painter}},\ and\ \bibinfo {author}
  {\bibfnamefont {D.~E.}\ \bibnamefont {Chang}},\ }\bibfield  {title} {\bibinfo
  {title} {Subradiant states of quantum bits coupled to a one-dimensional
  waveguide},\ }\href {https://doi.org/10.1088/1367-2630/ab0134} {\bibfield
  {journal} {\bibinfo  {journal} {New Journal of Physics}\ }\textbf {\bibinfo
  {volume} {21}},\ \bibinfo {pages} {025003} (\bibinfo {year}
  {2019})}\BibitemShut {NoStop}%
\bibitem [{\citenamefont {Finsterh\"{o}lzl}\ \emph {et~al.}(2020)\citenamefont
  {Finsterh\"{o}lzl}, \citenamefont {Katzer}, \citenamefont {Knorr},\ and\
  \citenamefont {Carmele}}]{Finsterhlzl2020}%
  \BibitemOpen
  \bibfield  {author} {\bibinfo {author} {\bibfnamefont {R.}~\bibnamefont
  {Finsterh\"{o}lzl}}, \bibinfo {author} {\bibfnamefont {M.}~\bibnamefont
  {Katzer}}, \bibinfo {author} {\bibfnamefont {A.}~\bibnamefont {Knorr}},\ and\
  \bibinfo {author} {\bibfnamefont {A.}~\bibnamefont {Carmele}},\ }\bibfield
  {title} {\bibinfo {title} {Using matrix-product states for open quantum
  many-body systems: Efficient algorithms for markovian and non-markovian
  time-evolution},\ }\href {https://doi.org/10.3390/e22090984} {\bibfield
  {journal} {\bibinfo  {journal} {Entropy}\ }\textbf {\bibinfo {volume} {22}},\
  \bibinfo {pages} {984} (\bibinfo {year} {2020})}\BibitemShut {NoStop}%
\bibitem [{\citenamefont {Masson}\ and\ \citenamefont
  {Asenjo-Garcia}(2019)}]{Masson2019AtomicWaveguideQE}%
  \BibitemOpen
  \bibfield  {author} {\bibinfo {author} {\bibfnamefont {S.~J.}\ \bibnamefont
  {Masson}}\ and\ \bibinfo {author} {\bibfnamefont {A.}~\bibnamefont
  {Asenjo-Garcia}},\ }\href@noop {} {\bibinfo {title} {Atomic-waveguide quantum
  electrodynamics}} (\bibinfo {year} {2019}),\ \Eprint
  {https://arxiv.org/abs/arXiv:1912.06234} {arXiv:1912.06234} \BibitemShut
  {NoStop}%
\bibitem [{\citenamefont {Wang}\ \emph {et~al.}(2020)\citenamefont {Wang},
  \citenamefont {Jaako}, \citenamefont {Kirton},\ and\ \citenamefont
  {Rabl}}]{PhysRevLett.124.213601}%
  \BibitemOpen
  \bibfield  {author} {\bibinfo {author} {\bibfnamefont {Z.}~\bibnamefont
  {Wang}}, \bibinfo {author} {\bibfnamefont {T.}~\bibnamefont {Jaako}},
  \bibinfo {author} {\bibfnamefont {P.}~\bibnamefont {Kirton}},\ and\ \bibinfo
  {author} {\bibfnamefont {P.}~\bibnamefont {Rabl}},\ }\bibfield  {title}
  {\bibinfo {title} {Supercorrelated radiance in nonlinear photonic
  waveguides},\ }\href {https://doi.org/10.1103/PhysRevLett.124.213601}
  {\bibfield  {journal} {\bibinfo  {journal} {Phys. Rev. Lett.}\ }\textbf
  {\bibinfo {volume} {124}},\ \bibinfo {pages} {213601} (\bibinfo {year}
  {2020})}\BibitemShut {NoStop}%
\end{thebibliography}%

\end{document}